\documentclass[useAMS,usenatbib,a4paper]{mn2e}
\usepackage{times}
\usepackage{amssymb}
\usepackage{amsmath}
\usepackage[normalem]{ulem}
\usepackage{graphicx}
\usepackage{subfigure}
\usepackage{mathrsfs}
\usepackage{multicol}
\usepackage{float}
\usepackage{cancel}

\newcommand{\gsim}{\mathrel{\hbox{\rlap{\lower.55ex \hbox {$\sim$}}
                   \kern-.3em \raise.4ex \hbox{$>$}}}}
\newcommand{\lsim}{\mathrel{\hbox{\rlap{\lower.55ex \hbox {$\sim$}}
                   \kern-.3em \raise.4ex \hbox{$<$}}}}

\newcommand{\Msolar}{{\rm M_{\odot}}}   
\newcommand{\Qmin}{Q_{\rm min}}  
\newcommand{\betacrit}{\beta_{\rm crit}}  
\newcommand{\alphaSPH}{\alpha_{\rm SPH}}  
\newcommand{\betaSPH}{\beta_{\rm SPH}}  



\title[Convergence of fragmenting self-gravitating discs]{On the convergence of the critical cooling timescale for the fragmentation of self-gravitating discs}
\author[Farzana Meru and Matthew R. Bate]{Farzana Meru$^{1,2,3}$\thanks{farzana.meru@phys.ethz.ch} and Matthew R. Bate$^3$\\
$^1$Institut f\"ur Astronomie, ETH Z\"urich, Wolfgang-Pauli-Strasse 27, 8093 Z\"urich, Switzerland\\
$^2$Institut f\"ur Astronomie und Astrophysik, Universit\"at T\"ubingen, Auf der Morgenstelle 10, 72076 T\"ubingen, Germany\\
$^3$School of Physics, University of Exeter, Stocker Road, Exeter, EX4 4QL, UK\\
}

\date{\today}
\begin{document}
\maketitle

\begin{abstract}
We carry out simulations of gravitationally unstable discs using a Smoothed Particle Hydrodynamics (SPH) code and a grid-based hydrodynamics code, {\sc fargo}, to understand the previous non-convergent results reported by \cite{Meru_Bate_resolution}.  We obtain evidence that convergence with increasing resolution occurs with both SPH and {\sc fargo} and in both cases we find that the critical cooling timescale is larger than previously thought.  We show that SPH has a first-order convergence rate while {\sc fargo} converges with a second-order rate.  We show that the convergence of the critical cooling timescale for fragmentation depends largely on the numerical viscosity employed in both SPH and {\sc fargo}.  With SPH, particle velocity dispersion may also play a role.  We show that reducing the dissipation from the numerical viscosity leads to larger values of the critical cooling time at a given resolution.  For SPH, we find that the effect of the dissipation due to the numerical viscosity is somewhat larger than had previously been appreciated.  In particular, we show that using a quadratic term in the SPH artificial viscosity ($\betaSPH$) that is too low appears to lead to excess dissipation in gravitationally unstable discs, which may affect any results that sensitively depend on the thermodynamics, such as disc fragmentation.  We show that the two codes converge to values of the critical cooling timescale, $\betacrit > 20$ (for a ratio of specific heats of $\gamma=5/3$), and perhaps even as large as $\betacrit \approx 30$.  These are approximately $3-5$ times larger than has been found by most previous studies.  This is equivalent to a maximum gravitational stress that a disc can withstand without fragmenting of $\alpha_{\rm GI,crit}\ \approx 0.013-0.02$, which is much smaller than the values typically used in the literature.  It is therefore easier for self-gravitating discs to fragment than has been concluded from most past studies.
\end{abstract}


\section{Introduction}

Historically, there have been two key quantities that have been used to determine whether a self-gravitating disc is likely to fragment.  The first is the stability parameter \citep{Toomre_stability1964},

\begin{equation}
  \label{eq:Toomre}
  Q=\frac{c_{\rm s}\kappa_{\rm ep}}{\pi\Sigma G},
\end{equation}
where $c_{\rm s}$ is the sound speed in the disc, $\kappa_{\rm ep}$ is the epicyclic frequency, which for Keplerian discs is approximately equal to the angular frequency, $\Omega$, $\Sigma$ is the surface mass density and $G$ is the gravitational constant.  \cite{Toomre_stability1964} showed that for an infinitesimally thin disc to fragment, the stability parameter must be less than a critical value, $Q_{\rm crit} \approx 1$.
 
\cite{Gammie_betacool} showed that in addition to the stability criterion above, the disc must cool at a fast enough rate.  Using shearing sheet simulations, he showed that if the cooling timescale can be parametrized as

\begin{equation}
  \label{eq:beta}
  \beta = t_{\rm cool}\Omega,
\end{equation}
where

\begin{equation}
  \label{eq:tcool}
  t_{\rm cool} = u \left ( \frac{{\rm d}u_{\rm cool}}{{\rm d}t} \right )^{-1},
\end{equation}
$u$ is the specific internal energy and ${\rm d}u_{\rm cool}/{\rm d}t$ is the total specific cooling rate, then for fragmentation we require $\beta \lesssim \beta_{\rm crit}$.  According to \cite{Gammie_betacool}, $\beta_{\rm crit} \approx 3$ for a ratio of specific heats, $\gamma = 2$.  \cite*{Rice_beta_condition} carried out three-dimensional simulations using a Smoothed Particle Hydrodynamics (SPH) code and showed that this cooling parameter is dependent on the equation of state.  They showed that $\beta_{\rm crit} \approx 6-7$ for discs with $\gamma = 5/3$ and $\beta_{\rm crit} \approx 12-13$ for discs with $\gamma = 7/5$.

\cite{Gammie_betacool} and \cite{Rice_beta_condition} also showed that in a steady state disc where the dominant form of heating is that due to gravitational instabilities, since the gravitational stress in a disc can be linked to the cooling timescale in the disc using
\begin{equation}
\label{eq:stress}
\alpha_{\rm GI} = \frac{4}{9} \frac{1}{\gamma (\gamma -1)} \frac{1}{\beta},
\end{equation}
the rapid cooling required for fragmentation, $\betacrit$, can also be interpreted as a maximum gravitational stress that a disc can support without fragmenting, which they showed to be $\alpha_{\rm GI,crit} \approx 0.06$.

Recently, \cite{Meru_Bate_resolution} showed using SPH calculations of gravitationally unstable discs similar to those that have been performed by \cite{Rice_beta_condition} that the previous results on the critical cooling timescale had not converged.  In particular, they found that the critical value of the cooling timescale, $\beta_{\rm crit}$, below which a disc would fragment increased linearly with increasing spatial resolution.  This implied that the critical cooling rate might be much greater than that found from past studies (which would, for example, have implications for where in a real disc planets may form by the gravitational instability method).  It also opened the question of whether or not a critical cooling rate indeed exists.  Instead, a self-gravitating disc that is subject to a fixed cooling rate might fragment regardless of the value, given sufficient resolution (i.e. a disc may never be able to settle into a self-regulated state).

\cite{Lodato_Clarke_resolution} suggested that the non-convergent results may be an artefact of SPH artificially smoothing the density enhancements, or may be due to artificial viscosity if its effect was much larger than expected.  \cite{Rice_cooling_convergence} suggested that the implementation of cooling in SPH may be to blame for the lack of convergence.  However, \cite*{Paardekooper_convergence} showed using the two-dimensional grid-based hydrodynamics code, {\sc fargo}, that the non-convergent problem was not specific to SPH.  The source of non-convergence therefore cannot be constrained to SPH or to three-dimensional codes.  \cite{Paardekooper_convergence} suggested that the boundary between the turbulent inner disc region and the smooth outer disc region (a consequence of starting the simulations with smooth initial conditions) may be the cause of the non-convergent results presented by \cite{Meru_Bate_resolution}.

\cite{Bate_disc_frag_res} carried out SPH simulations of the collapse of molecular clouds to form protostars and discs.  For particular initial conditions that lead to disc fragmentation, he noted that higher resolution simulations resulted in more fragments.  Unlike the simulations of gravitationally unstable isolated protoplanetary discs discussed above, these simulations were of very early stage discs that formed prior to stellar core formation and were subject to rapid accretion from the surrounding molecular envelope.  However, the interesting aspect here is that the fragmentation is more prevalent in higher resolution simulations of discs modelled using both isolated discs as initial conditions and using a parametrized cooling function \citep{Meru_Bate_resolution} as well as discs formed in molecular cloud collapse simulations using radiative transfer where no such smooth initial conditions are involved \citep{Bate_disc_frag_res}.

\cite{Meru_Bate_resolution} expressed a concern about a lack of convergence with numerical resolution.  However, even if convergence with numerical resolution is achieved, convergence between different numerical models is also important, i.e. the result is not believable if two different codes that can, in principle, model the same physical processes, produce physically different results. 

In this paper, we present additional SPH results to those presented by \cite{Meru_Bate_resolution}.  Rather than confining our investigations to a single hydrodynamics code, we also carry out a code comparison by performing further calculations using the grid-based Eulerian hydrodynamics code, {\sc fargo}.  We particularly focus on the dependence of the critical cooling timescale on the artificial viscosities employed in both codes.

In Section~\ref{sec:numerics} we describe the numerical methods adopted and discuss how numerical viscosity may affect the critical cooling timescale in discs in Section~\ref{sec:num_visc}.  We describe the simulations performed and present our results in Sections~\ref{sec:sim} and~\ref{sec:results}, respectively.  We discuss and make conclusions in Sections~\ref{sec:disc} and~\ref{sec:conc}, respectively.

\section{Numerical methods}
\label{sec:numerics}

Our SPH simulations are carried out using the exact same code as that used by \cite{Meru_Bate_resolution}, originally developed by \cite{Benz1990}, further developed by \cite*{Bate_Bonnell_Price_sink_ptcls} and \cite{Price_Bate_MHD_h} and parallelised using both OpenMP and MPI (see \citealp{Meru_Bate_fragmentation} for details).  Our simulations with a grid-based code are carried out using the Fast Advection in Rotating Gaseous Objects ({\sc fargo}) two-dimensional fixed polar hydrodynamics code \citep{Masset_FARGO,Baruteau_Masset_FARGO1,Baruteau_Masset_FARGO2}.

We include the heating effects in the disc due to work done on the gas and artificial viscosity.  The cooling in the disc is taken into account using the cooling parameter, $\beta$ (equation~\ref{eq:beta}), first proposed by \cite{Gammie_betacool} which cools the gas on a timescale given by equation~\ref{eq:tcool}.  For the SPH simulations carried out in this paper, we ensure that the timestepping is limited by the following timestep criterion (in addition to the Courant condition, the force condition and the viscous timestep condition; see \citealp{SPH_Monaghan}):

\begin{equation}
  \Delta t \leq C \frac{\beta}{\Omega},
  \label{eq:tcool_timestep}
\end{equation}
where $C = 0.3$.  \cite{Meru_Bate_fragmentation} show that this condition is adequate to ensure that the fragmentation results are not affected by the timestepping imposed.  For the simulations performed using {\sc fargo}, the timestep constraint (using $C = 1$) is also included for all simulations involving $\beta \lesssim 6$.  This constraint appears as an additional term in the denominator of equation 15 of \cite{Masset_FARGO}.  We have verified that for larger values of $\beta$ the effect of including this timestepping constraint is negligible.

To model the shocks in the discs, both codes use artificial viscosity.  The SPH code uses the artificial viscosity method described by \cite{Chow_Monaghan_newAV} and \cite{Price_Monaghan_v_dot_r_AV}, the implementation of which is summarised in equations~\ref{eq:mom_av},~\ref{eq:Pi} and~\ref{eq:mu_2} (see Appendix~\ref{appendixA} for details).  The artificial viscosity is controlled by the parameters $\alpha_{\rm SPH}$ and $\beta_{\rm SPH}$.  {\sc fargo} uses the \cite{Vonneumann_Richtmyer_av} artificial viscous pressure with parameter $q$.  We use the default values for the SPH and {\sc fargo} artificial viscosity parameters of $(\alpha_{\rm SPH}, \beta_{\rm SPH}) = (0.1, 0.2)$ and $q = 1.41$, respectively.  Where specified, we also vary the amount of SPH and {\sc fargo} artificial viscosities to investigate their effects on the fragmentation boundary.

{\sc fargo}'s grid is set up to use linear spacing in the azimuthal direction and logarithmic spacing in the radial direction.  We use open boundary conditions at both the inner and outer radial boundaries and use a fixed gravitational softening length of $3 \times 10^{-4} H$, where $H$ is the vertical scaleheight of the disc, in all the {\sc fargo} simulations.

\section{The effect of numerical viscosity}
\label{sec:num_visc}

In hydrodynamics codes, artificial viscosity is frequently applied to correctly capture shocks and to avoid post-shock oscillations.  In trying to understand the evolution of gravitationally unstable discs, equation~\ref{eq:stress} seeks to link the dissipation rate in the disc to the cooling rate of the disc.  However, equation~\ref{eq:stress} is derived for a steady-state disc which assumes that gravitational instability is the only heating source for the disc.  In reality, in numerical simulations, there will be additional heating due to numerical dissipation.

In SPH, the artificial viscosity typically includes both a linear term controlled by $\alpha_{\rm SPH}$ and a quadratic term controlled by $\beta_{\rm SPH}$.  The linear term provides a bulk viscosity which dissipates kinetic energy as particles approach each other to reduce particle oscillations following a shock, while the quadratic term is primarily important to stop particle interpenetration.  {\sc fargo}, on the other hand, only contains a quadratic term controlled by the artificial viscosity parameter, $q$, and provides a bulk viscosity.

The dissipation from the bulk viscosity in shocks (such as those generated by gravitational instability) is physical.  In a gravitationally unstable disc, this provides the $\alpha_{\rm GI}$ term.  However, in these discs the artificial viscosities will also provide some shear viscosity, the heating effects of which are not accounted for in equation~\ref{eq:stress}.

\subsection{SPH artificial viscosity}
\label{sec:SPH_av}

\cite{Monaghan1985} showed that in the continuum limit, the $\alpha_{\rm SPH}$ component mimics a Navier-Stokes viscosity with bulk and shear coefficients proportional to the resolution length (see \citealp{Meglicki_av_continuum} and Appendix~\ref{appendixA}).  This has been confirmed numerically \citep[e.g.][]{Artymowicz_Lubow_av,Lodato_Price_betaSPH}.  The shear viscosity contributions from the linear and quadratic SPH terms due to the artificial viscosity can also be compared to the \cite{SS_viscosity} viscosity of the form $\nu = \alpha_{\rm SS} c_{\rm s} H$.  For the SPH calculations discussed in this paper, where the viscosity is only applied between approaching particles, assuming a Keplerian disc it can be shown that (see Appendix~\ref{sec:recent_AV},~\ref{sec:new_alpha} and~\ref{appendixD})

\begin{equation}
\label{eqss}
\alpha_{\rm SS,lin} = \frac{31}{525} \alphaSPH \frac{h}{H}
\end{equation}
and

\begin{equation}
\label{eqssquad}
\alpha_{\rm SS,quad} = \frac{9}{70 \pi} \betaSPH \left ( \frac{h}{H} \right )^2
\end{equation}
where $\alpha_{\rm SS,lin}$ and $\alpha_{\rm SS,quad}$ are the contributions from the linear and quadratic terms, respectively, and $h$ is the particle smoothing length.  Note that the coefficients in this SPH code are marginally different to some other SPH implementations.  In SPH codes employing the older \cite{Monaghan_Gingold_art_vis} formalism, the coefficients would be $1/20$ and $3/(35 \pi)$ for the linear and quadratic terms, respectively (see Appendix~\ref{sec:original_AV},~\ref{sec:evaluating} and~\ref{appendixD} for details).

For a \cite{SS_viscosity} disc model, the dissipation rate per unit mass is given by $\frac{9}{4} \alpha_{\rm SS} c_{\rm s}^2 \Omega$.  In a purely gravitationally unstable disc the dissipation rate may be parametrized $\frac{9}{4} \alpha_{\rm GI} c_{\rm s}^2 \Omega$.  However, using SPH we expect an additional heating due to numerical dissipation given by $\frac{9}{4} (\alpha_{\rm SS, lin} + \alpha_{\rm SS, quad}) c_{\rm s}^2 \Omega$.  Thus, the combined heating rate per unit mass is expected to be

\begin{equation}
\frac{9}{4} \left ( \alpha_{\rm GI} + \frac{31}{525} \alphaSPH \frac{h}{H} + \frac{9}{70 \pi} \betaSPH \left ( \frac{h}{H} \right )^2 \right ) c_{\rm s}^2 \Omega.
\label{eq:av_dissipation}
\end{equation}
Note that while the dissipation due to the quadratic term is often ignored when comparing the viscosity in SPH simulations of $\alpha$-discs, we show in Sections~\ref{sec:diss_nonGI} and Appendix~\ref{appendixD} that its contribution can be non-negligible.  In a numerical simulation, it is this heating rate that must be balanced by the imposed cooling (equation~\ref{eq:tcool}) for the disc to settle into a quasi-steady state.  Thus, equation \ref{eq:stress} can be rewritten as

\begin{equation}
\beta = \frac{4}{9} \frac{1}{\gamma (\gamma -1)} \frac{1}{(\alpha_{\rm GI} + \alpha_{\rm SS,lin} + \alpha_{\rm SS,quad})}.
\end{equation}
For the disc to fragment, the combined heating must be insufficient to balance the cooling so that

\begin{equation}
\beta_{\rm crit} = \frac{4}{9} \frac{1}{\gamma (\gamma -1)} \frac{1}{(\alpha_{\rm GI, crit} + \alpha_{\rm SS, lin, crit} + \alpha_{\rm SS, quad, crit})},
\end{equation}
where $\alpha_{\rm GI,crit}$ is the \emph{true} value of the gravitational stress that a disc can support before fragmenting and $\alpha_{\rm SS, lin, crit}$ and $\alpha_{\rm SS, quad, crit}$ are the contributions to the heating from the artificial viscosity that allows a disc to fragment once $\beta \lesssim \betacrit$ for any one particular resolution.

Now, for a particular cooling self-gravitating disc calculation, let us suppose that there is some maximum gravitational stress that can be produced by the disc beyond which it will fragment.  In this case, $\alpha_{\rm GI, crit}$ will be a constant, but $\alpha_{\rm SS, lin, crit}$ and $\alpha_{\rm SS, quad, crit}$ will decrease with increasing resolution.  If $\alpha_{\rm SS, lin}$ and $\alpha_{\rm SS, quad}$ obey equations~\ref{eqss} and~\ref{eqssquad}, then for a set of simulations with increasing resolution

\begin{equation}
\beta_{\rm crit} = \frac{4}{9} \frac{1}{\gamma (\gamma -1)} \left( \alpha_{\rm GI, crit} + \eta \frac{31}{525} \alpha_{\rm SPH} \frac{h}{H} + \zeta \frac{9}{70 \pi} \betaSPH \left ( \frac{h}{H} \right )^2 \right)^{-1},
\label{eqnumerical}
\end{equation}
where $\eta$ and $\zeta$ are constants which we expect to equal 1.

In order to compare equation \ref{eqnumerical} with the results of SPH simulations, we need to determine $h/H$ just before the fragmentation sets in.  Assuming that the disc fragments when $Q \approx 1$, using equation \ref{eq:Toomre} and noting $H=c_{\rm s}/\Omega$ and $\Omega^2=G M_*/R^3$, we have

\begin{equation}
H \approx \frac{\pi \Sigma R^3}{M_*},
\label{eq:H}
\end{equation}
where $R$ is the radius in the disc and $M_*$ is the mass of the central object.  The smoothing length in an SPH simulation for a disc that is resolved (i.e. $h<H$, which is true for all the simulations presented here at the radius of fragmentation) can be estimated, using equation 3 of \cite{Price_Bate_MHD_h} with density, $\rho \approx \frac{\Sigma}{2 H}$, as

\begin{equation}
\label{eqh}
h \approx 1.2 \left( \frac{2 H m_{\rm p}}{\Sigma} \right)^{1/3},
\end{equation}
where $m_{\rm p}$ is the mass of an SPH particle.  We use constant mass SPH particles and so the mass of an SPH particle is the disc mass divided by the number of SPH particles, $m_{\rm p} = M_{\rm d}/N_{\rm part}$.  The ratio of the smoothing length to disc scaleheight can then be approximated to be

\begin{equation}
\frac{h}{H} \approx \frac{1.2}{\Sigma R^2} \left( \frac{2 M_*^2 M_{\rm d}}{\pi^2 N_{\rm part}} \right)^{1/3}.
\label{eq:h_H}
\end{equation}
In Section~\ref{sec:h_H} we verify that this is indeed the case for steady-state marginally stable discs that have a Toomre parameter, $Q \approx 1$.  We therefore expect that

\begin{align}
\label{eq:betacrit_hH}
\beta_{\rm crit} &= \frac{4}{9} \frac{1}{\gamma (\gamma -1)} \left( \alpha_{\rm GI, crit} + \eta \frac{31}{525} \frac{1.2}{\Sigma R^2} \left( \frac{2 M_*^2 M_{\rm d}}{\pi^2 N_{\rm part}} \right)^{1/3} \alpha_{\rm SPH} \right. \nonumber \\
&\qquad \left. + \zeta \frac{9}{70 \pi} \frac{(1.2)^2}{\Sigma^2 R^4} \left ( \frac{2 M_*^2 M_{\rm d}}{\pi^2 N_{\rm part}} \right )^{\frac{2}{3}} \betaSPH \right )^{-1},
\end{align}
i.e. we have three unknowns: $\alpha_{\rm GI,crit}$, $\eta$ and $\zeta$, since we can determine $\betacrit$ for any one resolution.  If $\eta$ and $\zeta$ are unity, then the second and third terms in the denominator in equation~\ref{eq:betacrit_hH} each have values of $\approx O(10^{-3})$ for the discs studied by \cite{Rice_beta_condition} and \cite{Meru_Bate_resolution} simulated with 250,000 particles (using the parameters described in Section~\ref{sec:sim}, and given that the fragmentation occurs in the outer parts of the disc -- see \citealp{Meru_Bate_fragmentation}).  The contribution from these terms are approximately a factor of $O(10)$ smaller than the original estimate of  $\alpha_{\rm GI,crit} \approx 0.06$ \citep{Gammie_betacool,Rice_beta_condition}.  Thus it was assumed in earlier SPH studies that the heating due to artificial viscosity would be negligible compared to the dissipation due to gravitational instabilities (see Appendix A of \citealp{Lodato_Rice_original}).  Note, however, that if the SPH artificial viscosity plays a significant role then the $\alphaSPH$ term scales linearly with the smoothing length, $h$, such that the convergence of $\beta_{\rm crit}$ towards the true value is expected to be first order in $h$ (i.e. very slow as the numerical resolution is increased).  On the other hand, if the dominant term is the $\betaSPH$ term, the convergence will be faster since the SPH artificial viscosity scales quadratically with the smoothing length.

\subsection{{\sc fargo} artificial viscosity}
\label{sec:FARGO_av}

Most grid-based hydrodynamical codes are second order and do not have a linear viscosity.  Therefore, one would expect that their rate of convergence towards the true value of $\beta_{\rm crit}$ will be second order in spatial resolution and thus possibly faster than SPH codes.  Indeed, {\sc fargo} uses the \cite{Vonneumann_Richtmyer_av} artificial viscous pressure given by \citep[see also][]{Numerical_Methods}

\begin{equation}
\label{eq:visc_press}
q^2 \rho (\Delta x)^2 \left | \frac{\partial v}{\partial x} \right |^2
\end{equation}
where $\Delta x$ is the cell size, $\rho$ is the density and $q = l/\Delta x$ is a constant which indicates the number of grid cells over which the shock is spread and whose value is dependent on the numerical scheme and is usually $0.05 \le q \le 2$ ($l$ indicates the strength of the artificial viscosity).  This is a bulk viscosity.  In a cylindrical code, if the gas travels in circles, there should be no shear viscosity at all.  However, in a gravitationally unstable disc, this will not be the case and there will be some shear viscosity (that arises from the bulk viscosity) which we expect will scale in roughly the same way, i.e. proportional to the square of the size of the grid cell.  Assuming the shear rate, $\left | \partial v / \partial x \right |$, is approximately Keplerian, then

\begin{equation}
\left | \frac{\partial v}{\partial x} \right | \sim \left | R \frac{{\rm d} \Omega}{{\rm d} R} \right |.
\label{eq:shear}
\end{equation}
Using equation~\ref{eq:shear} and equating equation~\ref{eq:visc_press} to the shearing force per unit area as defined in equation~\ref{eq:FoverA} yields a kinematic viscosity due to the artificial viscosity given by

\begin{equation}
\nu_{\rm av, {\sc fargo}} = q^2 (\Delta x)^2 \left | R \frac{{\rm d} \Omega}{{\rm d} R} \right | \approx \frac{3}{2} q^2 (\Delta x)^2 \Omega.
\label{eq:fargo_av}
\end{equation}
where the final approximation assumes a Keplerian flow.  This gives a \cite{SS_viscosity} type viscosity of the form $\nu = \alpha_{\rm SS} c_{\rm s} H$ with

\begin{equation}
\label{eqss_fargo}
\alpha_{\rm SS, {\sc fargo}} = \frac{3}{2} q^2 \left ( \frac{\Delta x}{H} \right )^2.
\end{equation}
Analogous to the derivation of equation~\ref{eqnumerical} for SPH artificial viscosity this yields

\begin{equation}
\label{eqgrid}
\beta_{\rm crit} = \frac{4}{9} \frac{1}{\gamma (\gamma -1)} \frac{1}{ \left ( \alpha_{\rm GI, crit} + \xi \frac{3}{2} q^2 \left(\frac{\Delta x}{H}\right)^2 \right ) },
\end{equation}
where we expect that $\xi$ is unity.  Substituting for the scaleheight using equation~\ref{eq:H} yields a formula for $\betacrit$ that is equivalent to equation~\ref{eq:betacrit_hH} for a grid-based code:

\begin{equation}
\beta_{\rm crit} = \frac{4}{9} \frac{1}{\gamma (\gamma -1)} \left( \alpha_{\rm GI, crit} + \xi \frac{3}{2} q^2 \left(\frac{\Delta x M_*}{\pi \Sigma R^3}\right)^2 \right )^{-1}.
\label{eq:betacrit_fargo}
\end{equation}
Since the {\sc fargo} simulations use a logarithmic grid we take the cell size at a radius of $R = 22$~au, i.e. close to the edge of the disc where we would expect fragmentation to occur when the cooling timescale is close to the critical value for any one particular resolution.  We find that the cell size at this radius scales as

\begin{equation}
\Delta x \approx \frac{125}{N_{\rm cells}^{\frac{1}{2}}} \rm ~au.
\end{equation}
where $N_{\rm cells}$ is the total number of cells used in the simulation.  We therefore have two unknowns: $\alpha_{\rm GI, crit}$ and $\xi$, since we can determine $\betacrit$ at any one resolution.

Equation~\ref{eqgrid} shows that if the artificial viscosity plays a significant role in the dissipation in the disc, the convergence is expected to be second order in spatial resolution, i.e. potentially faster than with SPH.

\subsection{Artificial viscosity effects with resolution}
\label{sec:av_res}

In an SPH code when the number of particles increases, $h/H$ is reduced and therefore as the resolution increases, $\alpha_{\rm SS, lin}$ and $\alpha_{\rm SS, quad}$, and thus the heating due to artificial viscosity, tend to zero.  Similarly, as the resolution is increased in a grid-based code, the cell size decreases for any one problem, and the contribution to the dissipation from the artificial viscosity decreases.  In the limit of infinite resolution, equations~\ref{eqnumerical} and~\ref{eqgrid} will return to equation \ref{eq:stress}.  But for a finite resolution, the value of $\beta_{\rm crit}$ {\it obtained from a numerical simulation should always be smaller than the true value}.

It is also important to note that simply reducing the value of $\alphaSPH$, $\betaSPH$ or $q$ is not necessarily a sufficient way in which to decrease the numerical dissipation and obtain the \emph{true} value of $\betacrit$.  By inspection of equations~\ref{eqnumerical} and~\ref{eqgrid} we might naively assume this to be the case.  However, reducing these values may mean that shocks are treated inaccurately.  For example, the shocks may not be spread over a large enough lengthscale to model them numerically and/or there may be post-shock oscillations that are eventually damped, resulting in dissipation.  \cite{Lodato_Price_betaSPH} showed using SPH that setting $\alphaSPH = 0$ counter-intuitively led to a larger amount of dissipation.  \cite{Price_Federrath_betaSPH} found that if an adequate value of $\betaSPH$ is not used, particle interpenetration may occur.  In their case, they stated that this makes very little difference to their dissipation rate since their linear term dominates almost everywhere.  However, their simulations explored a different regime to that being explored here and with a different artificial viscosity scheme.  Their simulations employed the \cite{MM_viscosity} artificial viscosity switch where the value of $\alphaSPH$ ranges between 0.05 and 1.0 (the higher value being implemented close to shocks).  Their simulations were of high Mach number shocks (${\mathscr M} = 10$) and so a large part of their simulations would require the use of the higher value of $\alphaSPH$ and thus this would dominate the dissipation.  The simulations performed by \cite{Rice_beta_condition} and \cite{Meru_Bate_resolution} used a fixed value of $\alphaSPH = 0.1$ - such a low value opens up the possibility of the quadratic term being important and therefore decreasing the value of $\betaSPH$ may affect the overall dissipation rate and thus the fragmentation outcome.

In {\sc fargo}, if $q$ is set to zero, there will be no controlled numerical dissipation (e.g. to capture shocks or other disturbances at the grid scale).  However, the code will still have some level of numerical diffusion and dissipation which is not controllable other than that it too should decrease with increasing resolution.

\section{Simulations}
\label{sec:sim}

The disc and star properties used to carry out the simulations in this paper are exactly the same as those used by \cite{Rice_beta_condition} and \cite{Meru_Bate_resolution}: a $0.1 \Msolar$ disc surrounding a $1 \Msolar$ star.  The SPH simulations span a radial range, $0.25 \le R \le 25$~au, while the {\sc fargo} simulations span a radial range, $1 \le R \le 25$~au (only marginally different to the SPH simulations for numerical reasons).

The initial surface mass density and temperature profiles are $\Sigma \propto R^{-1}$ and $T \propto R^{-1/2}$, respectively, and the temperature is normalised so that the minimum initial Toomre stability value at the outer edge of the disc, $\Qmin=2$.  The discs are modelled with a ratio of specific heats, $\gamma = 5/3$.

Table~\ref{tab:res_sim} shows a summary of the initial SPH simulations and the key fragmenting results carried out by \cite{Meru_Bate_resolution} (obtained from their Table 1) as well as those in this paper (bold text).  We supplement the \cite{Meru_Bate_resolution} results by carrying out an additional simulation using 2 million particles with $\beta = 9$ and three additional simulations using 16 million particles with $\beta = 12$, 15 and 20.

Table~\ref{tab:fargo_sim} summarises the initial simulations carried out using {\sc fargo} and the key fragmentation results.  We perform simulations, using $q = 1.41$, at five different resolutions and determine the critical value of $\beta$ at each of these resolutions.  The lowest resolution simulations are carried out using 768 and 256 grid cells in the azimuthal and radial directions, respectively.  We then increase the linear resolution by factors of 2, 4, 8 and 16 in both the azimuthal and radial directions.

The simulations were run either for at least 6 outer rotation periods (ORPs) or until the discs fragmented.  Fragments are defined as regions whose surface mass densities are at least two orders of magnitude denser than their surroundings.  In addition, we ensure that the fragments survive for at least one rotation to verify that they do not shear apart.

To investigate the effects of the different components of artificial viscosity in SPH, we carry out a number of simulations where we vary the values of $\alphaSPH$ and $\betaSPH$ (see Table~\ref{tab:alpha_beta_inves} for details).  We perform a suite of simulations using 250,000 particles.  Firstly, we set $\alphaSPH = 0.1$ and vary the value of $\betaSPH$ (Table~\ref{tab:alpha_beta_inves}, top section).  Secondly, we set $\betaSPH = 2.0$ and vary the value of $\alphaSPH$ (Table~\ref{tab:alpha_beta_inves}, middle section).  We then carry out simulations with $(\alphaSPH, \betaSPH) = (0.1, 2.0)$ using 31,250, 250,000 and 2 million particles to determine the effect that changing the value of $\betaSPH$ has on the fragmentation boundary at each of these resolutions (Table~\ref{tab:alpha_beta_inves}, bottom section).

In addition, we perform a number of SPH simulations without self-gravity (see Table~\ref{tab:non-GI} for details) using various values of $\alphaSPH$, $\betaSPH$ to compare the measured dissipation to the analytically expected values in equation~\ref{eq:av_dissipation} (also see Appendix~\ref{appendixD}).

Finally, we investigate the effects that artificial viscosity in {\sc fargo} has on the critical cooling timescale by varying the value of $q$ in equation~\ref{eq:visc_press} between 0 and 2.5 (see Table~\ref{tab:fargo_av}) for discs modelled using 786,432 grid cells (512 and 1536 cells in the radial and azimuthal directions, respectively).  We then carry out simulations using an artificial viscosity parameter, $q = 0.5$, at all but the lowest resolutions considered in this paper to determine the effect that this has on the fragmentation boundary.

\section{Results}
\label{sec:results}

\subsection{The convergence rate of $\betacrit$ with SPH}
\label{sec:nonconverge}

\begin{table}
\centering
  {\small
    \begin{tabular}{llll}
    \hline
    Simulation name & No of particles & $\beta$ & Fragmented?\\
   \hline
    \hline
    31k-beta2 & 31,250 & 2.0 & Yes\\
    31k-beta2.5 & 31,250 & 2.5 & Yes\\
    31k-beta3 & 31,250 & 3.0 & Yes\\
    31k-beta3.5 & 31,250 & 3.5 & No\\
    31k-beta4 & 31,250 & 4.0 & No\\
    250k-beta5 & 250,000 & 5.0 & Yes\\
    250k-beta5.5 & 250,000 & 5.5 & Yes\\
    250k-beta5.6 & 250,000 & 5.6 & No\\
    250k-beta6 & 250,000 & 6.0 & No\\
    250k-beta6.5 & 250,000 & 6.5 & No\\
    250k-beta7 & 250,000 & 7.0 & No\\
    250k-beta7.5 & 250,000 & 7.5 & No\\
    2m-beta5.5 & 2 million & 5.5 & Yes\\
    2m-beta6 & 2 million & 6.0 & Yes\\
    2m-beta6.5 & 2 million & 6.5 & Yes\\
    2m-beta7 & 2 million & 7.0 & Yes\\
    2m-beta8 & 2 million & 8.0 & Yes\\
    {\bf 2m-beta9} & {\bf 2 million} & {\bf 9.0} & {\bf No}\\
    2m-beta10 & 2 million & 10.0 & No\\
    2m-beta10.5 & 2 million & 10.5 & No\\
    2m-beta11 & 2 million & 11.0 & No\\
    2m-beta15 & 2 million & 15.0 & No\\
    16m-beta10 & 16 million & 10.0 & Yes\\
    {\bf 16m-beta12} & {\bf 16 million} & {\bf 12.0} & {\bf No}\\
    {\bf 16m-beta15} & {\bf 16 million} & {\bf 15.0} & {\bf No}\\
   16m-beta18 & 16 million & 18.0 & No\\
   {\bf 16m-beta10} & {\bf 16 million} & {\bf 20.0} & {\bf No}\\
   \hline
  \end{tabular}
}
  \caption{Table showing the SPH simulations carried out by \citet{Meru_Bate_resolution}, as well as the supplementary SPH simulations performed in this paper (bold text), and the key fragmenting results.  The simulations are performed using ($\alphaSPH$, $\betaSPH$) = (0.1,0.2)}
 \label{tab:res_sim}
\end{table}

\begin{figure}
\centering
  \includegraphics[width=1.0\columnwidth]{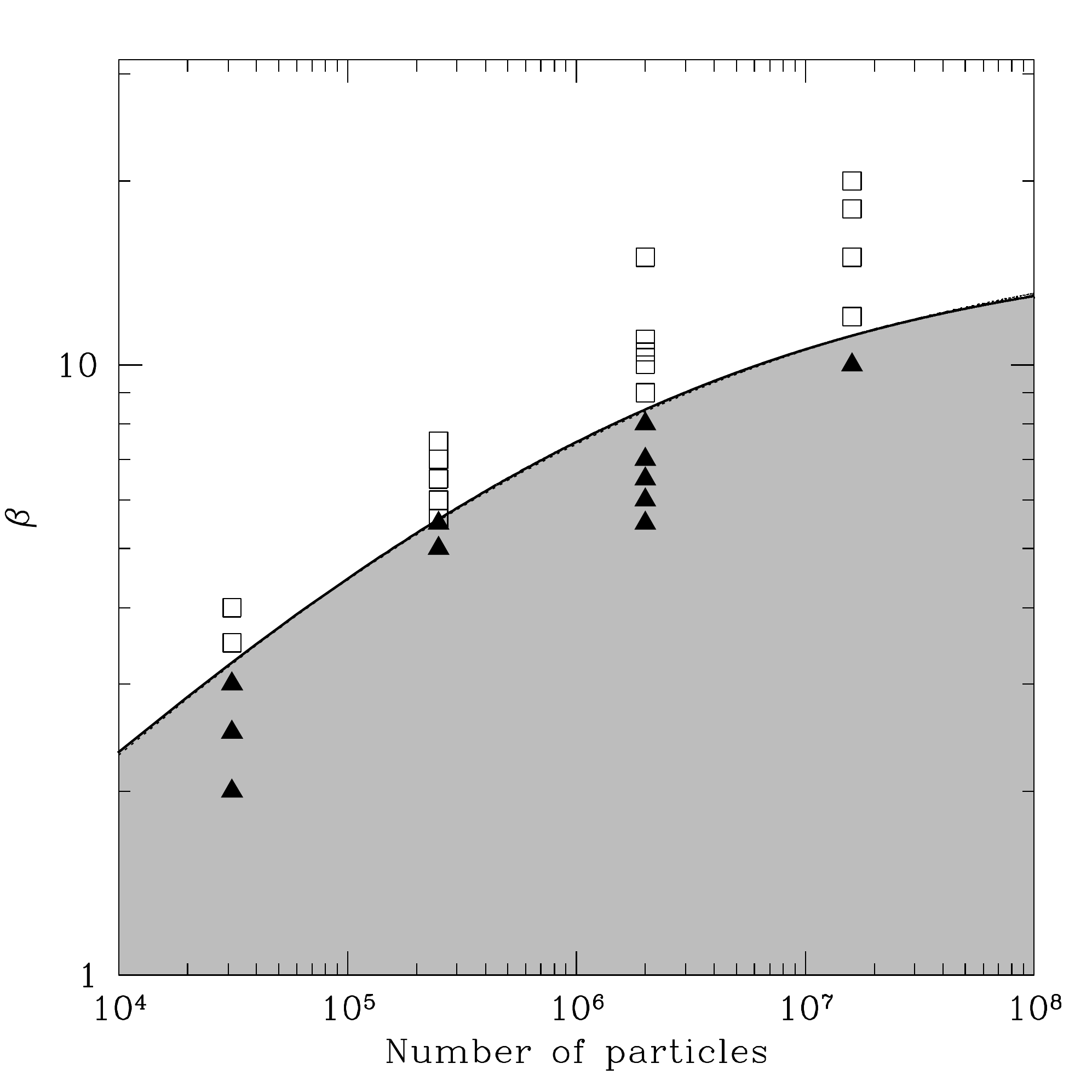}
  \caption{Graph of $\beta$ against resolution of the non-fragmenting (open squares) and fragmenting (solid triangles) SPH simulations.  This figure contains the results presented by \citet[][their Figure 3]{Meru_Bate_resolution} as well as the new simulations highlighted in Table~\ref{tab:res_sim}.  The solid line, obtained by fitting equation~\ref{eq:fit}, shows a dividing line between the fragmenting and non-fragmenting cases and the grey region is where fragmentation can take place.  The graph shows clear evidence of convergence of results with increased resolution.  These simulations are carried out with ($\alphaSPH$, $\betaSPH$) = (0.1, 0.2).  The convergence rate is first order with spatial resolution.  The dotted line (which coincides well with the solid line) is obtained by fitting equation~\ref{eq:betacrit_hH}.}
 \label{fig:res_beta}
\end{figure}

\begin{figure*}
\centering
  \includegraphics[width=2.0\columnwidth]{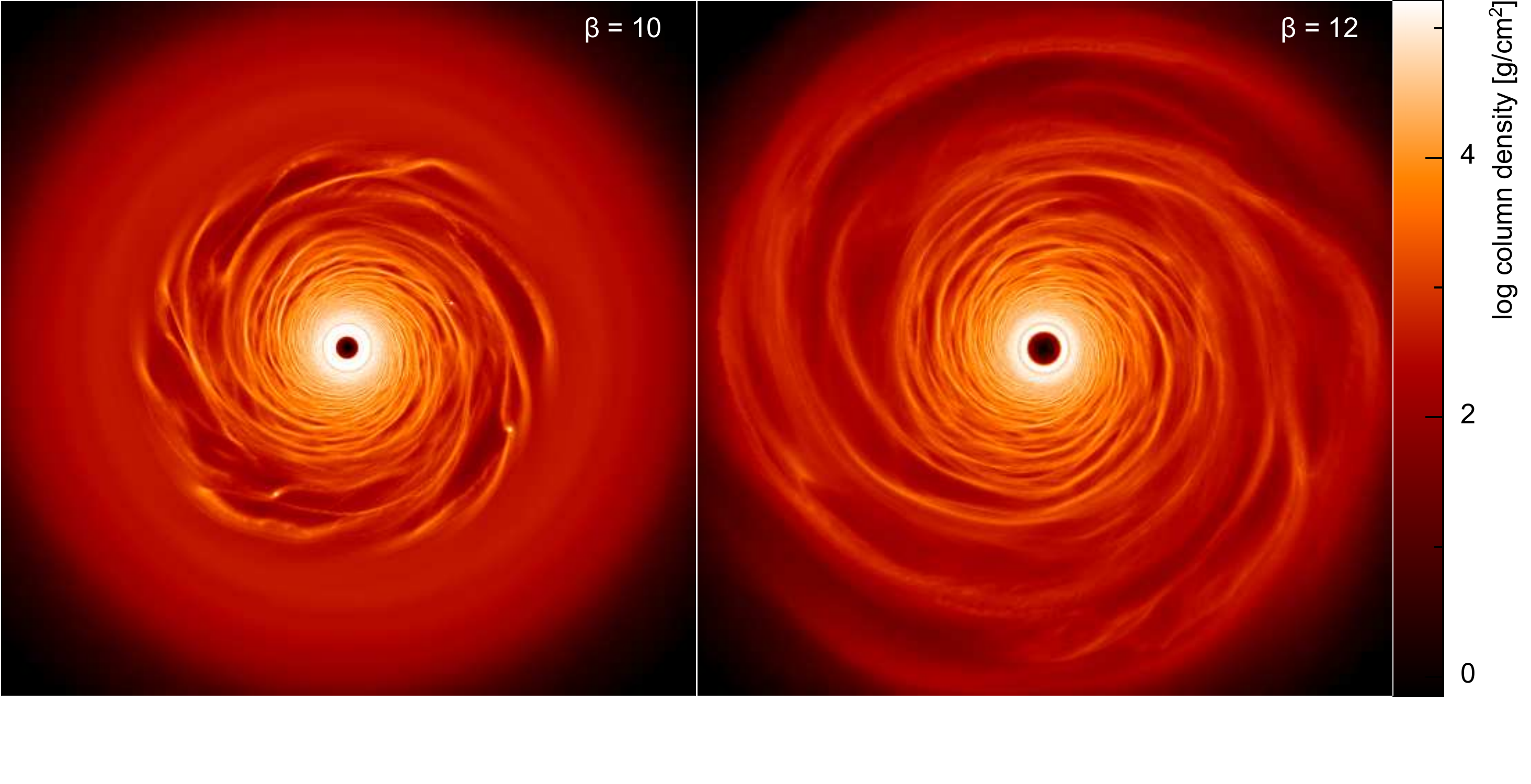}
  \caption{Surface mass density rendered image of two identical simulations carried out using $\beta = 10$ (left panel) and $\beta = 12$ (right panel), using 16m SPH particles.  Fragmentation occurs for $\beta = 10$ but not for $\beta = 12$.  These simulations are performed with ($\alphaSPH$, $\betaSPH$) = (0.1, 0.2).}
 \label{fig:image_SPH_beta0.2}
\end{figure*}

Table~\ref{tab:res_sim} summarises the results of the SPH simulations, using $\alphaSPH$, $\betaSPH$) = (0.1, 0.2), carried out by \cite{Meru_Bate_resolution} and those performed for this paper.  \cite{Meru_Bate_resolution} found \emph{borderline} simulations which they defined to be discs which showed signs of fragmentation but the fragments sheared apart rapidly (within 1 ORP) and no further signs of fragmentation were seen.  We find that \emph{borderline} simulations can in fact range a span of $\beta$ values.  However, since ultimately they are discs that do not end up fragmenting, we now simplify this terminology and refer to them as non-fragmenting simulations.  Figure~\ref{fig:res_beta} shows a summary of the SPH results.  This figure is the same as Figure 3 of \cite{Meru_Bate_resolution} but with the the addition of the new SPH results presented in this paper.  \cite{Meru_Bate_resolution} found no evidence for convergence, but the addition of the new high resolution calculations now provides evidence for a very slow convergence of $\beta_{\rm crit}$ with increasing resolution.  Figure~\ref{fig:image_SPH_beta0.2} shows two of the highest resolution SPH simulations (16 million particles) carried out using $\beta = 10$ and $\beta = 12$.  It can clearly be seen that at this resolution, fragmentation occurs for $\beta = 10$ but not for $\beta = 12$.  To estimate the rate of convergence we fit a formula of the form

\begin{equation}
\beta = \frac{\betacrit}{1 + \lambda l^{\sigma}}
\label{eq:fit}
\end{equation}
where $l$ is the linear spatial resolution, $\lambda$ is a constant and $\sigma$ is the convergence rate.  For SPH, we simply take $l \propto N_{\rm part}^{-\frac{1}{3}}$.  We fit this formula to the values of $\beta$ in Figure~\ref{fig:res_beta} that lie half way between the lowest non-fragmenting value of $\beta$ and the highest fragmenting value of $\beta$ for each numerical resolution, i.e. the fragmentation boundary.  We find that a good fit is obtained with $\betacrit = 15.6 \pm 1.0$ and $\sigma = 1.08 \pm 0.05$.  The value of $\sigma$ shows that the rate of convergence is first order in spatial resolution.  For the benefit of understanding the results presented in Section~\ref{sec:frag_bdry_optimumSPH} (which only have data points at the lowest three resolutions) we fit equation~\ref{eq:fit} to the data presented in Figure~\ref{fig:res_beta} but \emph{exclude} the highest resolution simulations.  We find $\betacrit = 17.4$ and $\sigma = 1.03$, so excluding the last point does not alter the fit significantly due to the slow convergence rate of SPH.

In addition, we also fit equation~\ref{eq:betacrit_hH} to this data.  We find that $\alpha_{\rm GI, crit} = 0.024 \pm 0.001$, $\eta = 21.1 \pm 1.3$ and $\zeta = 1.7 \pm 1.5$.  In the limit of infinite resolution, this value of the critical gravitational stress obtained is equivalent to a critical cooling timescale of $\betacrit \approx 17$.  However, we point out that while $\zeta$ is reasonably close to unity, the value of $\eta$ is very large.  This is suggestive of an additional source of dissipation present in the simulations over and above what we expect from artificial viscosity in a shear dominated disc.

\subsection{The convergence rate of $\betacrit$ with {\sc fargo}}
\label{sec:FARGO}

\begin{table}
\centering
  {\small
\begin{tabular}{lllll}
    \hline
    Simulation name & No of & No of & $\beta$ & Fragmented?\\
    & radial cells & azimuthal cells & &\\
   \hline
    \hline
    197k\_cells-beta0.5 & 256 & 768 & 0.5 & Yes\\
    197k\_cells-beta1 & 256 & 768 & 1 & Yes\\
    197k\_cells-beta2 & 256 & 768 & 2 & No\\
    197k\_cells-beta3 & 256 & 768 & 3 & No\\
    197k\_cells-beta4 & 256 & 768 & 4 & No\\
    786k\_cells-beta3 & 512 & 1536 & 3 & Yes\\
    786k\_cells-beta3.5 & 512 & 1536 & 3.5 & Yes\\
    786k\_cells-beta4.5 & 512 & 1536 & 4.5 & Yes\\
    786k\_cells-beta5 & 512 & 1536 & 5 & Yes\\
    786k\_cells-beta5.5 & 512 & 1536 & 5.5 & Yes\\
    786k\_cells-beta6 & 512 & 1536 & 6 & No\\
    786k\_cells-beta10 & 512 & 1536 & 10 & No\\
    3.1m\_cells-beta10 & 1024 & 3072 & 10 & Yes\\
    3.1m\_cells-beta12 & 1024 & 3072 & 12 & Yes\\
    3.1m\_cells-beta13 & 1024 & 3072 & 13 & No\\
    13m\_cells-beta11 & 2048 & 6144 & 11 & Yes\\
    13m\_cells-beta14 & 2048 & 6144 & 14 & Yes\\
    13m\_cells-beta15 & 2048 & 6144 & 15 & Yes\\
    13m\_cells-beta16 & 2048 & 6144 & 16 & Yes\\
    13m\_cells-beta18 & 2048 & 6144 & 18 & No\\
    50m\_cells-beta18 & 4096 & 12288 & 18 & Yes\\
    50m\_cells-beta20 & 4096 & 12288 & 20 & Yes\\
    50m\_cells-beta22 & 4096 & 12288 & 22 & Yes\\
    50m\_cells-beta24 & 4096 & 12288 & 24 & No\\
   \hline
  \end{tabular}
}
  \caption{Table showing the simulations carried out using {\sc fargo} and the key fragmenting results.  The simulations are performed using the artificial viscosity parameter, $q = 1.41$.}
 \label{tab:fargo_sim}
\end{table}

\begin{figure}
\centering
  \includegraphics[width=1.0\columnwidth]{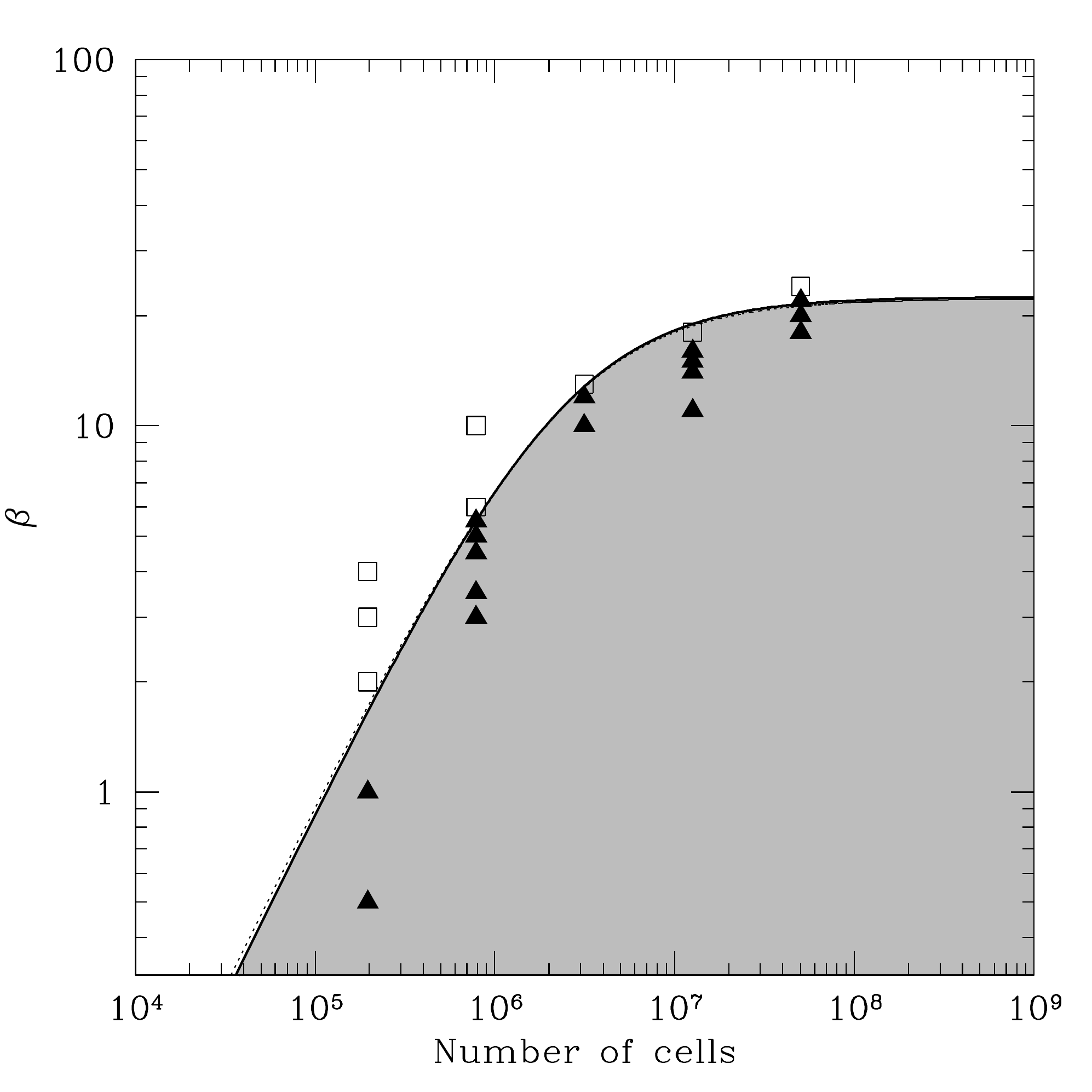}
  \caption{Graph of $\beta$ against resolution of the non-fragmenting (open squares) and fragmenting (solid triangles) {\sc fargo} simulations carried out using $q = 1.41$.  The solid line, obtained by fitting equation~\ref{eq:fit}, shows a dividing line between the fragmenting and non-fragmenting cases and the grey region is where fragmentation can take place.  The graph shows clear evidence of convergence of results with increased resolution.  The convergence rate is second order with spatial resolution.  The dotted line (which coincides well with the solid line) is obtained by fitting equation~\ref{eq:betacrit_fargo}.}
 \label{fig:res_beta_fargo}
\end{figure}

\begin{figure*}
\centering
\includegraphics[width=0.74\columnwidth,angle=-90]{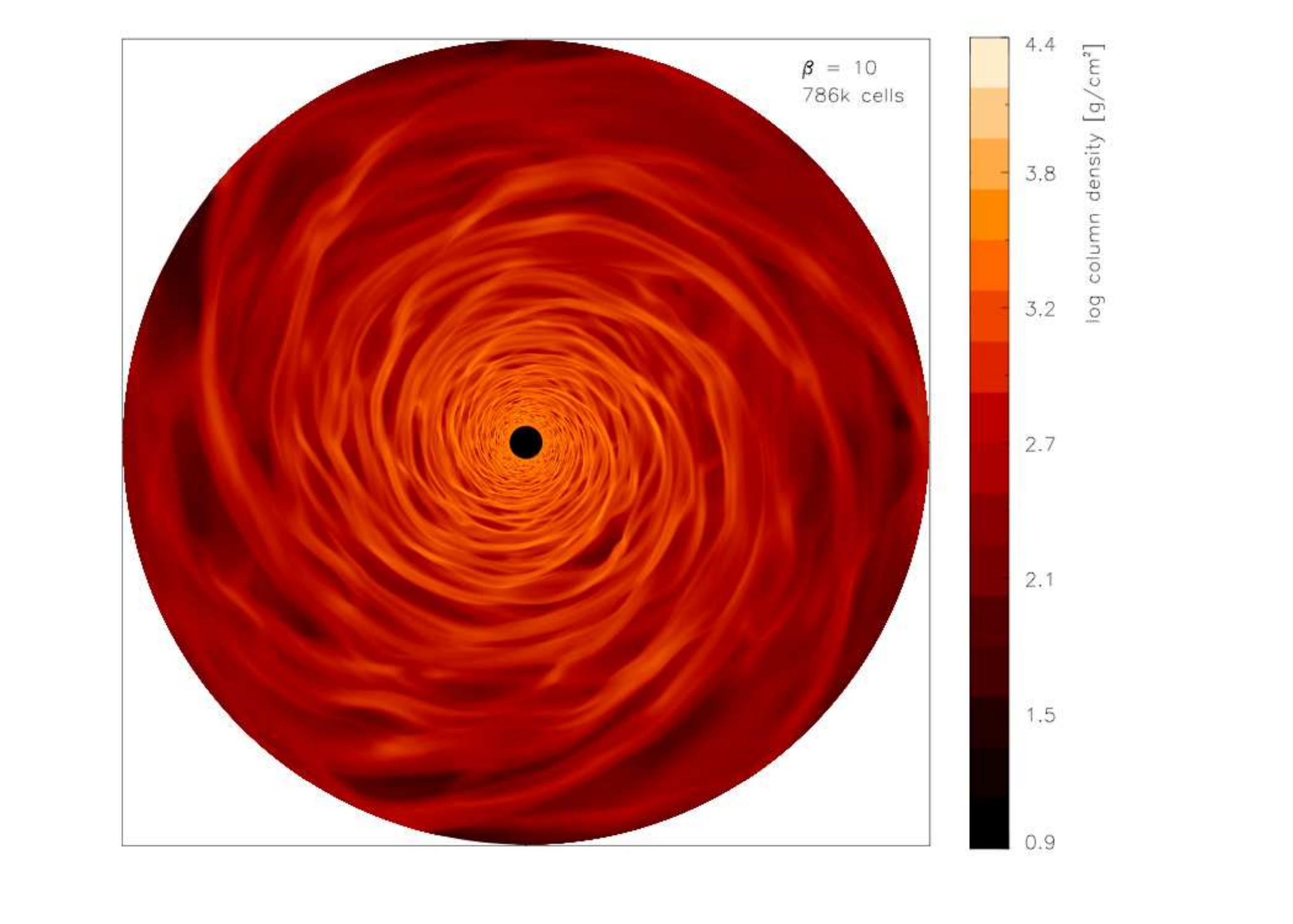}
\includegraphics[width=0.74\columnwidth,angle=-90]{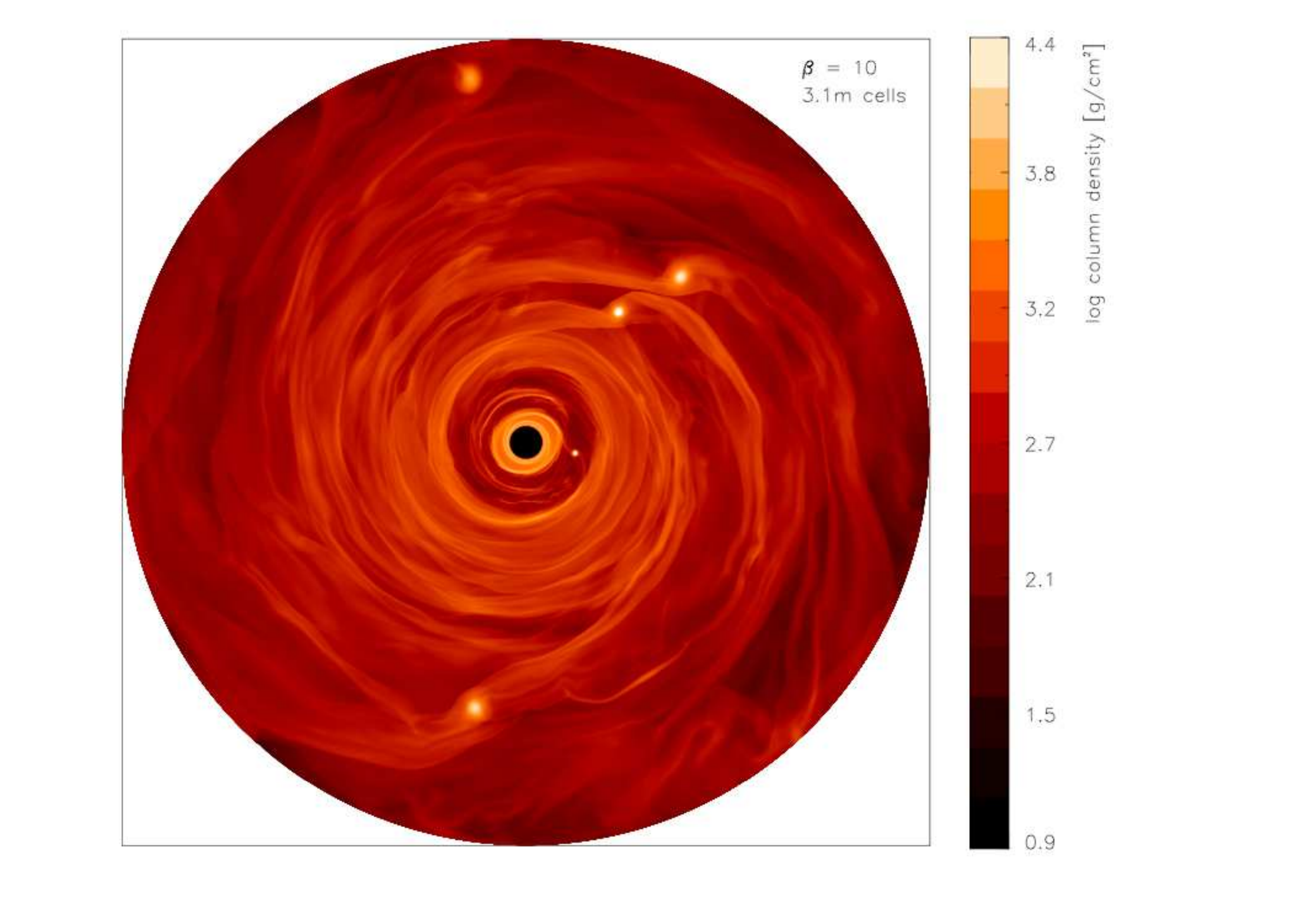}
\includegraphics[width=0.74\columnwidth,angle=-90]{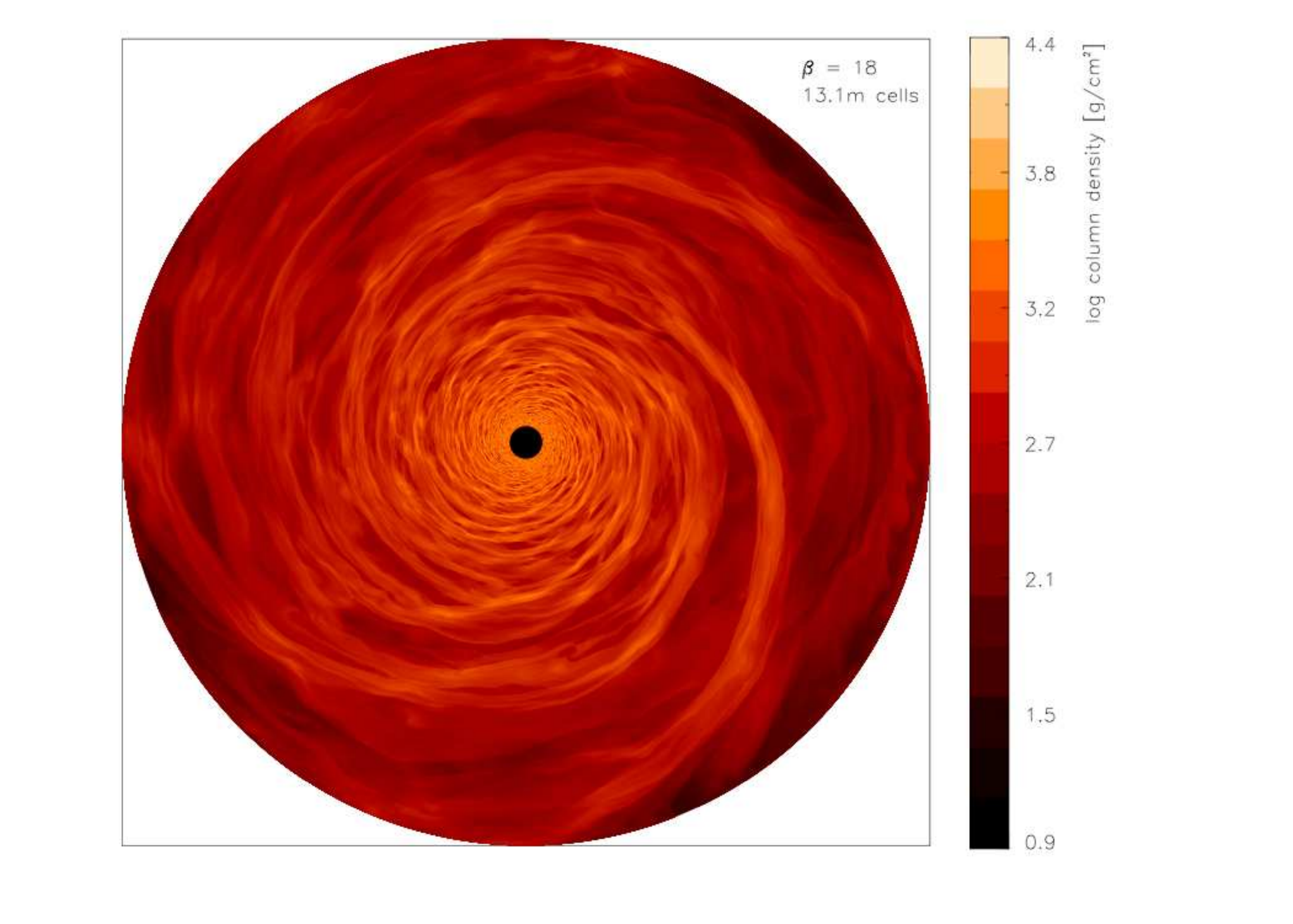}
\includegraphics[width=0.74\columnwidth,angle=-90]{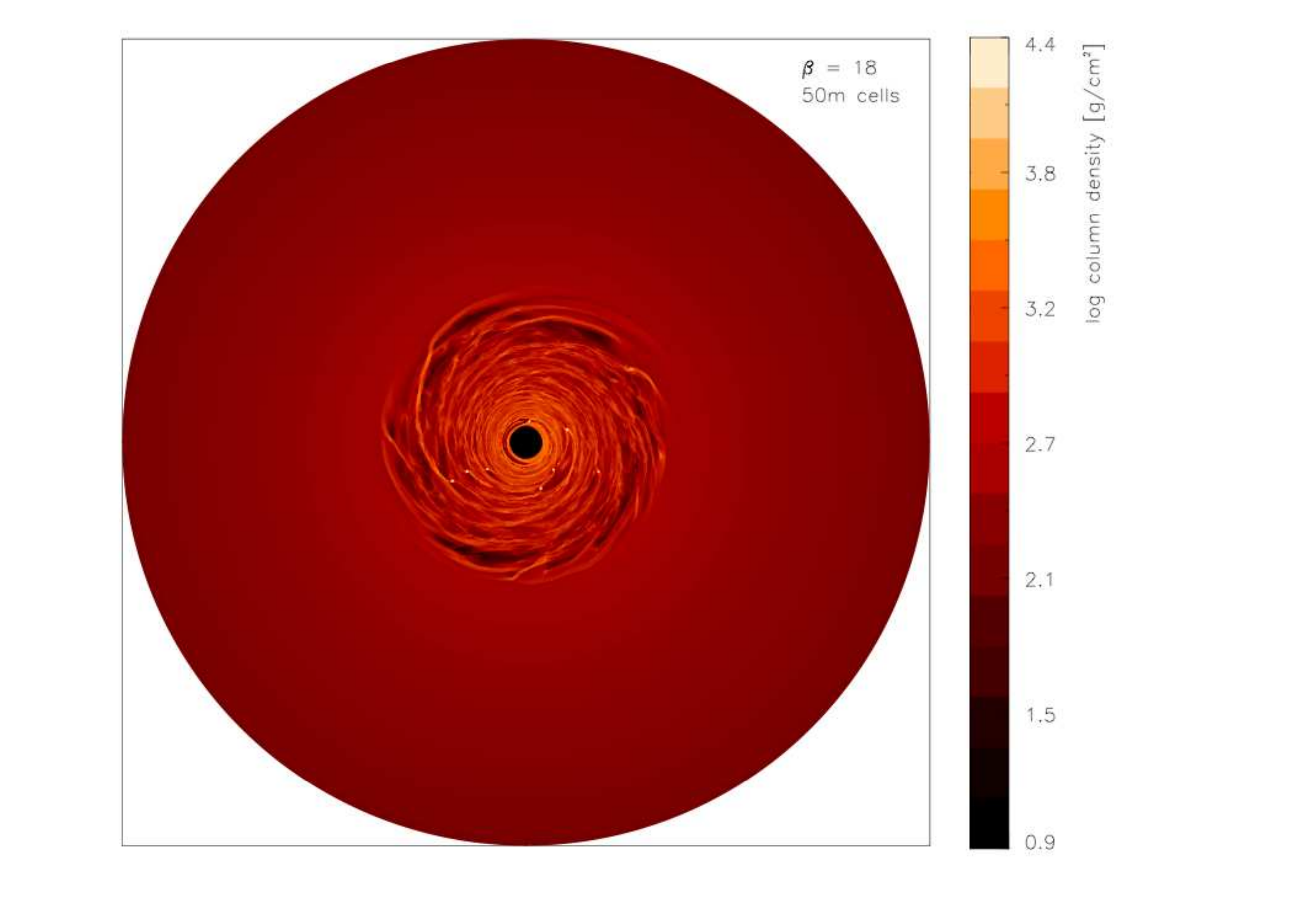}
  \caption{Surface mass density rendered images of four simulations carried out using {\sc fargo} with cooling timescales of $\beta = 10$ (top panel) and $\beta = 18$ (bottom panel) using $q = 1.41$.  For the simulation with 786,432 grid cells (upper left panel), fragmentation does not occur when modelled with $\beta = 10$ but fragmentation is seen when the resolution is increased to 3.1 million grid cells (upper right panel).  Similarly, when the resolution is increased further to 13 million grid cells (lower left panel), fragmentation is not seen using a cooling timescale of $\beta = 18$ whereas when the same simulation is carried out using 50 million grid cells (lower right panel) fragmentation is seen.}
 \label{fig:image_fargo_q1.41}
\end{figure*}

Table~\ref{tab:fargo_sim} and Figure~\ref{fig:res_beta_fargo} summarise the results using the grid-based code, {\sc fargo}.  As with the SPH results, we also see that as the resolution increases numerical convergence does appear to take place.  Figure~\ref{fig:image_fargo_q1.41} shows the surface mass density rendered images of the discs modelled using {\sc fargo} at resolutions of 786,432, 3.1 million, 13 million and 50 million grid cells.  The top two panels show that using a cooling timescale of $\beta = 10$, the discs do not fragment using 786,432 grid cells but when the resolution is increased to 3.1 million grid cells, fragmentation is seen.  Similarly, using a cooling timescale of $\beta = 18$, the disc modelled with 13 million grid cells does not fragment whereas that modelled at the higher resolution of 50 million grid cells does fragment.

Again, we use equation~\ref{eq:fit} to estimate the rate of convergence.  For {\sc fargo}, we simply take $l \propto N_{\rm cells}^{-\frac{1}{2}}$, where $N_{\rm cells}$ is the number of grid cells.  Note that the linear resolution is inversely proportional to the square root of the number of cells because the calculation is two dimensional.  We fit this formula to the fragmentation boundary in Figure~\ref{fig:res_beta_fargo}) as done for the SPH results.  We find that a good fit is obtained with $\betacrit = 22.3 \pm 2.3$ and $\sigma = 2.03 \pm 0.36$.  The value of $\sigma$
 shows that the rate of convergence is second order in spatial resolution.  We then fit the data using equation~\ref{eq:betacrit_fargo} and find that $\alpha_{\rm GI, crit} = 0.018 \pm 0.001$ and $\xi = 0.87 \pm 0.08$.  In the limit of infinite resolution, this value of $\alpha_{\rm GI, crit}$ is equivalent to a critical cooling timescale, $\betacrit \approx 22$ (using equation~\ref{eq:stress}), similar to the value obtained using equation~\ref{eq:fit}.

Thus the value of $\betacrit$ converges more rapidly using {\sc fargo} than SPH.  In Section~\ref{sec:FARGO_av} we note that if artificial viscosity plays a significant role in the determination of $\betacrit$ then {\sc fargo} might be expected to display second-order convergence since it only applies a quadratic artificial viscosity.  On the other hand SPH includes both linear and quadratic artificial viscosities.  If the linear term is dominant, this may lead to first-order convergence.  This implies that artificial viscosity may be significant in determining $\betacrit$.  We note that {\sc fargo} appears to converge to a higher value of $\betacrit$ than SPH, but this result may also be caused by the different artificial viscosities.  Therefore, in the following sections we investigate the dependence of $\betacrit$ on the artificial viscosities applied in both codes.

\subsection{The effect of SPH artificial viscosity on convergence}
\label{sec:SPH_av_effect}

In Section~\ref{sec:num_visc} we present analytical arguments that suggest that artificial viscosity may play a role in the numerically determined value of the critical cooling timescale.  We show that the contribution to the dissipation due to the artificial viscosity is expected to decrease with increasing resolution (Appendix~\ref{appendixD} and equation~\ref{eq:av_dissipation}).  Therefore, if the slow convergence can be attributed to SPH artificial viscosity, this may be the reason why \cite{Meru_Bate_resolution} found that $\betacrit$ increases with increasing resolution and is a plausible explanation as to why the results presented in Section~\ref{sec:nonconverge} show a slow convergence.  We test the role that artificial viscosity plays on the fragmentation of self-gravitating discs by varying the values of $\alphaSPH$ and $\betaSPH$ separately.

\subsubsection{The effect of $\betaSPH$ on the critical cooling timescale}
\label{sec:betaSPH_test}

\begin{table*}
\centering
  {\small
\begin{tabular}{llllll}
    \hline
    Simulation name & No of particles & $\alphaSPH$ & $\betaSPH$& $\beta$ & Fragmented?\\
    \hline
    \hline
    250k-betaSPH0.1-beta5 & 250,000 & 0.1 & 0.1 & 5.0 & Yes\\
    250k-betaSPH0.1-beta5.5 & 250,000 & 0.1 & 0.1 & 5.5 & Yes\\
    250k-betaSPH0.1-beta5.6 & 250,000 & 0.1 & 0.1 & 5.6 & No\\
    250k-betaSPH0.1-beta6 & 250,000 & 0.1 & 0.1 & 6.0 & No\\
    250k-beta5 & 250,000 & 0.1 & 0.2 & 5 & Yes\\
    250k-beta5.5 & 250,000 & 0.1 & 0.2 & 5.5 & Yes\\
    250k-beta5.6 & 250,000 & 0.1 & 0.2 & 5.6 & No\\
    250k-beta6 & 250,000 & 0.1 & 0.2 & 6.0 & No\\
    250k-beta6.5 & 250,000 & 0.1 & 0.2 & 6.5 & No\\
    250k-beta7 & 250,000 & 0.1 & 0.2 & 7.0 & No\\
    250k-beta7.5 & 250,000 & 0.1 & 0.2 & 7.5 & No\\
    250k-betaSPH0.4-beta6 & 250,000 & 0.1 & 0.4 & 6.0 & Yes\\
    250k-betaSPH0.4-beta6.5 & 250,000 & 0.1 & 0.4 & 6.5 & Yes\\
    250k-betaSPH0.4-beta6.8 & 250,000 & 0.1 & 0.4 & 6.8 & No\\
    250k-betaSPH0.4-beta7 & 250,000 & 0.1 & 0.4 & 7.0 & No\\
    250k-betaSPH1-beta6.5 & 250,000 & 0.1 & 1 & 6.5 & Yes\\
    250k-betaSPH1-beta6.8 & 250,000 & 0.1 & 1 & 6.8 & Yes\\
    250k-betaSPH1-beta7 & 250,000 & 0.1 & 1 & 7.0 & No\\
    250k-betaSPH2-beta4 & 250,000 & 0.1 & 2 & 4 & Yes\\
    250k-betaSPH2-beta5 & 250,000 & 0.1 & 2 & 5 & Yes\\
    250k-betaSPH2-beta6 & 250,000 & 0.1 & 2 & 6 & Yes\\
    250k-betaSPH2-beta7 & 250,000 & 0.1 & 2 & 7 & Yes\\
    250k-betaSPH2-beta8 & 250,000 & 0.1 & 2 & 8 & Yes\\
    250k-betaSPH2-beta8.5 & 250,000 & 0.1 & 2 & 8.5 & No\\
    250k-betaSPH2-beta9 & 250,000 & 0.1 & 2 & 9 & No\\
    250k-betaSPH2-beta10 & 250,000 & 0.1 & 2 & 10 & No\\
    250k-betaSPH4-beta8 & 250,000 & 0.1 & 4 & 8.0 & Yes\\
    250k-betaSPH4-beta8.5 & 250,000 & 0.1 & 4 & 8.5 & No\\
    \hline
    250k-alphaSPH0.05-beta7 & 250,000 & 0.05 & 2 & 7 & Yes\\
    250k-alphaSPH0.05-beta8 & 250,000 & 0.05 & 2 & 8 & No\\
    250k-alphaSPH0.05-beta9 & 250,000 & 0.05 & 2 & 9 & No\\
    250k-betaSPH2-beta4 & 250,000 & 0.1 & 2 & 4 & Yes\\
    250k-betaSPH2-beta5 & 250,000 & 0.1 & 2 & 5 & Yes\\
    250k-betaSPH2-beta6 & 250,000 & 0.1 & 2 & 6 & Yes\\
    250k-betaSPH2-beta7 & 250,000 & 0.1 & 2 & 7 & Yes\\
    250k-betaSPH2-beta8 & 250,000 & 0.1 & 2 & 8 & Yes\\
    250k-betaSPH2-beta8.5 & 250,000 & 0.1 & 2 & 8.5 & No\\
    250k-betaSPH2-beta9 & 250,000 & 0.1 & 2 & 9 & No\\
    250k-betaSPH2-beta10 & 250,000 & 0.1 & 2 & 10 & No\\
    250k-alphaSPH0.2-beta7 & 250,000 & 0.2 & 2 & 7 & Yes\\
    250k-alphaSPH0.2-beta7.5 & 250,000 & 0.2 & 2 & 7.5 & Yes\\
    250k-alphaSPH0.2-beta8 & 250,000 & 0.2 & 2 & 8 & No\\
    250k-alphaSPH0.5-beta6 & 250,000 & 0.5 & 2 & 6 & Yes\\
    250k-alphaSPH0.5-beta6.5 & 250,000 & 0.5 & 2 & 6.5 & Yes\\
    250k-alphaSPH0.5-beta7 & 250,000 & 0.5 & 2 & 7 & Yes\\
    250k-alphaSPH0.5-beta7.5 & 250,000 & 0.5 & 2 & 7.5 & No\\
    250k-alphaSPH1-beta5 & 250,000 & 1 & 2 & 5 & Yes\\
    250k-alphaSPH1-beta6 & 250,000 & 1 & 2 & 6 & Yes\\
    250k-alphaSPH1-beta6.5 & 250,000 & 1 & 2 & 6.5 & Yes\\
    250k-alphaSPH1-beta7 & 250,000 & 1 & 2 & 7 & No\\
 \hline
    31k-betaSPH2-beta3.5 & 31,250 & 0.1 & 2 & 3.5 & Yes\\
    31k-betaSPH2-beta4 & 31,250 & 0.1 & 2 & 4 & Yes\\
    31k-betaSPH2-beta4.5 & 31,250 & 0.1 & 2 & 4.5 & Yes\\
    31k-betaSPH2-beta5 & 31,250 & 0.1 & 2 & 5 & No\\
    250k-betaSPH2-beta4 & 250,000 & 0.1 & 2 & 4 & Yes\\
    250k-betaSPH2-beta5 & 250,000 & 0.1 & 2 & 5 & Yes\\
    250k-betaSPH2-beta6 & 250,000 & 0.1 & 2 & 6 & Yes\\
    250k-betaSPH2-beta7 & 250,000 & 0.1 & 2 & 7 & Yes\\
    250k-betaSPH2-beta8 & 250,000 & 0.1 & 2 & 8 & Yes\\
    250k-betaSPH2-beta8.5 & 250,000 & 0.1 & 2 & 8.5 & No\\
    250k-betaSPH2-beta9 & 250,000 & 0.1 & 2 & 9 & No\\
    250k-betaSPH2-beta10 & 250,000 & 0.1 & 2 & 10 & No\\
    2m-betaSPH2-beta12 & 2 million & 0.1 & 2 & 12 & Yes\\
    2m-betaSPH2-beta14 & 2 million & 0.1 & 2 & 14 & No\\
    \hline
  \end{tabular}
}
  \caption{Table showing the simulations carried out to investigate how the fragmentation boundary changes with the SPH artificial viscosity parameters (i) $\alphaSPH = 0.1$ and varying $\betaSPH$ using 250,000 particles (upper panel), (ii) $\betaSPH = 2.0$ and varying $\alphaSPH$ using 250,000 particles (middle panel) and (iii) ($\alphaSPH$, $\betaSPH$) = (0.1,~2.0) at different resolutions (lower panel).  The key fragmenting results are also indicated.}
\label{tab:alpha_beta_inves}
\end{table*}

\begin{figure}
\centering
  \includegraphics[width=1.0\columnwidth]{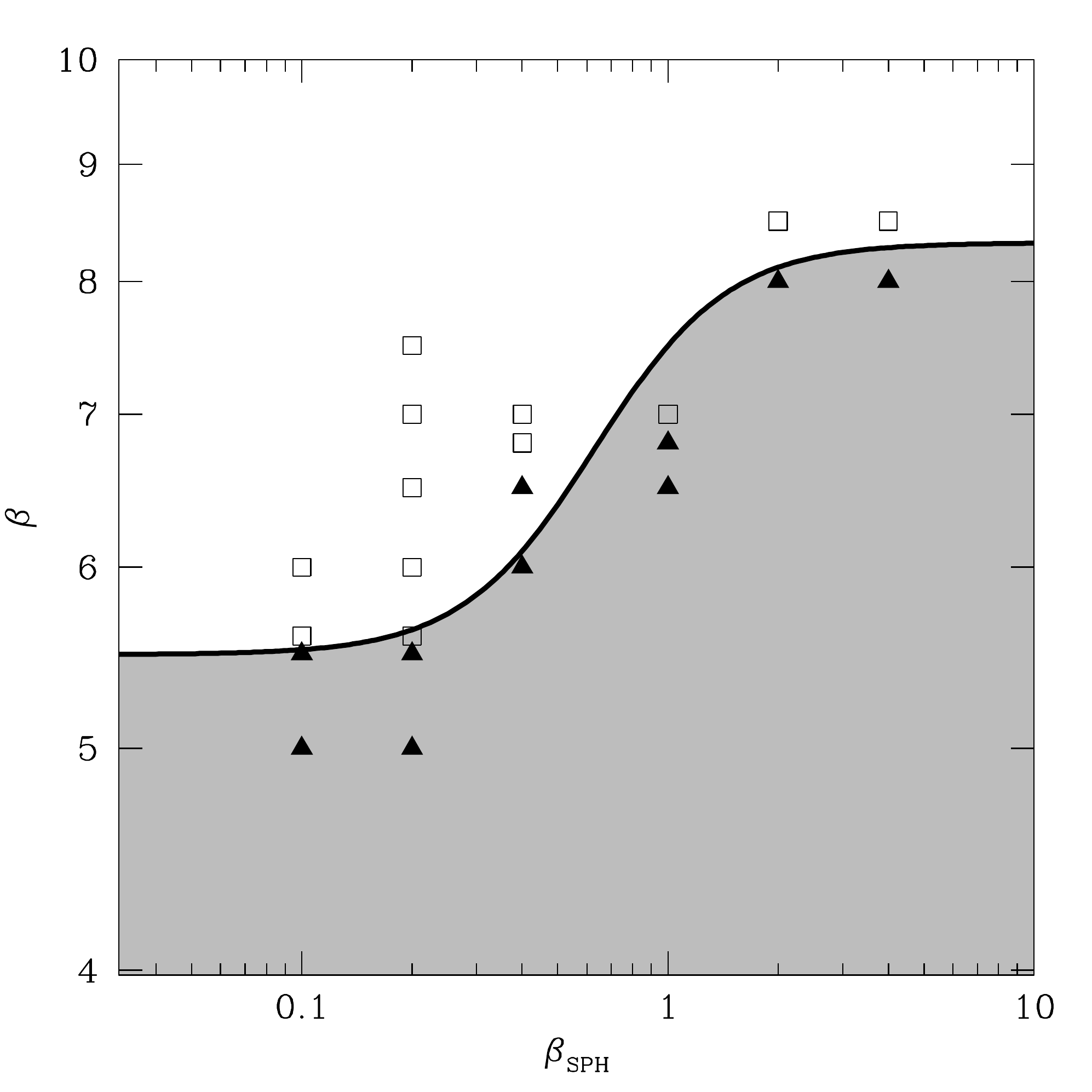}
  \caption{Graph of $\beta$ against $\betaSPH$ of the non-fragmenting (open squares) and fragmenting (solid triangles) simulations carried out using 250,000 particles and $\alphaSPH = 0.1$.  The solid line, included by eye, shows a dividing line between the fragmenting and non-fragmenting cases and the grey region is where fragmentation can take place.  The graph shows an S-shaped curve: any additional particle oscillation  that may exist is stopped using $\betaSPH \approx 2$ since the effect on the critical cooling timescale does not change above this value; at lower values of $\betaSPH$, the critical cooling timescale is smaller as the additional particle oscillation results in excess dissipation that needs to be overcome before fragmentation can take place.  At very low values of $\betaSPH$ either so much particle oscillation occurs (either at the edge of the shock front or due to particle interpenetration) or that the $\alphaSPH$ term dominates, that the effects of lowering $\betaSPH$ does not result in more dissipation.}
 \label{fig:S-shaped}
\end{figure}

In Section~\ref{sec:nonconverge} we show that the first-order convergence seen for the SPH results suggests that the $\alphaSPH$ term may be responsible.  However, as mentioned in Section~\ref{sec:av_res}, if the optimum value of $\betaSPH$ is not used (i.e. a value that minimises numerical dissipation) additional dissipation may occur and affect the fragmentation boundary.  Therefore, while not immediately obvious from the results in Section~\ref{sec:nonconverge}, the value of the $\betaSPH$ term may affect the fragmentation conclusions.  Table~\ref{tab:alpha_beta_inves} (top panel) and Figure~\ref{fig:S-shaped} summarise the results of the simulations carried out to investigate what effect the value of $\betaSPH$ has on the critical cooling timescale using 250,000 particles and maintaining a fixed value of $\alphaSPH = 0.1$.  It can be seen that the shape of the fragmenting/non-fragmenting boundary line appears to follow a somewhat S-shaped curve.  At high values of $\betaSPH$, any potential particle interpenetration is appropriately dealt with as changing the value of $\betaSPH$ from 2 to 4 has no effect on the critical cooling timescale.  As $\betaSPH$ is reduced, particle interpenetration and additional particle velocity dispersion can occur since the appropriate amount of the quadratic term of the artificial viscosity is not used.  Eventually, these oscillations are damped down by the $\alphaSPH$ term resulting in dissipation.  Consequently, a more rapid cooling is required to overcome the additional dissipation resulting in smaller critical cooling values.  At very low values of $\betaSPH$ the critical cooling timescale remains the same.  This may be due to one of two reasons: 1) a ``saturation'' of additional oscillations occurs such that reducing $\betaSPH$ further does not increase the dissipation - by this we mean that the cause of the oscillation (i.e. the incorrect modelling at the edge of the shock front and particle interpenetration) is so high that any reduction in $\betaSPH$ cannot cause more oscillation to occur; or 2) at such low values of $\betaSPH$, the linear artificial viscosity term dominates the dissipation such that any additional particle interpenetration does not increase the overall dissipation by much (an effect also noted by \citealp{Price_Federrath_betaSPH}).

In any case, Figure~\ref{fig:S-shaped} shows that the amount of $\betaSPH$ that is required to deal with the particle oscillations in this problem is $\approx 2$.  This value ensures that the dissipation resulting from artificial viscosity is as low as possible (since the value of $\betacrit$ that results is higher) and therefore is likely to give a result that is ``closer to the real answer''.  While these simulations are carried out at a single resolution (250,000 particles), \cite{Matthews_thesis} shows that $\betaSPH \approx 2$ is sufficient to stop particle interpenetration for Mach numbers across a shock of $\mathscr M \approx 3$.  In our simulations we find that the Mach numbers across the shock are up to $\approx 3$.  Note that the simulations presented by \cite{Rice_beta_condition} and \cite{Meru_Bate_resolution} were all carried out using $\betaSPH=0.2$.  Consequently, the calculations in both papers will have been affected by this and thus the converged value of $\betacrit$ is expected to be even higher than that suggested by Figure~\ref{fig:res_beta}.

\subsubsection{The effect of $\alphaSPH$ on the critical cooling timescale}
\label{sec:alphaSPH_test}

\begin{figure}
\centering
  \includegraphics[width=1.0\columnwidth]{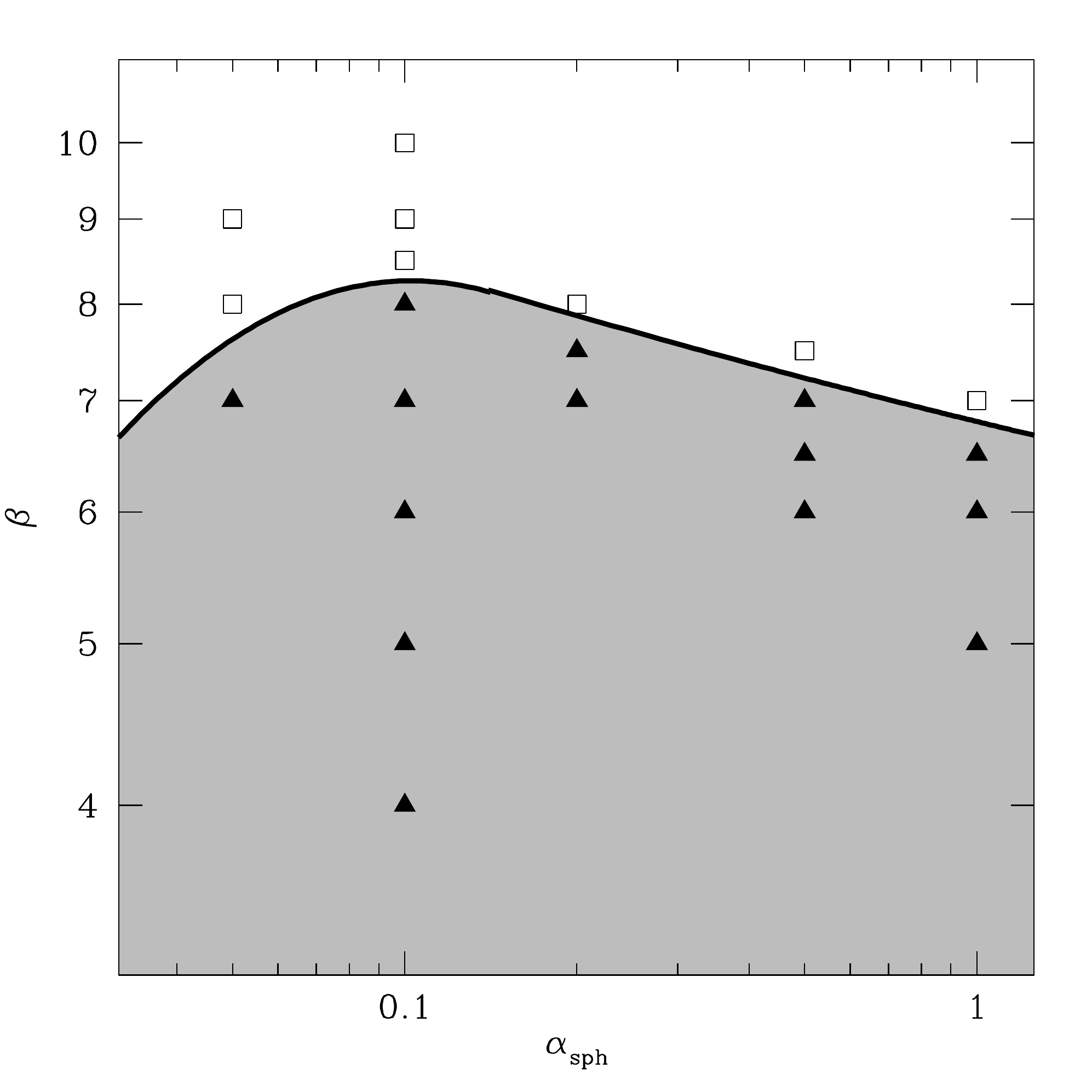}
  \caption{Graph of $\beta$ against $\alphaSPH$ of the non-fragmenting (open squares) and fragmenting (solid triangles) simulations carried out using 250,000 particles and with $\betaSPH = 2.0$.  The solid line, included by eye, shows a dividing line between the fragmenting and non-fragmenting cases and the grey region is where fragmentation can take place.  At high viscosities, the dissipation is higher (Figure~\ref{fig:diss_alphaSPH_betaSPH}, left panel), resulting in a faster cooling, i.e. a lower value of $\beta$, required to overcome the dissipation and cause fragmentation.  As $\alphaSPH$ is decreased, the dissipation also decreases requiring a slower cooling for fragmentation.  At very low values of $\alphaSPH$, additional dissipation occurs resulting in a lower value of $\betacrit$.  This may be due to additional particle oscillation as the shocks are not modelled adequately with such a low value of $\alphaSPH$.}
\label{fig:beta_alphaSPH}
\end{figure}

In Section~\ref{sec:SPH_av} we show that the dissipation due to the linear term in the artificial viscosity may play a part in the fragmentation results, and hence the value of $\betacrit$ at any one resolution.  In Section~\ref{sec:nonconverge} we show that this may indeed have been the case by considering the rate of convergence with increasing resolution.  In this section we maintain a fixed resolution using 250,000 particles (i.e. keep the value of $h/H$ constant) and use a fixed value of $\betaSPH = 2.0$, but vary the value of $\alphaSPH$ to confirm that equation~\ref{eqss} does indeed play a part in determining the fragmentation boundary.  Table~\ref{tab:alpha_beta_inves} (middle panel) and Figure~\ref{fig:beta_alphaSPH} summarise the simulations performed and the key fragmentation results.  At higher values of $\alphaSPH$ the dissipation due to the artificial viscosity is expected to increase.  Consequently, the cooling required to overcome this additional dissipation is larger and as a result, the critical cooling timescale for fragmentation is lower.  As the $\alphaSPH$ term is decreased, the amount of dissipation also decreases and thus the cooling does not have to be so rapid, resulting in a higher critical cooling timescale.  At values below $\alphaSPH \approx 0.1$, however, the dissipation increases once again as there is not enough artificial viscosity to remove the oscillations at shock fronts.  Examining the velocity dispersion of particles in the disc around their expected almost Keplerian values, we find that with very low viscosity, the velocity dispersion of the particles increases.  The particles are `jostled' by one another when the viscosity is lower and the relative motions grow larger (also see Section~\ref{sec:diss_nonGI}).  Therefore, even though the value of $\alphaSPH$ is decreased, the dissipation increases as the small amount of viscosity that is present tries to damp these larger velocities.  This suggests that $\alphaSPH \approx 0.1$ is a happy medium whereby it minimises the dissipation and avoids large oscillations at shock fronts.

\subsubsection{Determining the fragmentation boundary using optimum values of $\alphaSPH$ and $\betaSPH$}
\label{sec:frag_bdry_optimumSPH}

\begin{figure}
\centering
 \includegraphics[width=1.0\columnwidth]{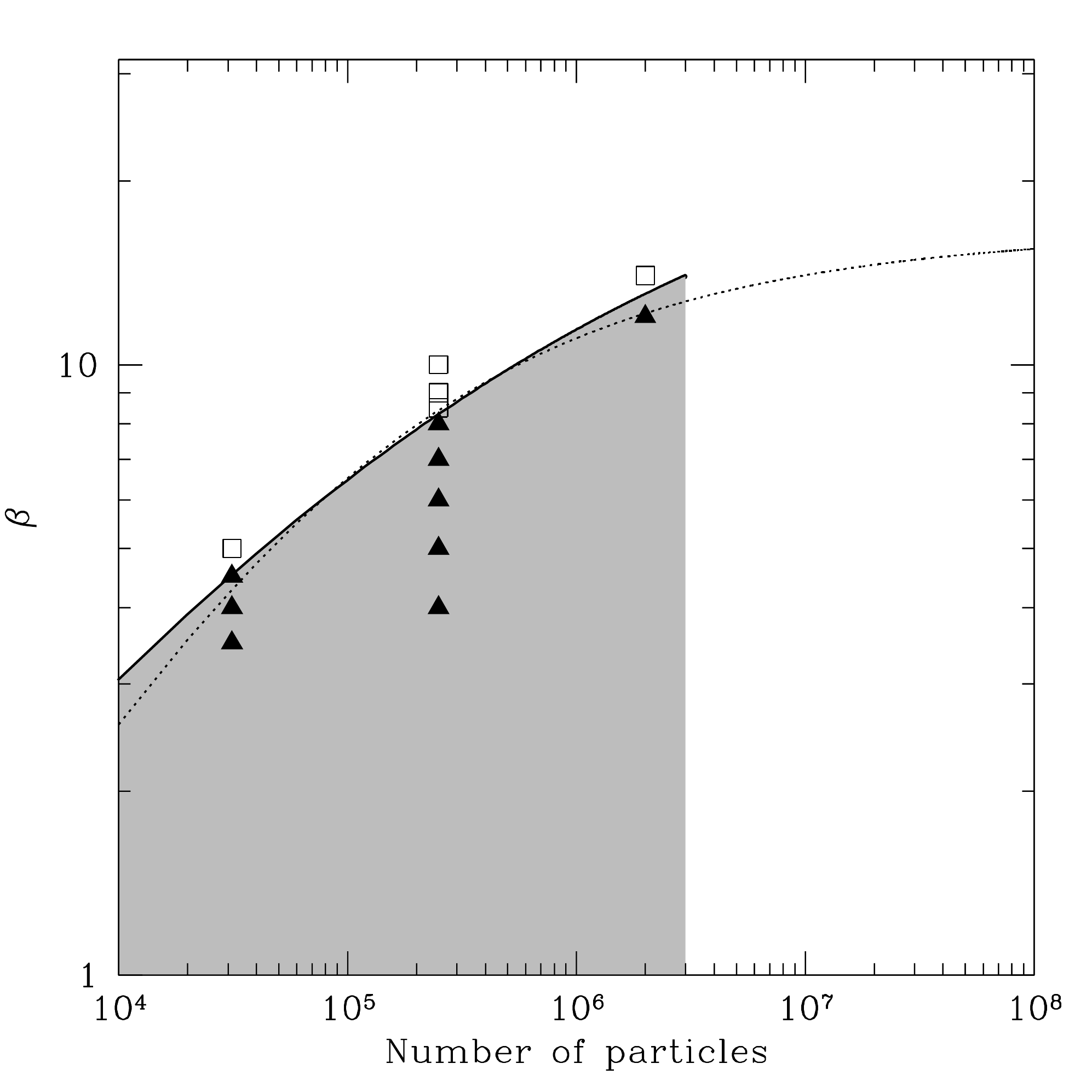}
  \caption{Graph of $\beta$ against resolution for non-fragmenting (open squares) and fragmenting (solid triangles) simulations carried out with ($\alphaSPH$, $\betaSPH$) = (0.1, 2.0) using 31,250, 250,000 and 2 million particles.  The solid line, obtained by fitting equation~\ref{eq:betacrit_hH}, shows a dividing line between the fragmenting and non-fragmenting cases and the grey region is where fragmentation can take place.  The region to the right of $\approx 2$~million particles has not been shaded in as it is unclear from these results alone what the shape of the dividing line would lie.  The dotted line shows the fit using equation~\ref{eq:betacrit_hH} when $\zeta$ is set to the minimum value it can be, i.e. unity.}
\label{fig:factor1.5}
\end{figure}

\begin{figure}
\centering
  \includegraphics[width=1.0\columnwidth]{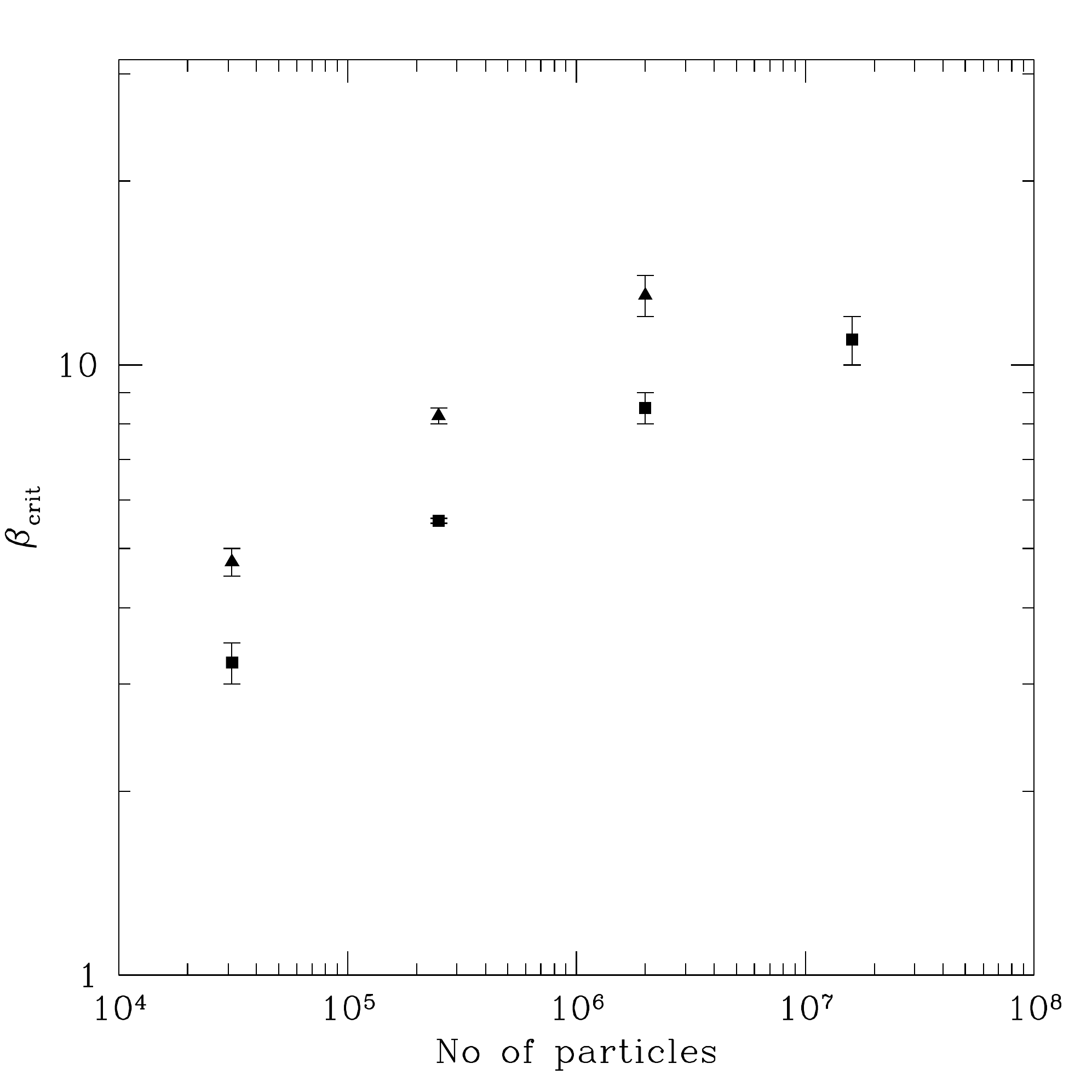}
  \caption{Graph of $\betacrit$ against resolution of the SPH simulations carried out using $\betaSPH = 0.2$ (squares) and $\betaSPH = 2.0$ (triangles).  The value of $\alphaSPH$ is 0.1.  It can be seen that the effect of increasing $\betaSPH$ to 2.0 (i.e. to a value that minimises the additional dissipation) is to increase the critical cooling timescale.}
\label{fig:SPH_betaSPH0.2_2_comparison}
\end{figure}

\begin{figure*}
\centering
\includegraphics[width=2.0\columnwidth]{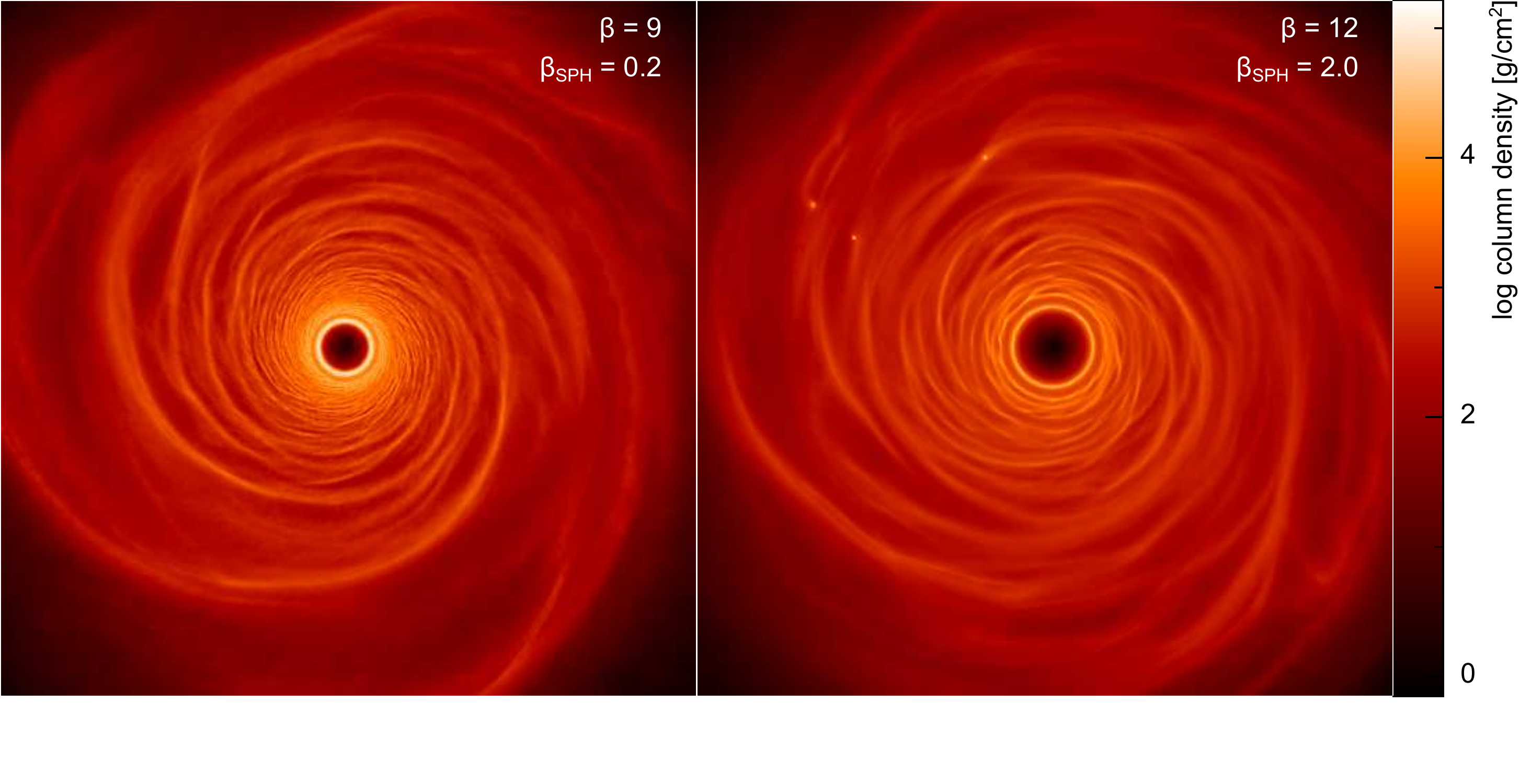}
 \caption{Surface mass density rendered image of discs modelled using 2 million particles and with $\alphaSPH = 0.1$.  The left image shows a disc modelled with $\betaSPH = 0.2$ with a cooling timescale, $\beta = 9$ while the right image shows a disc modelled with $\betaSPH = 2.0$ with a cooling timescale, $\beta = 12$.  The disc modelled using a lower amount of artificial viscosity does not fragment even though it is modelled with a faster cooling as counterintuitively, there is excess dissipation with a lower value of $\betaSPH$.}
\label{fig:disc_betaSPH}
\end{figure*}

The SPH artificial viscosity parameters that appear to produce a minimum excess dissipation for this problem are $(\alphaSPH, \betaSPH) \approx (0.1, 2.0)$.  However, given that the previous simulations did not use these optimum values \citep{Rice_beta_condition,Meru_Bate_resolution}, it is important to correct for this.  We therefore carry out a number of SPH simulations using 31,250, 250,000 and 2 million particles to determine what the critical cooling timescale is for the discs simulated using ($\alphaSPH$, $\betaSPH$) = (0.1, 2.0).  Table~\ref{tab:alpha_beta_inves} (bottom panel) and Figure~\ref{fig:factor1.5} summarise the results of these simulations.  It can immediately be seen that the critical cooling timescale is higher than the equivalent simulations with $\betaSPH = 0.2$ (also see Figure~\ref{fig:SPH_betaSPH0.2_2_comparison}).  Figure ~\ref{fig:disc_betaSPH} (left panel) shows an image of a fragmented disc modelled using 2 million SPH particles, ($\alphaSPH$, $\betaSPH$) = (0.1, 0.2) and a cooling timescale of $\beta = 9$ which fails to fragment.  Figure~\ref{fig:disc_betaSPH} (right panel) shows the equivalent disc modelled using ($\alphaSPH$, $\betaSPH$) = (0.1, 2.0) which fragments even though it is modelled using a slower cooling time of $\beta = 12$.  The results (Figures~\ref{fig:factor1.5} and~\ref{fig:SPH_betaSPH0.2_2_comparison}) still show that as the resolution increases, the critical cooling timescale increases, consistent with the results presented with a lower value of $\betaSPH$.  However, since there are only three data points with $\betaSPH = 2.0$, it is firstly not clear whether convergence exists and secondly, what function should be used to fit this data.  If we assume a functional form as given by equation~\ref{eq:fit}, we find that $\betacrit = 36.6 \pm 6.9$, assuming a first-order convergence rate (i.e. $\sigma = 1.0$) as indicated in Section~\ref{sec:nonconverge}.  Using the same convergence rate as found in Section~\ref{sec:nonconverge} (i.e. $\sigma = 1.08$) we find that $\betacrit = 29.2 \pm 2.4$.  This implies that the true value may well be as high as $\approx 30$.  We note that a fit assuming a second-order convergence rate gives a poor fit to the data.  In Section~\ref{sec:nonconverge}, omitting the 16 million particle data point makes no significant difference to the value of $\betacrit$ obtained due to the slow convergence rate.  Therefore we do not expect the absence of the 16 million particle data point here to significantly affect the fit.

Furthermore, we attempt to fit the analytical formula from equation~\ref{eq:betacrit_hH}.  Allowing all three parameters to vary ($\alpha_{\rm GI, crit}$, $\eta$ and $\zeta$) we find that $\alpha_{\rm GI, crit} = 0.015$, $\eta = 14.0$ and $\zeta = 0.3$.  There are several points to note here.  Firstly, this value of the critical gravitational stress is equivalent to a critical cooling timescale in the limit of infinite resolution of $\betacrit \approx 27$.  Secondly, the values of $\eta$ and $\zeta$ are much less than those obtained in Section~\ref{sec:nonconverge}.  This implies that when increasing the quadratic artificial viscosity term to $\betaSPH = 2.0$, not only does the total dissipation decrease but the level of excess dissipation also decreases.  However, we express caution here: the value of $\zeta$ obtained here is lower than unity which is not possible.  This is likely to be an artefact of using three data points to fit three unknowns.  We therefore refit the data by setting $\zeta$ to the minimum possible value it can be, i.e. unity.  In this case, we find that $\alpha_{\rm GI, crit} = 0.024 \pm 0.005$ and $\eta = 6 \pm 3$ (see dotted line in Figure~\ref{fig:factor1.5}).  Again, we emphasise that $\eta$ is smaller than previously obtained in Section~\ref{sec:nonconverge} suggesting that with $\betaSPH = 2.0$, the excess dissipation is significantly reduced.

We note that the analytical formula in equation~\ref{eqnumerical} is dependent on two aspects.  Firstly it assumes that the ratio of the smoothing length to the disc scaleheight, $h/H$, is given by equation~\ref{eq:h_H}.  Secondly, it assumes that the dissipation due to the artificial viscosity is indeed given by equation~\ref{eq:av_dissipation}.  To check the analytical arguments presented in Section~\ref{sec:SPH_av}, it is important to test these two aspects.

\subsubsection{Testing the analytical formula for $h/H$}
\label{sec:h_H}

\begin{figure*}
\centering
  \includegraphics[width=0.66\columnwidth]{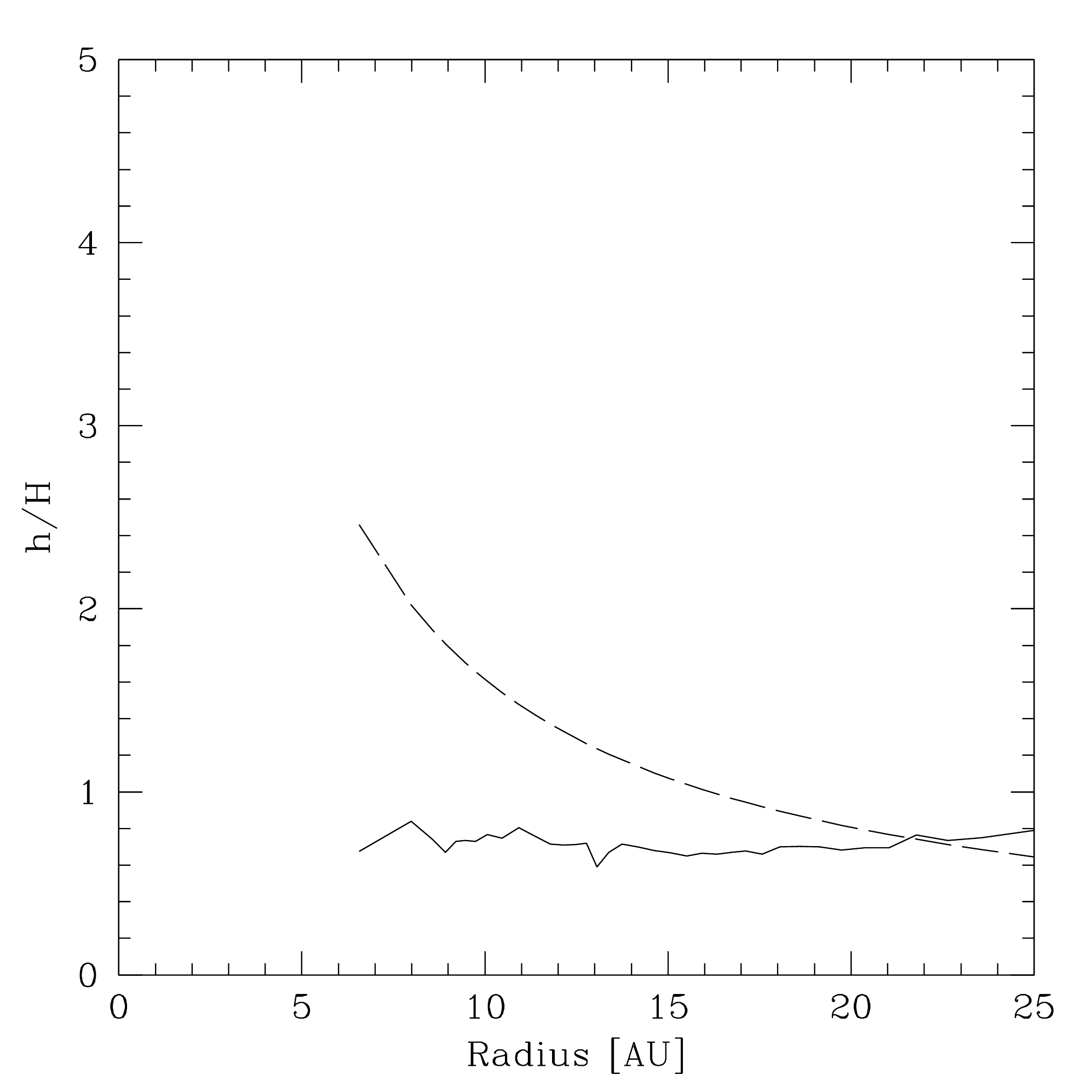}
  \includegraphics[width=0.66\columnwidth]{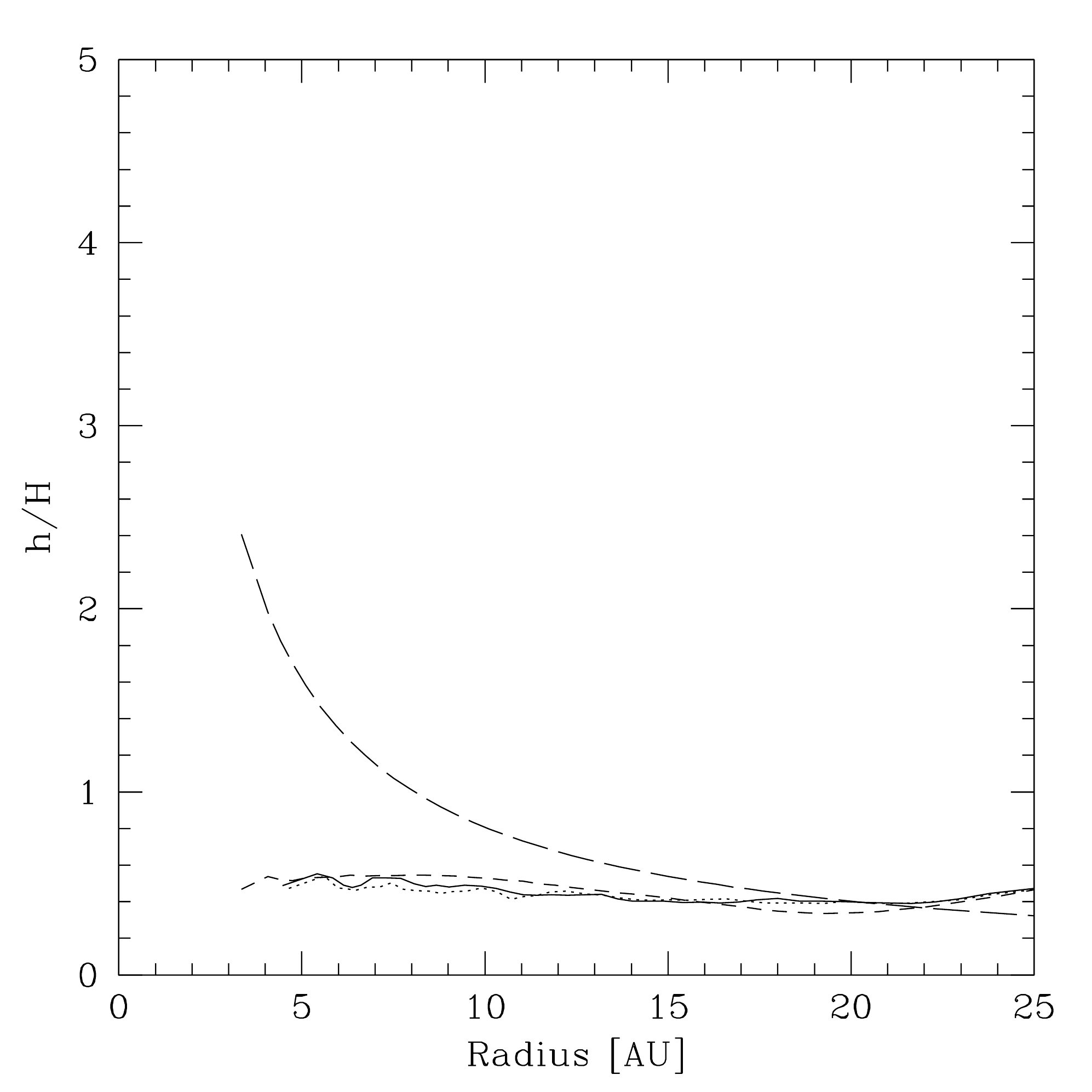}
  \includegraphics[width=0.66\columnwidth]{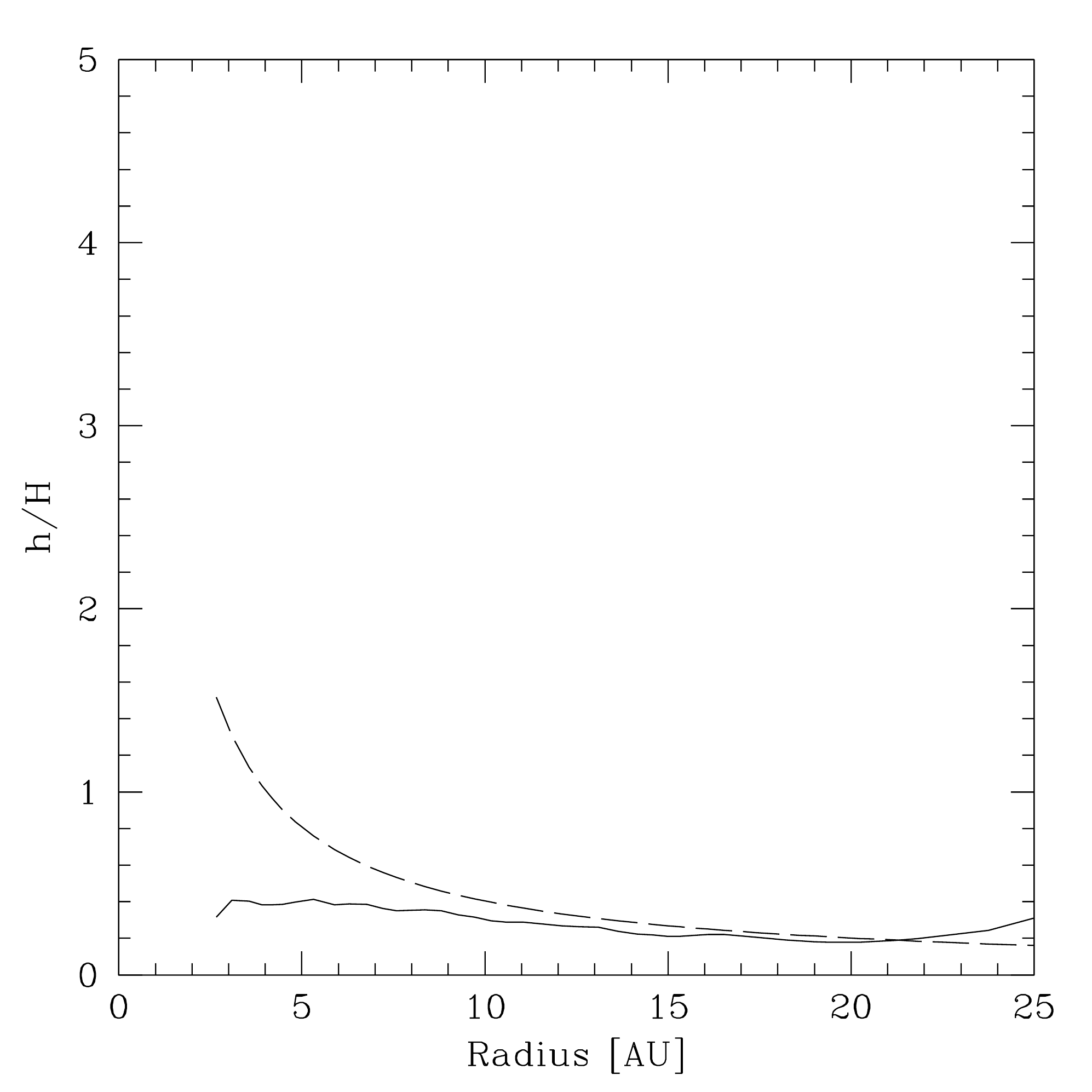}
  \caption{Graph of the azimuthally averaged and the analytically estimated (using Equation~\ref{eq:h_H}) radial profile of the ratio of the smoothing length to disc scaleheight, $h/H$, for the non-fragmenting (i.e. marginally stable, $Q \approx 1$) discs presented in Table~\ref{tab:alpha_beta_inves} (bottom panel) using 31,250 (left panel), 250,000 (middle panel) and 2 million (right panel) particles.  The analytically estimated radial profile is plotted using long dashed lines while all other lines are the simulation results.  In the outer parts of the disc where fragmentation will occur when the cooling is close to the fragmentation boundary, the azimuthally averaged measured values are very close to the expected values.}
\label{fig:h_H_profile}
\end{figure*}

Figure~\ref{fig:h_H_profile} shows the analytical estimate of the ratio of the smoothing length to the disc scaleheight, $h/H$, against the azimuthally averaged radial profile of $h/H$ for the non-fragmenting discs (i.e. marginally stable discs where $Q \approx 1$) carried out in Section~\ref{sec:frag_bdry_optimumSPH} (bottom section of Table~\ref{tab:alpha_beta_inves}).  It is very clear from this graph that at all resolutions considered, the analytical estimate of $h/H$ is a good approximation in the outer parts of the disc where the fragments generally form.  It is important to note, that the analytical formula assumes the initial surface mass density profile remains constant.  However, the discrepancy in the inner regions is due to the change in surface mass density profile as the disc evolves into a state of mechanical equilibrium on a viscous timescale.  The change in surface mass density profile thus changes the value of $h/H$ (equation~\ref{eq:h_H}).  Since the viscous timescale is shorter at small radii, the disc evolves more rapidly there and thus the discrepancy is larger.  However, it is important to note that for a cooling timescale close to the critical one, fragmentation occurs in the outer parts of the discs for surface mass density profiles shallower than $\Sigma \propto R^{-2}$ (\cite{Meru_Bate_fragmentation}) since the resolution increases with radius (equation~\ref{eq:h_H}).  The outer parts are where the agreement is best between the analytically expected and azimuthally averaged values of $h/H$.  We also note that the agreement between the analytical formula for $h/H$ and the simulation data is better with increasing resolution as the viscosity decreases (equations~\ref{eqss} and~\ref{eqssquad}) and so the effective viscous time is larger resulting in a slower evolution of the surface mass density profile.  We therefore conclude that the reason why the value of $\eta$ in Section~\ref{sec:frag_bdry_optimumSPH} is not unity cannot therefore be put down to a mismatch between the analytical and actual values of $h/H$ in the region where fragmentation will occur.

\subsubsection{Testing the analytical formula for $\betacrit$ using non-self-gravitating discs}
\label{sec:diss_nonGI}

\begin{table}
\centering
  {\small
\begin{tabular}{lll}
    \hline
    $\alphaSPH$ & $\betaSPH$ & No of particles\\
    \hline
    \hline
    0.01 & 2.0 & 250,000\\
    0.025 & 2.0 & 250,000\\
    0.03 & 2.0 & 250,000\\
    0.05 & 2.0 & 250,000\\
    0.1 & 2.0 & 250,000\\
    0.25 & 2.0 & 250,000\\
    0.5 & 2.0 & 250,000\\
    1.0 & 2.0 & 250,000\\
    3.0 & 2.0 & 250,000\\
    10.0 & 2.0 & 250,000\\
    \hline
    0.1 & 0.2 & 250,000\\
    0.1 & 0.4 & 250,000\\
    0.1 & 0.6 & 250,000\\
    0.1 & 0.8 & 250,000\\
    0.1 & 1.0 & 250,000\\
    0.1 & 1.2 & 250,000\\
    0.1 & 1.4 & 250,000\\
    0.1 & 1.6 & 250,000\\
    0.1 & 1.8 & 250,000\\
    0.1 & 2.0 & 250,000\\
    \hline
  \end{tabular}
}
  \caption{Table showing the simulations carried out \emph{without} self-gravity to determine if the dissipation due to the artificial viscosity is the same as that expected in a shear-dominated disc.  The value of $\alphaSPH$ is changed while maintaining a fixed value of $\betaSPH = 2.0$ (top panel).  The bottom panel shows the simulations carried out with $\alphaSPH = 0.1$ and varying the value of $\betaSPH$.}
\label{tab:non-GI}
\end{table}

\begin{figure*}
\centering
  \includegraphics[width=1.0\columnwidth]{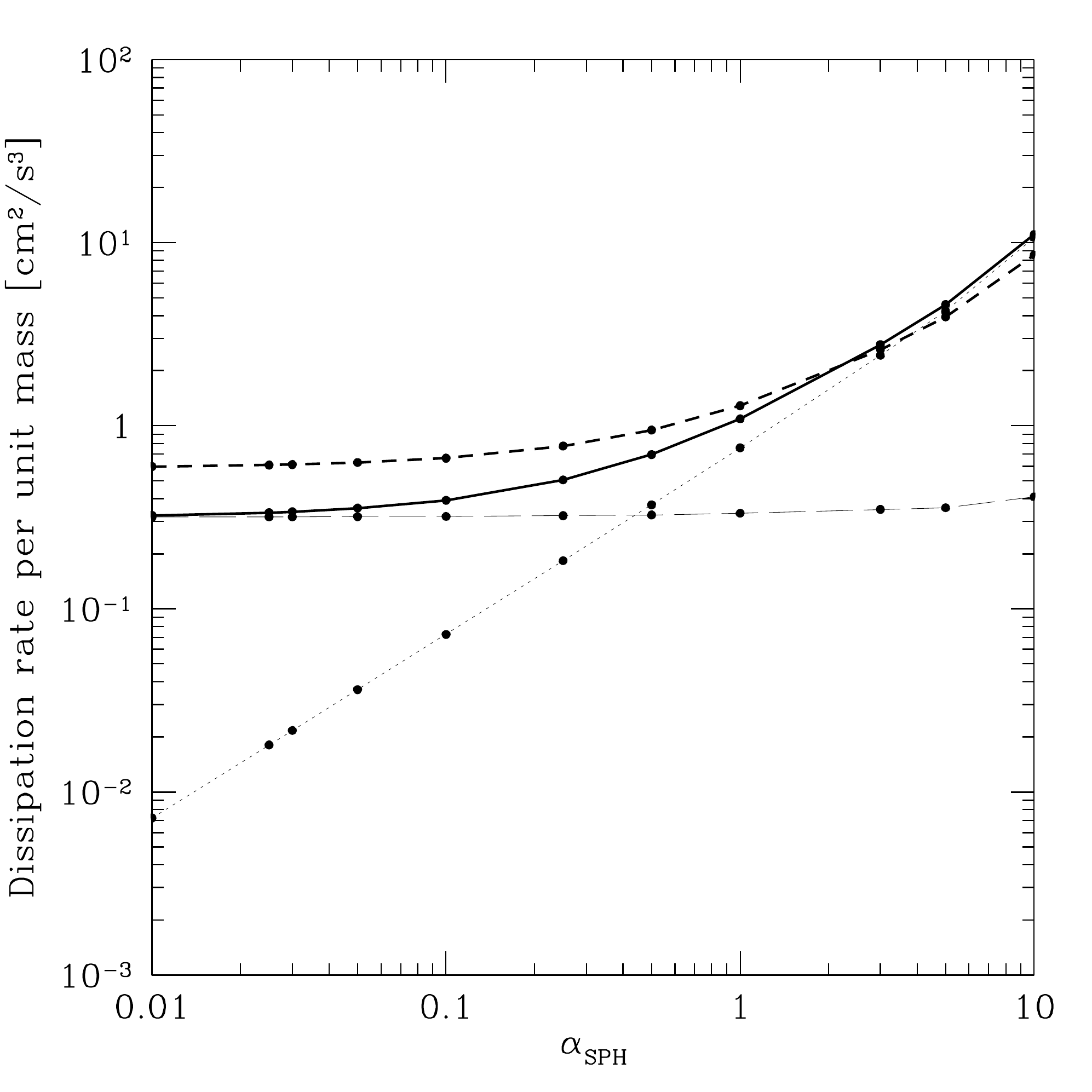}
  \includegraphics[width=1.0\columnwidth]{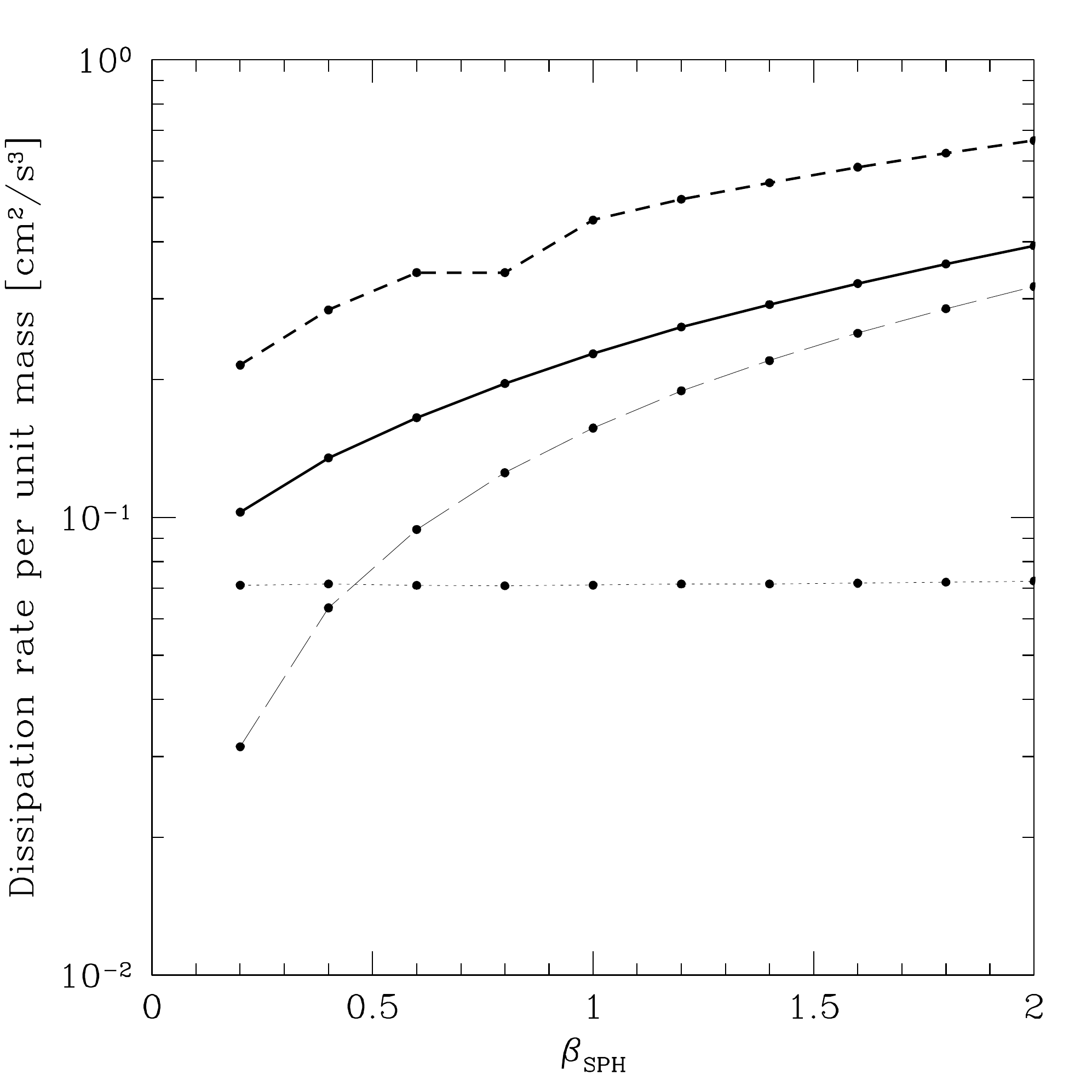}
 \caption{Graph of measured dissipation rate per unit mass (short dashed line) in non-self-gravitating discs against $\alphaSPH$ (left panel, using $\betaSPH = 2.0$) and $\betaSPH$ (right panel, using $\alphaSPH = 0.1$), modelled using 250,000 particles.  The expected dissipation rate per unit mass due to the $\alphaSPH$ (dotted line) and $\betaSPH$ (long dashed line) terms and the combined total expected dissipation due to the artificial viscosity (solid line) are also plotted (using equation~\ref{eq:orig_dissipation}).  The actual measured dissipation rate is higher than the analytical estimates of the dissipation due to the shear in all cases other than when $\alphaSPH$ is high.  In addition, the dissipation due to the $\betaSPH$ term is not always negligible, as is often presumed to be the case.}
\label{fig:diss_alphaSPH_betaSPH}
\end{figure*}

\begin{figure}
 \centering
  \includegraphics[width=1.0\columnwidth]{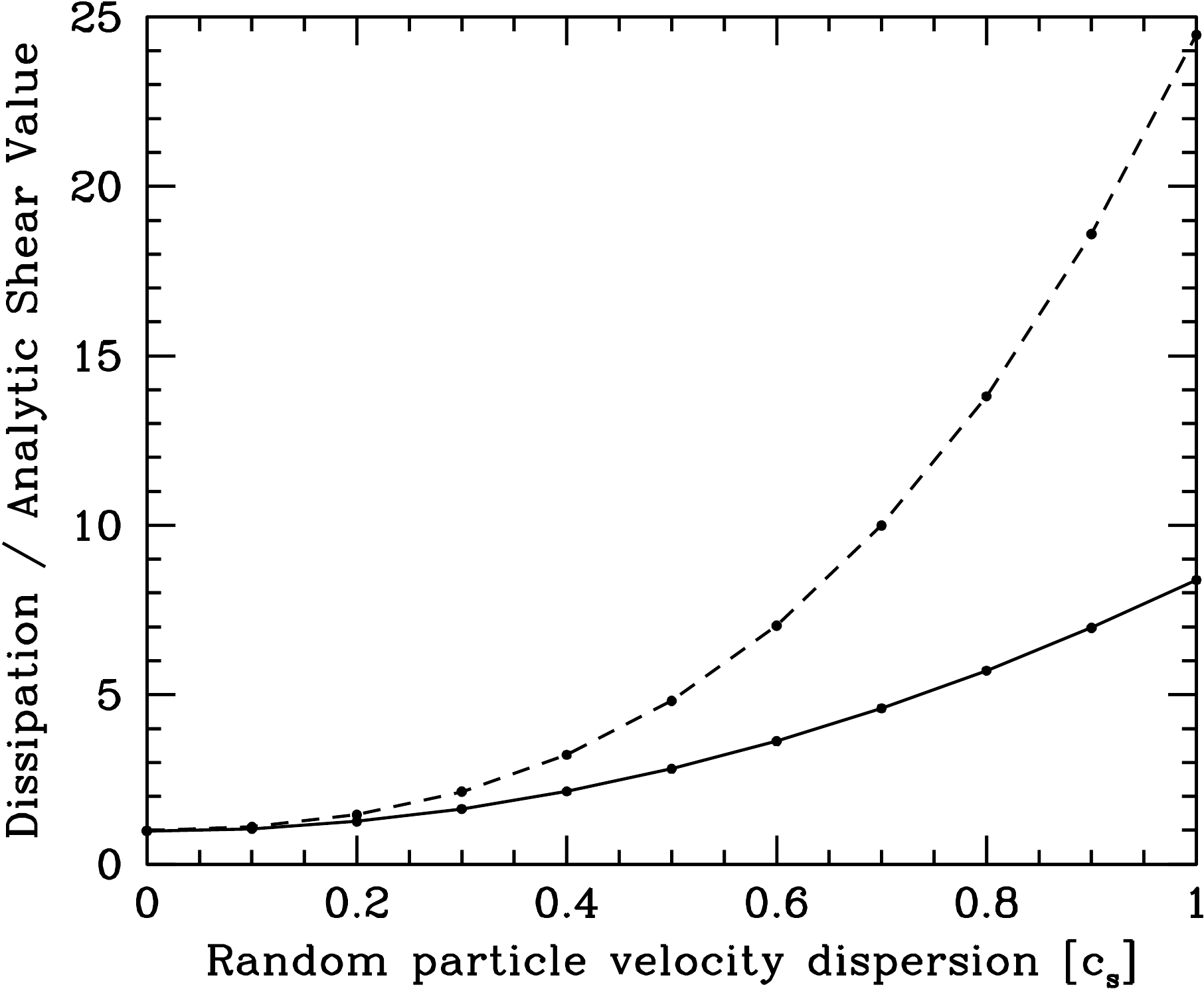}
 \caption{Setting up an initial disc model with purely Keplerian shear flow, we measure the instantaneous viscous dissipation averaged over $\approx 800$ particles in a thin radial extent.  We then add increasing random particle velocities to the disc setup and measure the dissipation.  In this figure, we plot the measured values of the dissipation divided by the analytic values expected for purely Keplerian flow due to the $\alpha_{\rm SPH}$ (solid line) and $\beta_{\rm SPH}$ (dashed line) terms separately versus the magnitude of the random velocity dispersion in units of the sound speed (i.e. we plot $\eta$ (solid line) and $\zeta$ (dashed line) as defined in equation~\ref{eqnumerical}).  The excess dissipation due to small-scale particle velocity dispersion can be substantially higher than that produced by a pure shear flow.}
  \label{fig:dissipation}
\end{figure}

We carry out a number of simulations of discs with 250,000 particles with the same physical parameters as described in Section~\ref{sec:sim} but \emph{without} self-gravity.  Table~\ref{tab:non-GI} shows a summary of these simulations.  The goal of this exercise is to see what effect the change in the SPH artificial viscosity parameters has on the dissipation in a \emph{laminar} disc \emph{which should only be due to shear} and whether this is as we would expect from the analytical formulae.  To do this, we must start from exactly the same disc.  However, the effects of the initial conditions must also be removed as this may affect the amount of dissipation.  Therefore, we run a disc using a cooling time, $\beta = 20$, for 1.5 ORPs using the artificial viscosity parameters $(\alphaSPH, \betaSPH) = (0.1, 2.0)$ (i.e. the values that we expect would minimise the additional dissipation).  This is equivalent to $\approx 3$ orbital periods at 15~au (where this analysis is done).  The cooling time (equation~\ref{eq:beta}) is also $\approx 3$ orbital periods at 15~au.  Since the initial evolution time and the cooling time are approximately equal, the disc is then settled such that the heating matches the cooling.  We then change the artificial viscosity parameters in the disc according to what is shown in Table~\ref{tab:non-GI} and run the simulations for a short period of time (0.1 ORPs or $\approx 0.2$ orbital periods at 15~au) and measure the total dissipation rate due to the artificial viscosity in the radial range $14.9 \le R \le 15.1$~au and compare these with the expected dissipation due to the artificial viscosity using equation~\ref{eq:orig_dissipation} (using the actual values of the sound speed and smoothing length obtained from the simulation rather than the initial values).  It is important to note that in order to make this comparison, we only calculate the dissipation over a short period of time as we are comparing the instantaneous expected dissipation rate with the instantaneous actual (azimuthally averaged) dissipation rate.  If we allow the discs to run for a very long time before measuring the dissipation rate, the discs will evolve considerably and a like-for-like comparison is then not possible.  This subsequent evolution takes place over a much smaller timescale than the orbital timescale.  Therefore, there is no time for $h/H$ or the velocity field of the particles to change.  Therefore any change in the disc's dissipation \emph{must} be due to the change in artificial viscosity parameters.

Figure~\ref{fig:diss_alphaSPH_betaSPH} (left panel) shows a graph of how the dissipation rate (measured and expected) changes with $\alphaSPH$ (using a fixed $\betaSPH = 2.0$).  The expected dissipation rate is the sum of the dissipation due to the $\alphaSPH$ and $\betaSPH$ terms.  It can immediately be seen that at low values of $\alphaSPH$, the expected dissipation due to the $\betaSPH$ term is very important (though its contribution is often thought to be negligible in comparison to the $\alphaSPH$ term).  At high values of $\alphaSPH$ the measured dissipation matches the expected values very well.  Most strikingly, the total dissipation is \emph{higher} than the expected dissipation from the analytical formula at low values of $\alphaSPH$.  The discrepancy is about a factor of two for $\alphaSPH \lesssim 0.1$.

Figure~\ref{fig:diss_alphaSPH_betaSPH} (right panel) shows the measured and expected dissipation rates against $\betaSPH$ (using a fixed $\alphaSPH = 0.1$).  In this case the total dissipation is always approximately a factor of 2 higher than the expected values in these non-self-gravitating calculations.  However, we note that there is no obvious additional dissipation at small values of $\betaSPH$ compared to large values, in contrast to the self-gravitating calculations in Figure~\ref{fig:S-shaped} which shows a definite difference in results between low and high $\betaSPH$ values.  The $\betaSPH$ viscosity was originally introduced into SPH to stop particle interpenetration at shocks in supersonic flows. Shocks are not present in the non-self-gravitating calculations, so particle interpenetration is not an issue, but shocks play a significant role in the self-gravitating calculations.  Thus, the apparent reduction of the dissipation in the self-gravitating calculations when the value of $\betaSPH$ is increased is likely to be because particle interpenetration is stopped more effectively with a higher value of $\betaSPH$.  Regardless, both panels in Figure~\ref{fig:diss_alphaSPH_betaSPH} clearly show that the expected contribution to the dissipation from the quadratic artificial viscosity term can be larger than that from the linear term.  In particular, for the simulation using $(\alphaSPH, \betaSPH) = (0.1, 0.2)$, which were the values used by \cite{Rice_beta_condition} and \cite{Meru_Bate_resolution} as well as many others, the dissipation is more than three times larger than the analytically expected value from the $\alphaSPH$ viscosity alone.

The level of dissipation expected from the SPH artificial viscosity given in Appendix~\ref{appendixD} is lower than that measured from the actual simulations.  What is the source of the excess dissipation that we find?  The key is that the derivation assumes that the only contribution from the artificial viscosity to the thermal energy is due to shear flow in a purely Keplerian disc.  Any other motions will add to this dissipation.  In a gravitationally unstable disc, we also expect heating from the bulk component of the artificial viscosity due to shocks generated in the disc.  Indeed, this is the assumed source of heating that is supposed to allow a gravitationally unstable disc to achieve a quasi-steady state when an imposed cooling timescale, $\beta$, is applied.  However, it is exactly this maximum heating rate that the disc can provide without fragmenting that we are trying to measure when we try to determine $\beta_{\rm crit}$. The fact that the convergence rate of $\beta_{\rm crit}$ with increasing resolution is slow (first order) and that increasing $\beta$ from 0.2 to 2.0 increases the critical cooling timescale significantly implies that there is a third source of heating.  Furthermore, as demonstrated earlier in this section, even when we measure the dissipation in a calculation without self-gravity, we still find some excess dissipation beyond what Appendix~\ref{appendixD} predicts.  Figure A3 of \cite{Lodato_Rice_original} shows a calculation of the Reynolds stress in a non-self-gravitating disc with $(\alphaSPH, \betaSPH) = (0.1, 0.2)$.  They find that the $\alpha_{\rm SS}$ parameter due to this is a few $\times 10^{-3}$ in the range $0 \le R \le 25$~au.  As a check to ensure we are consistent with previous results, we calculate the Reynolds stress in the same way as \cite{Lodato_Rice_original} (though we only average over 0.1ORPs) and also find the $\alpha_{\rm SS}$ value to be a few $\times 10^{-3}$ over the same radial range.  Furthermore, our results are also consistent with \cite{Forgan_alpha} who also find their $\alpha_{\rm SS}$ parameter to be a few $\times 10^{-3}$ (in the inner parts of their self-gravitating discs where the effects of self-gravity are not very important; see their Figure 4).

In a non-self-gravitating calculation there are essentially only two possible contributions to the viscous heating.  The first is from the Keplerian shear flow. The second is any additional particle motions.  Without self-gravity, these can only come from `random' particle motions.  It is well known that in a typical SPH simulation the particles `jostle' one another, resulting in a particle velocity dispersion.  This velocity dispersion results from errors in the pressure gradients due to the finite number of particles within a smoothing kernel.  In a compressible SPH simulation, such motions are typically at the level of some fraction of the sound speed.  In order to determine the effect of these motions on the dissipation in a disc we perform a simple toy experiment whereby for illustrative purposes, we introduce different amoounts of particle velocity dispersion to see its effects on the dissipation in the disc.  Setting up a purely Keplerian disc, we compute the instantaneous average values of the dissipation for particles in a small radial extent, due to the $\alpha_{\rm SPH}$ and $\beta_{\rm SPH}$ terms separately.  Comparing these values to those expected from the $\alphaSPH$ and $\betaSPH$ terms respectively (Appendix~\ref{appendixD}), i.e.

\begin{equation}
D_{\alpha} = \frac{93}{700} \eta \alphaSPH c_s h \Omega^2
\label{eq:Dalpha}
\end{equation}
and

\begin{equation}
D_{\beta} = \frac{81}{280} \zeta \betaSPH h^2 \Omega^3,
\label{eq:Dbeta}
\end{equation}
respectively.  We find that as expected, $\eta=\zeta=1$ to a high level of precision ($\approx 1-3$ per cent when averaging over $\approx 800$ particles).  We then experiment with adding different levels of random velocities in addition to the underlying Keplerian motion.  The results are displayed in Figure~\ref{fig:dissipation}, where the magnitude of the particle velocity dispersion is given as a fraction of the local sound speed in the disc.  We see that if random motions at the level of, e.g. 30 per cent of the sound speed are present, the dissipation increases by factors of $\eta=1.7$ and $\zeta=2.2$.  This provides us with an explanation for the excess dissipation in the non-self-gravitating calculations.  When the level of artificial viscosity is low, the velocity dispersion of the particles in the disc generates a non-negligible fraction of the dissipation.  When the viscosity is high (in particular the linear $\alpha_{\rm SPH}$ term), this velocity dispersion is damped, and since the contribution to the dissipation from the shear flow is larger, no significant dissipation beyond that expected from the shear flow is found.  We can see from Figure~\ref{fig:diss_alphaSPH_betaSPH} that $\alpha_{\rm SPH}=0.1$ is too low to effectively damp the particle velocity dispersion, resulting in dissipation rates that are approximately a factor of two larger than expected.  This does not require a high level of particle velocity dispersion -- from Figure~\ref{fig:dissipation} we see that a velocity dispersion of only $\approx 25$ per cent of the sound speed is enough to boost the dissipation due to the $\beta_{\rm SPH}$ viscosity by a factor of two.  Thus, although we show in Sections~\ref{sec:betaSPH_test} and~\ref{sec:alphaSPH_test} that the minimum dissipation is obtained for $(\alpha_{\rm SPH}, \beta_{\rm SPH}) = (0.1, 2.0)$, this minimum dissipation is still larger than that expected from the analytic derivation.

In the self-gravitating disc calculations the situation is more complex.  Here there are gravitational forces from the fluid, shocks and local pressure gradients in the disc which can stir up the particles.  A low level of artificial viscosity (particularly $\beta_{\rm SPH}$) will allow particle penetration in shocks and a low value of $\alpha_{\rm SPH}$ will be ineffective at damping post-shock oscillations and other small-scale particle motions.  If the random motions become a substantial fraction of the sound speed, which they may well do since we find the Mach numbers across a shock to be up to $\approx 3$, the factors can become very large ($\eta=3-9$ and $\zeta=7-24$ for random velocities of $60-100$ per cent of the sound speed).  Thus, if random particle motions are indeed also playing a part in the self-gravitating calculations, it should be no surprise that we infer a level of dissipation that is well beyond that expected from equation~\ref{eq:orig_dissipation}, i.e. the counter-intuitive nature of $\betaSPH$ that leads to Figure~\ref{fig:S-shaped}.  This leaves us with a problem with the SPH simulations.  In order to obtain a level of dissipation that is close to that predicted by a purely Keplerian flow we can infer from Figure~\ref{fig:diss_alphaSPH_betaSPH} that we would need to use $\alpha_{\rm SPH} \gsim 1$ and from Figure~\ref{fig:S-shaped} that we require $\beta_{\rm SPH} \ge 2$.  This should cut the particle penetration at shocks, post shock oscillations, and other particle velocity dispersion to low levels, thus making the analytic predictions of the dissipation accurate.  On the other hand, the higher level of viscosity would increase the dissipation generated by the shear flow, thus reducing the measured value of $\beta_{\rm crit}$ at a given resolution (c.f. Figure~\ref{fig:beta_alphaSPH}).  Thus, although the simulations may be better behaved, an even higher numerical resolution would be needed to determine the converged value of the critical cooling timescale, $\beta_{\rm crit}$.  We discuss other options in Section~\ref{sec:disc}.

In summary, in a purely Keplerian disc, with no random particle velocity dispersion, the dissipation is as expected from the analytical values in equation~\ref{eq:orig_dissipation}.  However, in the simulations of non-self-gravitating discs, particularly with low values of $\alphaSPH$, the dissipation is somewhat higher than expected.  We attribute this to random particle velocity dispersion, since there is no other source of heating in such discs over and above the viscous heating due to Keplerian shear flow.  In self-gravitating discs additional particle dispersion will be present, which may well result in the counter-intuitive nature of the artificial viscosity that leads to Figure~\ref{fig:S-shaped}, i.e. more dissipation with lower $\betaSPH$.

\subsection{The effect of the {\sc fargo} artificial viscosity on the critical cooling timescale}
\label{sec:FARGO_av0.5}

\begin{table}
\centering
  {\small
\begin{tabular}{lllll}
    \hline
    Simulation name & q & $\beta$ & Fragmented?\\
   \hline
    \hline
    786k\_cells-q0-beta10 & 0 & 10 & Yes\\
    786k\_cells-q0-beta11 & 0 & 11 & No\\
    786k\_cells-q0-beta12 & 0 & 12 & No\\
    786k\_cells-q0.01-beta10 & 0.01 & 10 & Yes\\
    786k\_cells-q0.01-beta11 & 0.01 & 11 & Yes\\
    786k\_cells-q0.01-beta12 & 0.01 & 12 & No\\
    786k\_cells-q0.05-beta10 & 0.05 & 10 & Yes\\
    786k\_cells-q0.05-beta11 & 0.05 & 11 & Yes\\
    786k\_cells-q0.05-beta12 & 0.05 & 12 & No\\
    786k\_cells-q0.1-beta8 & 0.1 & 8 & Yes\\
    786k\_cells-q0.1-beta9 & 0.1 & 9 & Yes\\
    786k\_cells-q0.1-beta10 & 0.1 & 10 & Yes\\
    786k\_cells-q0.1-beta10.5 & 0.1 & 10.5 & Yes\\
    786k\_cells-q0.1-beta11 & 0.1 & 11 & No\\
    786k\_cells-q0.1-beta12 & 0.1 & 12 & No\\
    786k\_cells-q0.5-beta7 & 0.5 & 7 & Yes\\
    786k\_cells-q0.5-beta8 & 0.5 & 8 & Yes\\
    786k\_cells-q0.5-beta10 & 0.5 & 10 & Yes\\
    786k\_cells-q0.5-beta10.5 & 0.5 & 10.5 & No\\
    786k\_cells-q0.5-beta11 & 0.5 & 11 & No\\
    786k\_cells-q1-beta6 & 1.0 & 6 & Yes\\
    786k\_cells-q1-beta7 & 1.0 & 7 & Yes\\
    786k\_cells-q1-beta8 & 1.0 & 8 & No\\
    786k\_cells-beta3 & 1.41 & 3 & Yes\\
    786k\_cells-beta3.5 & 1.41 & 3.5 & Yes\\
    786k\_cells-beta4 & 1.41 & 4 & Yes\\
    786k\_cells-beta4.5 & 1.41 & 4.5 & Yes\\
    786k\_cells-beta5 & 1.41 & 5 & Yes\\
    786k\_cells-beta5.5 & 1.41 & 5.5 & Yes\\
    786k\_cells-beta6 & 1.41 & 6 & No\\
    786k\_cells-beta10 & 1.41 & 10 & No\\
    786k\_cells-q2-beta4 & 2 & 4 & Yes\\
    786k\_cells-q2-beta5 & 2 & 5 & No\\
    786k\_cells-q2.5-beta3 & 2.5 & 3 & Yes\\
    786k\_cells-q2.5-beta4 & 2.5 & 4 & No\\
   \hline
  \end{tabular}
}
  \caption{Table showing the simulations carried out using {\sc fargo} and the key fragmenting results to test what the effect of changing the amount of artificial viscosity has on the critical cooling timescale.  The artificial viscosity coefficient, $q$, is defined in equation~\ref{eq:visc_press}.}
\label{tab:fargo_av}
\end{table}

\begin{figure}
\centering
  \includegraphics[width=1.0\columnwidth]{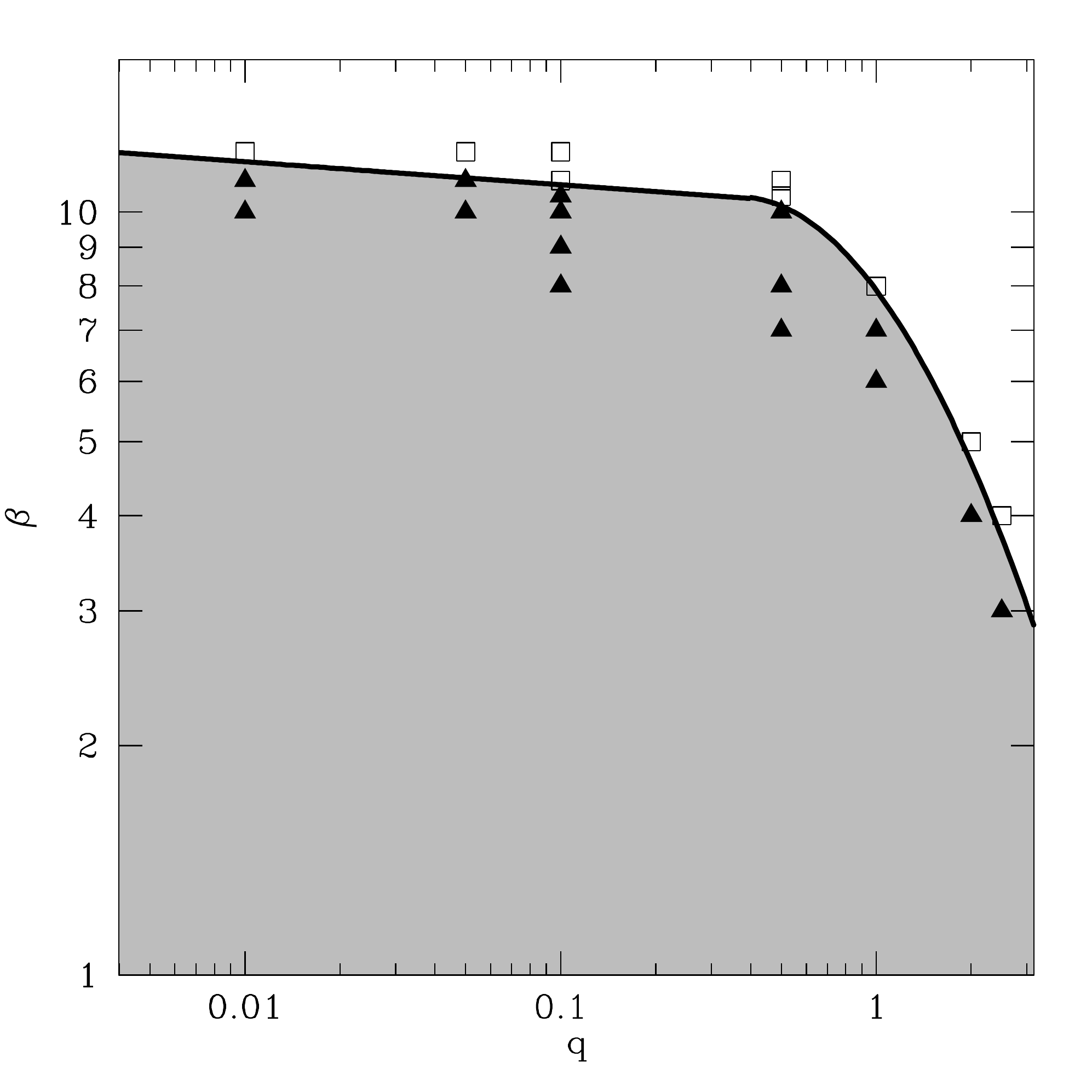}
  \caption{Graph of $\beta$ against the {\sc fargo} artificial viscosity parameter, $q$, of the non-fragmenting (open squares) and fragmenting (solid triangles) simulations carried out using 512 and 1536 cells in the radial and azimuthal directions, respectively.  The solid line, included by eye, shows a dividing line between the fragmenting and non-fragmenting cases and the grey region is where fragmentation can take place.  For low values of $q$, the dissipation due to artificial viscosity is low (and is most likely dominated by the intrinsic numerical diffusion) resulting in fragmentation occurring with high values of $\beta$.  As the artificial viscosity is increased, a faster cooling is required to overcome the additional dissipation, resulting in lower values of $\betacrit$.}
\label{fig:beta_q_fargo}
\end{figure}

In Section~\ref{sec:FARGO_av} we show that the dissipation due to artificial viscosity present in {\sc fargo} may play a part in the critical cooling timescale.  Table~\ref{tab:fargo_av} and Figure~\ref{fig:beta_q_fargo} summarise the results of the simulations carried out to investigate this.  As the artificial viscosity parameter, $q$, is increased, it becomes harder for the disc to fragment due to the extra heating.  Consequently, the critical value of $\beta$ required to overcome this and allow the disc to fragment decreases.

At lower values of $q$, the effect of artificial viscosity on the fragmentation boundary is much less obvious.  Note from Table~\ref{tab:fargo_av} that when the artificial viscosity parameter is set to zero, the fragmentation boundary decreases to a lower value of $\beta$ as with the SPH results in Figure~\ref{fig:beta_alphaSPH} but the effect of reducing the viscosity is much less pronounced in {\sc fargo} than in SPH.  However, the reasoning is likely to be different because there is no dissipation associated with the artificial viscosity term since it is set to zero.  We note from Figure~\ref{fig:beta_q_fargo} that $\betacrit$ increases rapidly as $q$ is decreased to $q \approx 0.5$ and then plateaus - this is most likely because at such low values of the artificial viscosity the dissipation is dominated by intrinsic dissipation in the code.

For fragmentation to occur, the dissipation associated with high values of the artificial viscosity parameter needs to be overcome with a faster cooling.  As with the SPH results presented in Sections~\ref{sec:betaSPH_test} and~\ref{sec:alphaSPH_test}, artificial viscosity clearly plays a part in whether these discs, modelled using a grid-based code, fragment or not.  Figure~\ref{fig:beta_q_fargo} suggests that a value of $q \approx 0.5$ may be sufficient to avoid any excess dissipation.

\subsubsection{Determining the fragmentation boundary using the optimum value of the {\sc fargo} artificial viscosity parameter}

\begin{table}
\centering
  {\small
\begin{tabular}{lllll}
    \hline
    Simulation name & No of & No of & $\beta$ & Fragmented?\\
    & radial cells & azimuthal cells & &\\
   \hline
    \hline
    786k\_cells-q0.5-beta3 & 512 & 1536 & 7 & Yes\\
    786k\_cells-q0.5-beta3.5 & 512 & 1536 & 8 & Yes\\
    786k\_cells-q0.5-beta4 & 512 & 1536 & 10 & Yes\\
    786k\_cells-q0.5-beta4.5 & 512 & 1536 & 10.5 & No\\
    786k\_cells-q0.5-beta5 & 512 & 1536 & 11 & No\\
    3.1m\_cells-q0.5-beta10 & 1024 & 3072 & 14 & Yes\\
    3.1m\_cells-q0.5-beta12 & 1024 & 3072 & 15 & Yes\\
    3.1m\_cells-q0.5-beta13 & 1024 & 3072 & 16 & Yes\\
    3.1m\_cells-q0.5-beta13 & 1024 & 3072 & 18 & Yes\\
    3.1m\_cells-q0.5-beta13 & 1024 & 3072 & 20 & No\\
    13m\_cells-q0.5-beta11 & 2048 & 6144 & 20 & Yes\\
    13m\_cells-q0.5-beta14 & 2048 & 6144 & 22 & Yes\\
    13m\_cells-q0.5-beta14 & 2048 & 6144 & 24 & Yes\\
    13m\_cells-q0.5-beta15 & 2048 & 6144 & 26 & No\\
    50m\_cells-q0.5-beta24 & 4096 & 12288 & 24 & Yes\\
    50m\_cells-q0.5-beta26 & 4096 & 12288 & 26 & Yes\\
    50m\_cells-q0.5-beta28& 4096 & 12288 & 28 & No\\
    50m\_cells-q0.5-beta32 & 4096 & 12288 & 32 & No\\
    \hline
  \end{tabular}
}
  \caption{Table showing the simulations carried out using {\sc fargo} and the key fragmenting results.  The simulations are performed using the artificial viscosity parameter, $q = 0.5$.}
 \label{tab:fargo_sim_q0.5}
\end{table}

\begin{figure}
\centering
  \includegraphics[width=1.0\columnwidth]{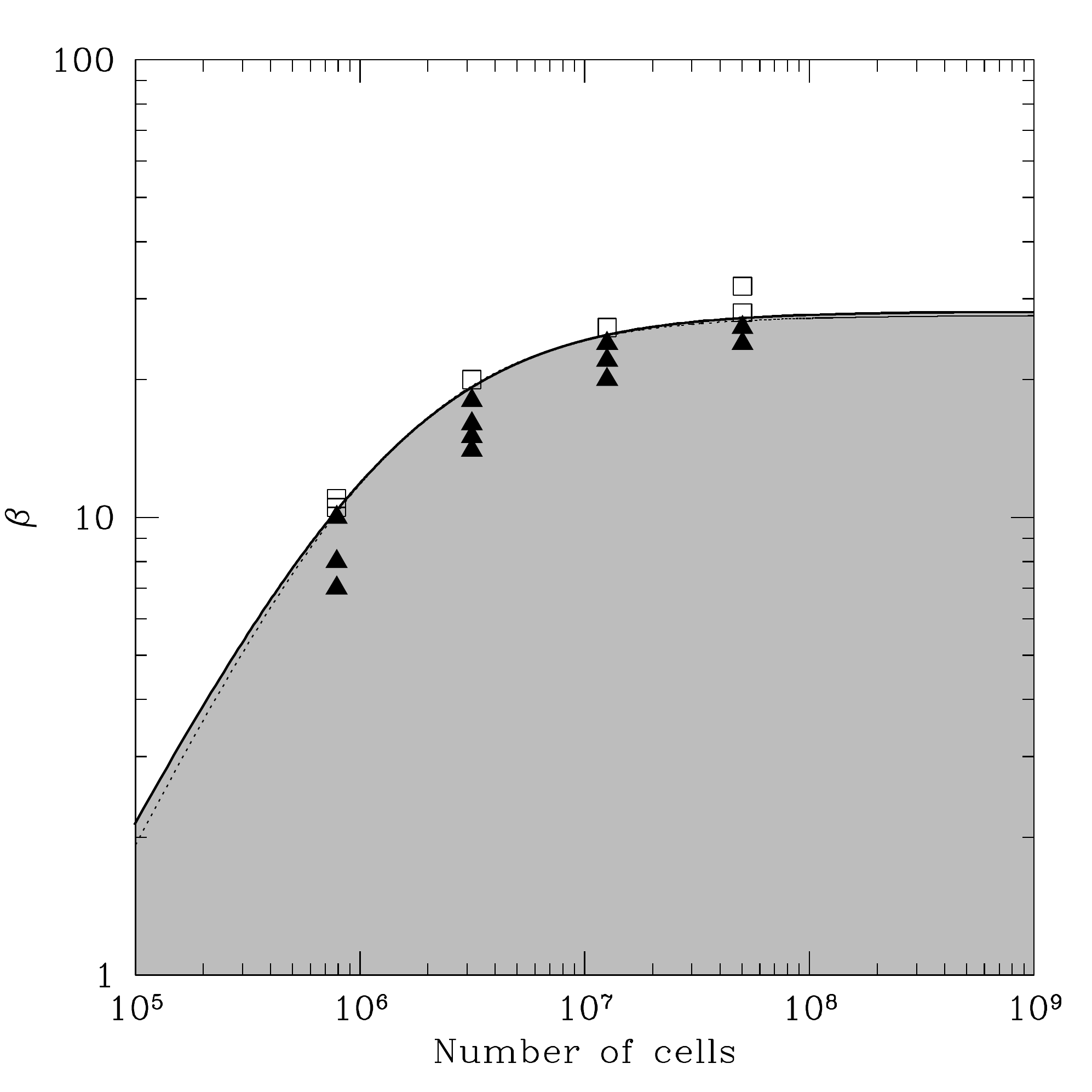}
  \caption{Graph of $\beta$ against resolution of the non-fragmenting (open squares) and fragmenting (solid triangles) {\sc fargo} simulations.  These simulations are carried out with an artificial viscosity parameter, $q = 0.5$.  The solid line, obtained by fitting equation~\ref{eq:fit}, shows a dividing line between the fragmenting and non-fragmenting cases and the grey region is where fragmentation can take place.  The graph shows clear evidence of convergence of results with increased resolution.  The convergence rate is second order with spatial resolution.  The dotted line (which coincides well with the solid line) is obtained by fitting equation~\ref{eq:betacrit_fargo}.}
 \label{fig:res_beta_fargo_q0.5}
\end{figure}

\begin{figure}
\centering
  \includegraphics[width=1.0\columnwidth]{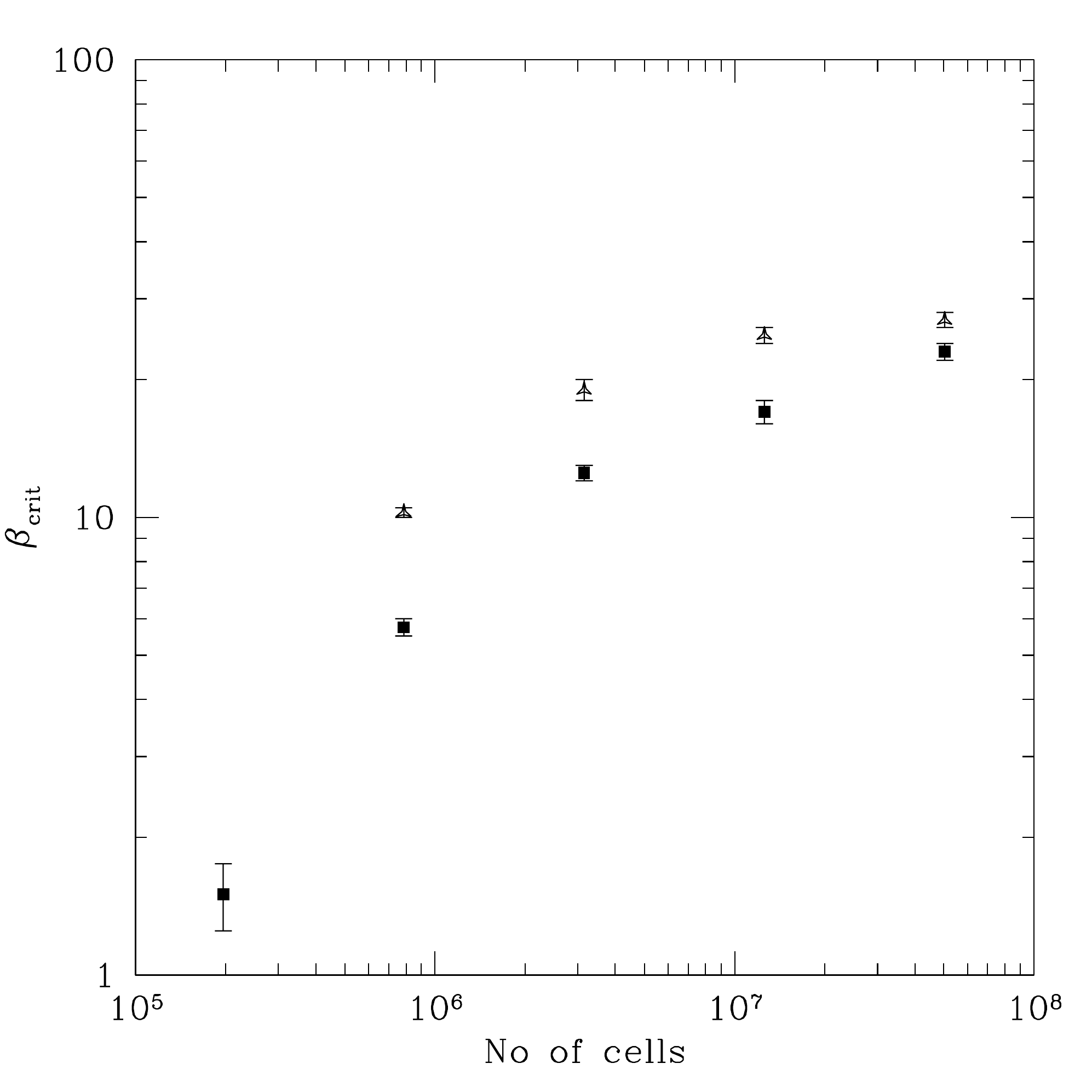}
  \caption{Graph of $\betacrit$ against resolution of the {\sc fargo} simulations carried out with an artificial viscosity parameter, $q = 1.41$ (squares) and $q = 0.5$ (triangles).  It can be seen that the effect of reducing the artificial viscosity parameter to $q = 0.5$ (i.e. to a value that minimises the additional dissipation) is to increase the critical cooling timescale, with the effect being much greater at lower resolution.}
 \label{fig:fargo_q1.41_q0.5_comparison}
\end{figure}

\begin{figure}
\centering
\includegraphics[width=0.85\columnwidth,angle=-90.0]{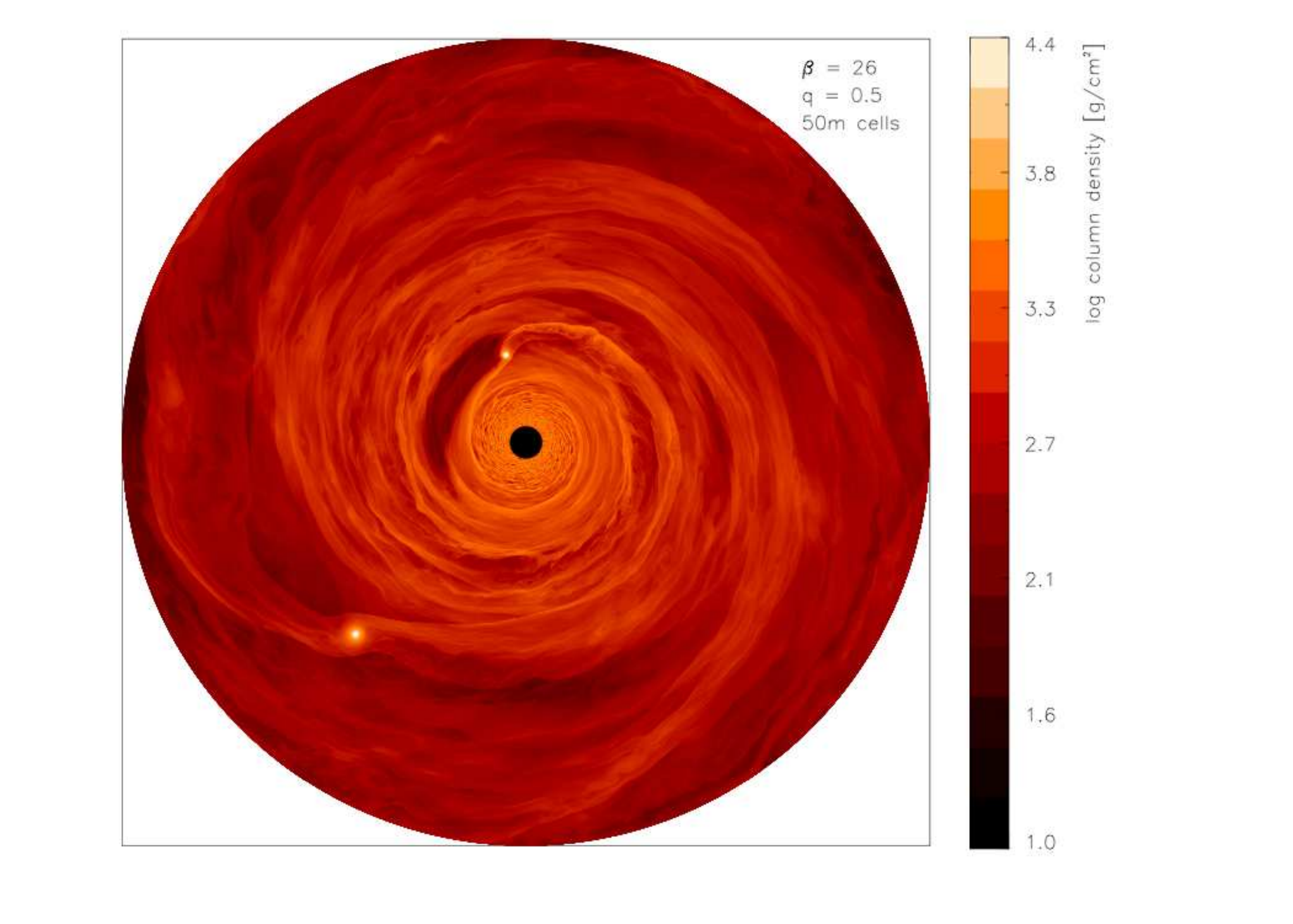}
  \caption{Surface mass density rendered image of a disc modelled using {\sc FARGO} with 50 million grid cells and using $q = 0.5$.  The disc is modelled with a cooling timescale as high as $\beta = 26$ and still fragments.}
\label{fig:disc_fargo_q}
\end{figure}

In Section~\ref{sec:FARGO} we show that convergence appears to be reached at higher resolution with {\sc fargo} and that the convergence is second order in spatial resolution.  However, these simulations use a value of the artificial viscosity parameter, $q = 1.41$, which we show in Section~\ref{sec:FARGO_av0.5} does not minimise the dissipation.  We carry out simulations of self-gravitating discs using a value of $q = 0.5$ at various different resolutions.  Table~\ref{tab:fargo_sim_q0.5} and Figure~\ref{fig:res_beta_fargo_q0.5} summarise the simulations carried out to investigate this, and the key fragmenting results.  It can be seen that the effect of using a lower value of $q$ is that $\betacrit$ is higher than obtained in Section~\ref{sec:FARGO}.  Figure~\ref{fig:disc_fargo_q} shows a surface mass density rendered image of one of the highest resolution discs (modelled using 50 million grid cells) and shows clear fragmentation with a cooling time as high as $\beta = 26$.  However, despite the critical cooling time being larger, we can see from Figure~\ref{fig:res_beta_fargo_q0.5} that convergence is still being achieved.  We firstly fit the data using equation~\ref{eq:fit} and find that $\betacrit = 28.0 \pm 0.2$ and $\sigma = 1.89 \pm 0.05$ showing that the convergence rate is second order with spatial resolution.  We then fit the data using equation~\ref{eq:betacrit_fargo} and find that $\alpha_{\rm GI, crit} = 0.0145 \pm 0.0001$ and $\xi = 3.16 \pm 0.04$.  In the limit of infinite resolution, this value of $\alpha_{\rm GI, crit}$ is equivalent to a critical cooling timescale, $\betacrit \approx 28$ (using equation~\ref{eq:stress}).

Figure~\ref{fig:fargo_q1.41_q0.5_comparison} shows the fragmentation boundary (with error bars) using $q = 1.41$ (as in Section~\ref{sec:FARGO}) and $q = 0.5$.  As the resolution increases, the difference between the two sets of results decreases: since the convergence with {\sc fargo} is fast, i.e. second-order, the effect of using different values of the artificial viscosity parameter (i.e. $q = 0.5$ versus $q = 1.41$) becomes negligible with 50 million grid cells compared to a lower resolution.  This further corroborates that at a higher resolution, the artificial viscosity plays less of a part in the fragmentation boundary.  As suggested by the analytics in Section~\ref{sec:FARGO_av} it is expected that the artificial viscous dissipation should decrease both when the resolution is increased and when $q$ is decreased (until the numerical dissipation becomes dominated by intrinsic grid dissipation).  Indeed, the higher value of $\xi$ obtained here in comparison to that in Section~\ref{sec:FARGO} suggests that the dissipation due to the artificial viscosity has been minimised and that the intrinsic grid dissipation is becoming more important, consistent with Figure~\ref{fig:beta_q_fargo}.  We note that the value $q = 0.5$ effectively means that the shock is spread over approximately half a grid cell which will affect the treatment of shocks.  We emphasise that we choose this value since it gives the lowest artificial heating rate that is possible with {\sc fargo}, as done so with the SPH simulations.  More importantly, we show that the choice of the value of $q$ has much less of an effect at higher resolution than at low resolution.

\section{Discussion}
\label{sec:disc}

The non-convergence of results concerning the fragmentation of self-gravitating discs has opened up a number of questions concerning both the physics and the numerics involved in determining whether a disc will fragment into bound objects.  Consequently, \cite{Meru_Bate_resolution} presented the possibility that either the critical cooling timescale was larger than originally thought, or the extreme possibility that the physics behind the fragmentation of discs needs to be reconsidered.  In this paper we find that using both SPH and the {\sc fargo} codes i.e. a three-dimensional particle-based Lagrangian code and a two-dimensional grid-based Eulerian code, respectively, the artificial viscosity that is used to accurately model shocks plays a significant part in the convergence rate.  Not only do we show that convergence can occur, but we also show that the rate at which it occurs is as expected from analytical arguments involving artificial viscosity (first-order with linear resolution for SPH and second order for {\sc fargo} i.e. a faster convergence with {\sc fargo}).  This affects the results on the fragmentation boundary in both SPH and grid-based calculations.

In particular, we conclude that oscillations at the shock front and particle interpenetration may not have been adequately accounted for in the previous SPH simulations (via the quadratic artificial viscosity term, $\betaSPH$).  Those SPH simulations that used a value of $\betaSPH$ that was too low, may have counterintuitively resulted in more dissipation, causing fragmentation to have been \emph{underestimated}.  After minimising the additional dissipation associated with the artificial viscosity employed in both codes, we find that the critical cooling timescale is at least as high as 20 and perhaps even as much as $\approx 30$, for a ratio of specific heats, $\gamma = 5/3.$

Previous simulations that investigate the effects of artificial viscosity show that the strength of the gravitational instabilities are weakened \citep{Pickett_thermal_AV} and clump formation is reduced \citep{Boss_AV} when artificial viscosity is employed.  These results are in the same sense as our results, i.e. excess effective viscosity reduces the propensity for fragmentation.

\cite{Mayer_AV} carried out a test on the fragmentation and disc evolution resulting from the inclusion of artificial viscosity in their three-dimensional SPH code.  \cite{Pickett_Durisen_AV} carried out a similar test using a three-dimensional grid-based code.  Both sets of authors perform their tests on isothermal simulations i.e. they only include the effects of artificial viscosity in the momentum equation and did not consider its heating effects in the energy equation.  They suggest that artificial viscosity may reduce or even prevent clump formation from occurring.   While our results are consistent with theirs with respect to preventing clump formation when artificial viscosity is increased, we stress that the dissipation associated with the artificial viscosity plays a key role in the fragmentation results.

Another possible numerical parameter that may affect the results is the gravitational softening used in the two-dimensional grid code.  \cite{Muller_smoothing} show that an incorrect value of the gravitational softening length in two-dimensional disc simulations can significantly affect the fragmentation conclusions: a low value causes the gravitational forces on short distances to be over-estimated, resulting in the conclusion that fragmentation does occur, when the converse conclusion is reached for larger values of the softening parameter.  Indeed, it is well known that three-dimensional discs are more stable than two-dimensional discs since the vertical component dilutes the effect of gravity \citep{Toomre_stability1964}.  Thus, incorrectly taking into account the effects of the vertical direction in a two-dimensional simulation may cause the disc to be more prone to fragmentation than its three-dimensional equivalent.  \cite{Muller_smoothing} show that a value of $\approx 0.6 H$ is required to model the gravitational forces correctly (though they do say that a comparison with 3D simulations is required).  Since we are using a softening length of $3 \times 10^{-4} H$, our {\sc fargo} simulations may overestimate fragmentation.

For the SPH simulations, although we see evidence for convergence of the critical cooling timescale, the convergence rate is only first order with increasing resolution.  This is partly due to the larger dissipation than that predicted by the continuum limit of the SPH equations in a shear flow.  We argue that the excess dissipation is due to small-scale particle velocity dispersion.  In non-self-gravitating discs, this results from pressure gradient errors due to discretisation, but in the self-gravitating discs there are other potential sources: primarily particle penetration at shock fronts and post-shock oscillations (particularly when the levels of artificial viscosity are low), but perhaps also discretisation errors in the self-gravity.  To achieve well-behaved dissipation (i.e. that which is close to that predicted by the continuum limit of the SPH equations) requires $\alpha_{\rm SPH}\gsim 1$ and $\beta_{\rm SPH}\gsim 2$. However, the dissipation from the shear flow is then relatively high meaning that even higher resolution would be necessary to obtain a converged value of the critical cooling timescale, $\beta_{\rm crit}$.

However, there are many possibilities that might improve the SPH performance.  We have employed the most basic form of SPH artificial viscosity in this paper (i.e. constant values of $\alpha_{\rm SPH}$ and $\beta_{\rm SPH}$).  An obvious aspect to investigate is whether a viscosity switch such as those proposed by \cite{MorrisMonaghan1997} or \cite{CullenDehnen2010} which increase $\alpha_{\rm SPH}$ and $\beta_{\rm SPH}$ in the presence of a shock and allow them to decay away from shocks can improve the convergence rate.  This could potentially provide high viscosity to avoid particle penetration at shocks and post-shock oscillations (i.e. reducing small-scale particle velocity dispersion), but retain low viscosity (and thus low heating rates) in the bulk of the disc, thus minimising the heating due to the shear flow.  Since some of the particle velocity dispersion originates from pressure gradient errors, another possibility is to try a more accurate kernel.  The quintic spline kernel generally performs better than the cubic spline kernel and \cite{MorrisFoxZhu1997} reported that it significantly reduced velocity field noise in their calculations.  Testing these variants is beyond the scope of this paper, but we expect that these and other SPH variations may be able to significantly improve the performance of SPH on this problem.  We stress that these possible improvements to the SPH convergence rate will not decrease the value of the critical cooling timescale and thus the values obtained in this paper indicate lower limits for $\betacrit$.

Since \cite{Meru_Bate_resolution} published their results highlighting the convergence problem, a number of authors have attempted to explain the non-convergence.  \cite{Lodato_Clarke_resolution} speculated that the cause may be the artificial smoothing of the density enhancements in SPH or a larger than expected level of artificial viscosity.  Our results clearly show that artificial viscosity plays a major role in numerical determinations of the critical cooling rate.

\cite{Paardekooper_convergence} suggested that the boundary between the turbulent inner disc region and the laminar outer disc region (a natural consequence of starting with smooth initial conditions) may cause an edge in the disc that becomes more and more pronounced at higher resolutions, making it easier to fragment.  They suggested that if the smooth initial conditions were removed, convergence could be achieved.  In light of the new results presented in this paper, the effect of edges should be considered in more detail.  If edge effects do play a part, it is unclear whether they should always continue to become sharper at higher resolution (and hence inconsistent with the results presented here), or whether they should ``saturate'' at higher resolution (and thus consistent with these results).  It is important to note, however, that \cite{Bate_disc_frag_res} performed radiative transfer calculations of molecular cloud collapses and found that disc fragmentation is more prevalent in higher resolution calculations.  These discs did not begin with smooth initial conditions, and yet a similar resolution dependence was seen.

More recently, \cite{Paardekooper_shearing_sheet} carried out shearing sheet simulations, similar to those performed by \cite{Gammie_betacool} where no such edge effects should play a part.  He found that as the resolution was increased, the critical cooling timescale also increased, showing that the convergence issue is not restricted to global simulations, but also affects local simulations.  He found fragmentation for at least as large as $\betacrit \approx 20$.  However, this was for simulations carried out with a ratio of specific heats, $\gamma = 2$.  This is equivalent to a maximum gravitational stress as least as small as $\alpha_{\rm GI, crit} \approx 0.011$, consistent with the value of the gravitational stress found using our global simulations.

It is important to note that many of the previous simulations that have attempted to explain the convergence problem highlighted by \cite{Meru_Bate_resolution} have tried to do so by carrying out simulations with resolutions in the \emph{non-convergent} region of the resolution space shown here in this paper.  It is therefore hard to interpret those results since they may have been affected by artificial viscosity.  For those codes that do not use artificial viscosity, other sources of numerical diffusion relating to the specific implementation may be important and the effect of these need to be thoroughly understood.  It would be interesting to try to understand the convergence problem with a Godunov scheme that does not implement an artificial viscosity, or to apply a fixed Navier-Stokes viscosity.

While this work focusses on the effects on fragmentation of self-gravitating discs, the key conclusion that artificial viscosity can play a significant role in the physical interpretation of simulations is more general.  We emphasise that any simulations whose outcome is highly dependent on the thermodynamics of a problem should ensure that the effects of artificial viscosity in their code are well understood as well as highlighting the importance of convergence of results with both resolution and numerical method.

\subsection{Implications for the fragmentation of real discs}
A critical cooling timescale of $\betacrit \approx 20$ or $\approx 30$ is equivalent to a maximum gravitational stress of $\alpha_{\rm GI, crit} \approx 0.02$ or $\approx 0.013$, respectively.  \cite{Clarke2009_analytical} produced an analytical model for the structure of a gravitationally unstable disc which is subject to realistic cooling.  She showed that for optically thick discs that are sufficiently low in temperature that they are dominated by ice grains,

\begin{equation}
  \alpha_{\rm GI} = 0.4~\bigg(\frac{R}{100~\rm{au}}\bigg)^{\frac{9}{2}},
\end{equation}
for a disc with interstellar opacities and surrounding a $1 \Msolar$ star, where $R$ is the radius being considered.  This relationship shows that for a maximum value of the gravitational stress, a critical radius, $R_{\rm crit}$, can be found outside of which fragmentation can occur (for a disc with a shallow surface mass density profile).  While the previously accepted result of $\alpha_{\rm GI,max} \approx 0.06$ gives a critical radius of $R_{\rm crit} \approx 68$~au, the values of $\betacrit$ obtained here moves the critical radius inwards to $R_{\rm crit} \approx 47-51$~au (for a disc around a $1 \Msolar$ star using interstellar opacities).  The core accretion scenario is thought to occur out to $\approx 10$~au, while gravitational instability is historically thought to operate outside of $\approx 70-120$~au \citep{Rafikov_SI,Clarke2009_analytical}.  Therefore, an intermediate radial region exists where no one in situ formation method adequately seems to describe the formation of planets.  Our results show that this gap can at least partly be bridged if the true critical cooling timescale is as much as $\betacrit \approx 20-30$.

We point out that equation~\ref{eq:stress} is derived by assuming that the dominant form of heating in a disc is that due to the gravitational instabilities.  In a real disc, there may be a contribution to the stress from the magnetorotational instability (MRI), $\alpha_{\rm MRI}$.  In this case, equation~\ref{eq:stress} may be written as

\begin{equation}
\beta = \frac{4}{9} \frac{1}{\gamma (\gamma -1)} \frac{1}{(\alpha_{\rm GI} + \alpha_{\rm MRI})}.
\end{equation}
Therefore, while the critical cooling timescale for a purely gravitationally unstable disc is quite large, if the contribution to the gravitational stress from the MRI (or in fact other heating sources) becomes important, a faster cooling will be required to overcome this additional heating and allow the disc to fragment.  Since we find that the critical stress may be as low as $\alpha_{\rm GI, crit} \approx 0.01$, the heating due to MRI will certainly be expected to be important if it provides an effective stress of approximately this level or higher.  Even if $\alpha_{\rm MRI}$ is a factor of 10 smaller, this will still make a 10\% difference to the heating which can change the critical cooling timescale required for fragmentation.

\section{Conclusions}
\label{sec:conc}

We perform hydrodynamical simulations using a three-dimensional Smoothed Particle Hydrodynamics code and a two-dimensional Eulerian grid-based code of self-gravitating discs to investigate how the presence of artificial viscosity may affect fragmentation results.  We present additional SPH results to those presented by \cite{Meru_Bate_resolution} as well as perform similar simulations using the grid-based hydrodynamics code, {\sc fargo}, and show that convergence with resolution of the critical cooling timescale can be achieved with both codes.  We show that the previous non-convergent results are largely due to the effects of artificial viscosity that play a more prominent role at lower resolution.  We find that the convergence rate of the critical cooling timescale required for fragmentation is first order in spatial resolution using SPH and second order using {\sc fargo}.  Furthermore, we find that the dissipation from the artificial viscosities in SPH is exactly as we would expect in a purely laminar disc. However, if random particle motions are present, they can produce dissipation due to artificial viscosity that is larger than expected.  In self-gravitating discs, using a value of the quadratic artificial viscosity term that is too low can result in counterintuitively high dissipation.  This may be caused by additional random particle velocity dispersion due to the presence of shocks, causing the dissipation to significantly deviate from that expected from the SPH continuum limit equations.  In addition, the dissipation due to the $\betaSPH$ term may not be as small as previously assumed and should not be ignored.

We show using analytical arguments and numerical simulations that as the resolution is increased, the artificial viscosity term becomes less important and the rate of convergence is as expected from the analytical arguments.  With the particular setup adopted here, which is the same as that used by \cite{Rice_beta_condition} and \cite{Meru_Bate_resolution}, we find that once the effects of artificial viscosity have been minimised, the critical cooling timescale converges with increasing resolution to a value at least as high as $\betacrit \approx 20$ and perhaps even as high as $\betacrit \approx 30$.  However, a convergence between the two codes has not yet been achieved.  We conclude that this is much more of a problem than had previously been supposed, in part due to the slow convergence rate of SPH and in part due to the enormous resolution required to obtain convergence.  The critical cooling timescale is a factor of $\approx 3-5$ times larger than the value of $\betacrit \approx 6$ that has been used in the past, and is equivalent to a maximum gravitational stress of $\alpha_{\rm GI, crit} \approx 0.013-0.02$ that a disc can handle before it fragments (in contrast to the previously obtained value of $\alpha_{\rm GI, crit} \approx 0.06$).

We show that using values of the artificial viscosity that do not minimise the additional dissipation caused by it can significantly affect fragmentation results and we expect that any other results that sensitively depend on the thermodynamics of a problem, e.g. collapse of AGN discs and molecular clouds into stars may also be affected.  This highlights the importance of ensuring that the artificial viscosity does not play a significant role when carrying out numerical simulations.  We show that fragmentation of self-gravitating discs can be suppressed if the effects of artificial viscosity are not carefully considered.  This suggests that fragmentation of discs into bound objects (e.g. for the formation of planets, binary companions and stars formation in galaxy simulations) is easier than previously thought.

\section*{Acknowledgments}

We thank the referee, Cathie Clarke, for her thorough review and insightful comments.  We thank Daniel Price, Jim Pringle, Neal Turner, Giuseppe Lodato, Ken Rice, Cl\'ement Baruteau, Sijme-Jan Paardekooper and Tom Quinn for helpful discussions.  Most of the calculations reported here were performed using the University of Exeter's SGI Altix ICE 8200 supercomputer.  We gratefully thank the bwGRiD project for some of the computational resources.  bwGRiD (http://www.bw-grid.de) is a member of the German D-Grid initiative, funded by the Ministry for Education and Research (Bundesministerium fuer Bildung und Forschung) and the Ministry for Science, Research and Arts Baden-Wuerttemberg (Ministerium fuer Wissenschaft, Forschung und Kunst Baden-Wuerttemberg).  Some of the calculations reported here were performed using the {\sc brutus} cluster at ETH Z\"urich.  Some of the figures were produced using the publicly available {\sc splash} visualisation software \citep{SPLASH}.  MRB is grateful for the support of a EURYI Award which also funded FM.  This work, conducted as part of the award ``The formation of stars and planets: Radiation hydrodynamical and magnetohydrodynamical simulations" made under the European Heads of Research Councils and European Science Foundation EURYI (European Young Investigator) Awards scheme, was supported by funds from the Participating Organisations of EURYI and the EC Sixth Framework Programme.  FM acknowledges the support of the German Research Foundation (DFG) through grant KL 650/8-2 within the Collaborative Research Group FOR 759: {\it The formation of Planets: The Critical First Growth Phase}.  FM was also supported by the ETH Zurich Postdoctoral Fellowship Programme as well as by the Marie Curie Actions for People COFUND program.

\bibliographystyle{mn2e}
\bibliography{allpapers}

\onecolumn
\appendix
\section{Analytic Derivation of the Shear Viscosity Present in SPH}
\label{appendixA}

In the following sections, for the sake of clarity, we make the simplifying assumptions that the SPH particle smoothing length $h$, the sound speed, $c_{\rm s}$, and the density, $\rho$, are all slowly varying (i.e. are constant).

\subsection{The original SPH artificial viscosity}
\label{sec:original_AV}

The standard SPH artificial viscosity method described by \cite{Monaghan_Gingold_art_vis}, which is a time-independent fixed artificial viscosity, adds the following term to the momentum equation:

\begin{equation}
\frac{{\rm d} {\bf v}_i}{{\rm d} t} = - \sum_{j} {m_j \Pi_{ij} {\bf \nabla}_i W_{ij}}
\label{eq:mom_av}
\end{equation}
where
\begin{equation}
\Pi_{ij} =
\left\{
\begin{array}{l l}
\displaystyle    \frac{- \alpha_{\rm SPH}~c_{{\rm s}}~\mu_{ij}~+~\beta_{\rm SPH}~\mu_{ij}^2}{\rho} & {\bf v}_{ij} \cdot {\bf r}_{ij} < 0 \\[1.25ex]
   0 & {\bf v}_{ij} \cdot {\bf r}_{ij} > 0,
\end{array}
\right.
\label{eq:Pi}
\end{equation}
\begin{equation}
\mu_{ij} = \frac{h {\bf v}_{ij} \cdot {\bf r}_{ij}}{{\bf r}_{ij}^2 + \eta^2},
\label{eq:mu}
\end{equation}
$m_j$ is the mass of particle $j$, $W_{ij}$ is the smoothing kernel adopted, $h$ is the smoothing length and ${\bf v}_{ij}={\bf v}_i - {\bf v}_j$ is the velocity difference between particles $i$ and $j$.  The quantity $\eta^2=0.01h^2$ is included to avoid divergence for small separations between neighbouring particles.  The viscosity involves two terms, the strengths of which are controlled by the parameters $\alphaSPH$ and $\betaSPH$.  Note that the artificial viscosity is only applied when particles approach each other and is turned off when they recede from each other.

It has been recognised for some time that in the continuum limit, the $\alpha_{\rm SPH}$ viscosity term applies both a bulk and a shear viscosity which has the form of a Navier-Stokes type viscosity \citep{Monaghan1985, Pongracic1988,Meglicki_av_continuum}.  In particular, \cite{Meglicki_av_continuum} provide a clear derivation showing that the viscous acceleration due to the $\alpha_{\rm SPH}$ viscosity is given by
\begin{equation}
\label{eq:viscous_force}
\frac{{\rm d}{\bf v}}{{\rm d}t} = \frac{\alpha_{\rm SPH} h \kappa}{2 \rho} \left[ \nabla \cdot (c_s \rho {\bf S})  + \nabla (c_s \rho \nabla \cdot {\bf v}) \right],
\end{equation}
where 
\begin{equation}
S_{ij} = \frac{\partial v^i}{\partial x^j} +  \frac{\partial v^j}{\partial x^i},
\end{equation}
is the deformation tensor.  The first term in equation \ref{eq:viscous_force} is a shear viscosity, while the second term is a bulk viscosity.  The constant, $\kappa$, depends on the number of spatial dimensions and the kernel used by the SPH code.  If the $\alpha_{\rm SPH}$ viscosity is applied in three dimensions and between both approaching and receding particles (unlike equation \ref{eq:Pi}) then (see Appendix~\ref{sec:evaluating})
\begin{equation}
\kappa = - \frac{4 \pi}{15} \int r^3 \frac{{\rm d}W}{{\rm d}r}~{\rm d}r.
\end{equation}
The value of this integral depends on the kernel that is being used.  We use the standard cubic spline kernel
\begin{equation}
W(q,h) = \frac{\sigma}{h^d}
\left\{
\begin{array}{l l}
   1-\frac{3}{2}q^{2}+\frac{3}{4}q^{3} & {\rm for}  \ 0 \leq q <  1, \\[1.25ex]
   \frac{1}{4}(2-q)^{3} & {\rm for} \ 1 \leq q < 2, \\[1.25ex]
   0 & {\rm otherwise},
\end{array}
\right.
\label{eq:kernel_1}
\end{equation}
where $d$ is the number of dimensions, $\sigma$ is the normalisation constant equal to 2/3, $10/(7\pi)$ and $1/\pi$ in one, two and three dimensions respectively, and $q = r/h$.  In this case, the integral has the value $-3/(4\pi)$, such that $\kappa=1/5$.  We find that in general the shear viscosity contribution to the momentum equation in two and three dimensions can be written
\begin{equation}
\frac{{\rm d}{\bf v}}{{\rm d}t}  = \nu \nabla \cdot {\bf S} = \frac{1}{2(2+d)} \alpha_{\rm SPH} c_s h \nabla \cdot {\bf S},
\end{equation}
where $\nu$ is the kinematic shear viscosity.  Thus, for example, when simulating accretion discs, \cite{Artymowicz_Lubow_av} used $\nu = \frac{1}{8} \alpha_{\rm SPH} c_s h$ for their two dimensional SPH simulations, while \cite{Lodato_Price_betaSPH} used $\nu = \frac{1}{10} \alpha_{\rm SPH} c_s h$ for their three dimensional SPH simulations.

However, as expressed in equation \ref{eq:Pi}, it is usual in SPH simulations to only apply the artificial viscous force between approaching particles.  Thus, in general the kinematic shear viscosity in SPH simulations is a factor of two smaller and, for three dimensional calculations, is
\begin{equation}
\nu = \frac{1}{20} \alpha_{\rm SPH} c_s h.
\label{eq:orig_nu}
\end{equation}

\subsection{A more recent variation of SPH artificial viscosity}
\label{sec:recent_AV}

Recently, a slightly different form of artificial viscosity has been applied in SPH codes \citep{Chow_Monaghan_newAV,Price_Monaghan_v_dot_r_AV} where, essentially, equation \ref{eq:mu} is replaced by
\begin{equation}
\mu_{ij} = {\bf v}_{ij} \cdot {\bf \hat{r}}_{ij}.
\label{eq:mu_2}
\end{equation}
This is the form of the artificial viscosity that we use and is constructed in analogy to dissipative terms in Riemann methods.  This form also avoids the arbitrary quantity $\eta^2$ which was included to avoid numerical divergences.  It also means that the magnitude of $\mu_{ij}$ differs by a factor of $r_{ij}/h$ from the original form of the viscosity.  Note that different implementations can differ slightly in the way that the coefficients $\alpha_{\rm SPH}$ and $\beta_{\rm SPH}$ enter the equations.  For example, in some implementations the coefficient of the quadratic viscosity term is actually given by the product of $\alpha$ and $\beta$, while in others the coefficient is fixed to be $2\alpha$.  Some implementations may also differ in the value of $\Pi_{ij}$ by factors of two, so care needs to be taken when evaluating the continuum limit of the viscosity in any particular SPH code.

It can be shown (see Appendix \ref{appendixB}), that when using equations \ref{eq:Pi} and \ref{eq:mu_2} and the standard cubic spline kernel in three dimensions that the shear viscosity is actually 18\% larger than for the original artificial viscosity such that
\begin{equation}
\nu = \frac{31}{525} \alpha_{\rm SPH} c_s h.
\label{eq:new_nu}
\end{equation}

\subsection{The $\beta_{\rm SPH}$ viscosity}

Although many past studies have considered the continuum limit of the linear $\alpha_{\rm SPH}$ viscosity in SPH, to our knowledge, nobody has considered the contribution of the quadratic $\beta_{\rm SPH}$ viscosity.  By inspection of equations~\ref{eq:Pi} and~\ref{eq:mu} we can determine how the shear viscosities due to both the $\alpha_{\rm SPH}$ and $\beta_{\rm SPH}$ should scale.  For a pure shear flow, $\mu_{ij}$ provides an estimate of the shear rate of the fluid multiplied by $h$.  Thus, since the kinematic viscosity is the ratio of the shear stress (given by $\rho \Pi_{\rm ij}$) to the shear rate, we expect the kinematic viscosity due to the $\alpha_{\rm SPH}$ term to scale as
\begin{equation}
\nu_{\alpha} = q_{\alpha} \alphaSPH c_{\rm s} h
\label{eq:nu_alpha}
\end{equation}
where $q_{\alpha}$ is a constant of proportionality.  This is consistent with the analysis given above.  An alternative method of arriving at this equation is to use the fact that from kinetic theory, kinematic viscosity is proportional to the characteristic speed of interchange between particles and to the characteristic distance over which interchange occurs.  In this case, the characteristic speed of particle interchange is given by $c_s$ and the distance over which particles interact is a smoothing length, $h$.

The \cite{Vonneumann_Richtmyer_av} type viscosity (SPH $\beta$-viscosity) is a second order viscosity with the viscous forces depending on the square of the relative speed of the particles.  Therefore, although the characteristic distance over which the viscosity acts is still a smoothing length, the characteristic speed is now the relative speed between particles over a smoothing length.  In a flow directed in the $x$-direction which is sheared in the $y$-direction this is given by

\begin{equation}
h \left( \frac{{\rm d}v_x}{{\rm d}y}\right)
\end{equation}
and thus the $\beta$-viscosity is expected to scale as

\begin{equation}
\nu_{\beta} = q_{\beta} \betaSPH h^2 \left( {{\rm d}v_x \over {\rm d}y} \right)
\label{eq:nu_beta}
\end{equation}
where $q_{\beta}$ is a constant of proportionality.

\section{SPH artificial viscosity in the continuum limit}
\label{appendixB}

One way to evaluate the constants of proportionality, $q_{\alpha}$ and 
$q_{\beta}$, is to use the defining equation for kinematic viscosity.  This can be obtained by 
considering the shearing force produced by a viscous fluid on a plane 
running parallel to the direction of motion of the fluid.  If the fluid 
flows in the $x$-direction and there is a velocity gradient across the flow
in the $y$-direction, then the kinematic viscosity of the fluid is defined by

\begin{equation}
{F \over A} = \nu \rho {{\rm d}v_x \over {\rm d}y}
\label{eq:FoverA}
\end{equation}
which gives the force per unit area exerted on the plane
surface as a function of the kinematic viscosity, $\nu$, the density of
the fluid, $\rho$, and the shear rate of the fluid.  

The force exerted on a volume element of fluid is determined by considering
two planes parallel to the flow.  If the two planes are parallel to the 
$x$-$z$ plane and they are separated by a distance $\delta y$, then the
net force on the fluid element is

\begin{equation}
F_1 - F_2 = \left[{\nu_1 \rho_1 \left({{\rm d}v_x \over {\rm d}y}\right)_1 - \nu_2 \rho_2 \left({{\rm d}v_x \over {\rm d}y}\right)_2 \over \delta y}\right]  \delta x \delta y \delta z
\end{equation}
so that the force per unit volume is given by

\begin{equation}
{F \over V} = \nu \rho {{\rm d}^2v_x \over {\rm d}y^2}
\end{equation}
in the limit that $\delta z \rightarrow 0$, assuming the density and 
kinematic viscosity are constant, which is a valid assumption to make if the two planes are sufficiently close to each other.  The force per unit volume for a fluid element is
simply the acceleration of the fluid element multiplied by its density. Thus,
the equation

\begin{equation}
{{\rm d}v_x \over {\rm d}t} = \nu {{\rm d}^2v_x \over {\rm d}y^2}
\end{equation}
is obtained, which can be compared directly to the SPH momentum equation~\ref{eq:mom_av}.  Using equations~\ref{eq:nu_alpha} and~\ref{eq:nu_beta} we therefore find that the specific force can be expressed as

\begin{equation}
\frac{{\rm d}v_x}{{\rm d}t} \bigg |_{\alphaSPH} = q_{\alpha} \alphaSPH c_{\rm s} h \bigg ( \frac{{\rm d}^2 v_x}{{\rm d}y^2} \bigg )
\label{eq:mom_alphaSPH}
\end{equation}
and

\begin{equation}
\frac{{\rm d}v_x}{{\rm d}t} \bigg |_{\betaSPH} = q_{\beta} \betaSPH h^2 \bigg ( \frac{{\rm d}v_x}{{\rm d}y} \bigg ) \bigg ( \frac{{\rm d}^2 v_x}{{\rm d}y^2} \bigg ),
\label{eq:mom_betaSPH}
\end{equation}
for the $\alphaSPH$ and $\betaSPH$ terms, respectively.

\subsection{Evaluating the constant of proportionality, $q_{\alpha}$, with the original form of the SPH artificial viscosity}
\label{sec:evaluating}

To evaluate the constant of proportionality for the $\alpha_{\rm SPH}$ term in the artificial viscosity, we can derive the continuum limit and compare it to equation~\ref{eq:mom_alphaSPH} to determine the magnitude of $q_\alpha$.  Using equations~\ref{eq:mom_av} -~\ref{eq:mu}, the force per unit mass due to artificial viscosity on particle $i$ is

\begin{align}
\frac{{\rm d}{{\bf v}_i}}{{\rm d}t} &= -
\sum_{j} \frac{m_j}{\rho} \left [ - \alphaSPH h c_{\rm s} 
\frac{{\bf v}_{ij} \cdot {\bf r}_{ij}}{({\bf r}_{ij}^2 + \eta^2)} + \betaSPH h^2 \frac{({\bf v}_{ij} \cdot {\bf r}_{ij})^2}{({\bf r}_{ij}^2 + \eta^2)^2} \right ] \nabla_i W_{ij} \nonumber \\
&\approx \int_{\rm Kernel} \left [\alphaSPH h c_{\rm s} \frac{{\bf v}_{ij} \cdot {\bf r}_{ij}}{{\bf r}_{ij}^2} - \betaSPH h^2 \frac{({\bf v}_{ij} \cdot {\bf r}_{ij})^2}{({\bf r}_{ij}^2)^2} \right ] \nabla_i W_{ij} {\rm d}^3x.
\label{eq:F_full}
\end{align}

The artificial viscosity in the continuum limit due to the linear term ($\alphaSPH$) is,
\begin{equation}
\frac{{\rm d}{{\bf v}_i}}{{\rm d}t} \approx \int_{\rm Kernel} \alphaSPH h c_{\rm s}  \frac{{\bf v}_{ij} \cdot {\bf r}_{ij}}{({\bf r}_{ij}^2 + \eta^2)} \nabla_i W_{ij} {\rm d}^3x.
\label{eq:F_alpha}
\end{equation} 
Following Appendix A of \cite{Meglicki_av_continuum}, we expand ${\bf v}_j$ around ${\bf r}_i$ to give
\begin{equation}
{\bf v}_{ij} \cdot {\bf r}_{ij} = \left [ \Delta x^p \frac{\partial v_i}{\partial x^p} + \frac{\Delta x^p \Delta x^q}{2} \frac{\partial^2 v_i}{\partial x^p \partial x^q} + \frac{\Delta x^p \Delta x^q \Delta x^a}{6} \frac{\partial^3 v_i}{\partial x^p \partial x^q \partial x^a} \right ] \Delta x^r .
\end{equation}
In addition,
\begin{equation}
\nabla_i W_{ij} = -\frac{{\rm d}W_{ij}}{{\rm d}r} \frac{\Delta x^k}{r}
\label{eq:grad_W}
\end{equation}
where $r = |{\bf r}_{ij}| = |r_i - r_j|$.  Inserting equation \ref{eq:grad_W} into the first part of equation~\ref{eq:F_alpha} and retaining terms of order $O(h^4)$, which are the lowest non-vanishing terms, gives
\begin{equation}
\frac{{\rm d}{v^k_i}}{{\rm d}t} \approx - \int_{\rm Kernel} \frac{\alphaSPH h c_s}{r^3} \left[ \frac{\partial v_i^r}{\partial x^p} + \frac{1}{2} \Delta x^q \frac{\partial^2 v_i^r}{\partial x^p \partial x^q} \right] \Delta x^p \Delta x^r \Delta x^k \frac{{\rm d}W_{ij}}{{\rm d}r} {\rm d}^3 x + O(h^6).
\label{eq:Falpha_full}
\end{equation} 
We note that integrating a term with $(\Delta x)^t$, where $t$ is odd, over a symmetric kernel yields a zero result.  Therefore, the result of the integration of the first term in equation~\ref{eq:Falpha_full} is zero.  Simplifying equation~\ref{eq:Falpha_full} gives
\begin{equation}
\frac{{\rm d}{v^k_i}}{{\rm d}t} \approx - \frac{\alphaSPH h c_{\rm s}}{2}  \frac{\partial^2 v_i^r}{\partial x^p \partial x^q}   \int_{\rm Kernel} \Delta x^p \Delta x^r \Delta x^k \Delta x^q \frac{1}{r^3} \frac{{\rm d}W_{ij}}{{\rm d}r} {\rm d}^3 x.
\label{eq:Falpha_simplified}
\end{equation}
The integral is a fourth order symmetric isotropic tensor which can be written in the form
\begin{equation}
- \int_{\rm Kernel} \Delta x^p \Delta x^r \Delta x^k \Delta x^q \frac{1}{r^3} \frac{{\rm d}W_{ij}}{{\rm d}r} {\rm d}^3 x = \kappa_{\alpha} ( \delta_{pq} \delta_{rk} + \delta_{pr} \delta_{qk} + \delta_{pk} \delta_{rq} )
\label{eq:fourth_order}
\end{equation}
where 
\begin{equation}
\kappa_{\alpha} = A \int_{\rm Kernel} r^3 \frac{{\rm d}W_{ij}}{{\rm d}r} ~{\rm d}r.
\label{eq:kappa}
\end{equation}
Equation~\ref{eq:Falpha_simplified} can therefore be written
\begin{equation}
\frac{{\rm d}{v^k_i}}{{\rm d}t} \approx \frac{\kappa_\alpha \alphaSPH h c_{\rm s}}{2} \frac{\partial^2 v_i^r}{\partial x^p \partial x^q}  ( \delta_{pq} \delta_{rk} + \delta_{pr} \delta_{qk} + \delta_{pk} \delta_{rq} ).
\end{equation}
Contracting with the delta terms yields
\begin{equation}
\frac{{\rm d}{v^k_i}}{{\rm d}t}  \approx \frac{\kappa_\alpha \alphaSPH h c_{\rm s}}{2} \left [
\frac{\partial^2 v_i^k}{\partial x^p \partial x^p} +
\frac{\partial^2 v_i^p}{\partial x^p \partial x^k} +
\frac{\partial^2 v_i^r}{\partial x^k \partial x^r}  \right ]
\label{eq:contraction}
\end{equation}
If we now assume that locally there is a constant flow in one direction (e.g. the $x$-direction) with a velocity gradient in an orthogonal direction  (e.g. the $y$-direction) so that $\partial v_i^k / \partial x^k = 0$ and $\partial / \partial x^k (\partial v_i^k / \partial x^p) = 0$, as we would expect for a shear flow, then the above equation can be simplified to give
\begin{equation}
\frac{{\rm d}{v^k_i}}{{\rm d}t}  \approx \frac{\kappa_\alpha \alphaSPH c_{\rm s} h}{2} \left (
\frac{\partial^2 v_i^k}{\partial x^p \partial x^p}
\right )
\label{eq:Falpha_continuum}
\end{equation}
Comparing this to equation~\ref{eq:mom_alphaSPH} yields $\kappa_\alpha / 2 = q_{\alpha}$.  

Equation~\ref{eq:fourth_order} defines $\kappa_\alpha$.  To calculate $\kappa_\alpha$ in three dimensions, we need to sum over all possible combinations of $r$, $p$ and $q$ in equation~\ref{eq:fourth_order}.  The simplest case that yields a non-zero value of the right hand side of equation~\ref{eq:fourth_order} involves $k = r = p = q$ such that equation~\ref{eq:fourth_order} gives

\begin{equation}
- \int_{\rm Kernel} (\Delta x^k)^4 \frac{1}{r^3} \frac{{\rm d}W_{ij}}{{\rm d}r} {\rm d}^3 x = 3 \kappa_{\alpha}
\label{eq:kappa_alpha1}.
\end{equation}
There are three additional cases where $k$ is equivalent to one other letter while the remaining two letters are equal but in an orthogonal direction to $k$.  Since in three dimensions there are two orthogonal directions to $k$, equation~\ref{eq:fourth_order} gives
\begin{equation}
- 6 \int_{\rm Kernel} (\Delta x^k)^2 (\Delta x^q)^2 \frac{1}{r^3} \frac{{\rm d}W_{ij}}{{\rm d}r} {\rm d}^3 x = 6 \kappa_{\alpha}
\label{eq:kappa_alpha2}.
\end{equation}
Summing equations~\ref{eq:kappa_alpha1} and~\ref{eq:kappa_alpha2} together yields
\begin{equation}
- \int_{\rm Kernel} \left [ (\Delta x^k)^4 + 6 (\Delta x^k)^2 (\Delta x^q)^2 ) \right ] \frac{1}{r^3} \frac{{\rm d}W_{ij}}{{\rm d}r} {\rm d}^3 x = 9 \kappa_{\alpha}.
\label{eq:kappa_alpha_sum}
\end{equation}
Without loss of generality, we use $x = r \sin \theta \cos \phi = \Delta x^k$, $y = r \sin \theta \sin \phi = \Delta x^q$, and $d^3x = r^2 \sin \theta ~{\rm d}r ~{\rm d}\theta ~{\rm d}\phi$ and substitute into equation~\ref{eq:kappa_alpha_sum}.  For the $\theta$-component we integrate over $\theta = [0, \pi]$.  However, for the $\phi$-component, care must be taken to integrate over the correct range since we only consider particles that are approaching each other.  Figure~\ref{fig:schematic} shows that in the frame of the particle being considered the ranges $\phi = [0, \pi/2]$ and $\phi = [\pi, (3\pi)/2]$ involve particles approaching each other whereas outside this range particles recede from each other.  Integrating yields

\begin{equation}
\kappa_{\alpha} = - \frac{2 \pi}{15} \int_{\rm Kernel} r^3 \frac{{\rm d}W}{{\rm d}r}~{\rm d}r.
\label{eq:kappa_alpha_full}
\end{equation}
Using the standard cubic spline kernel given above (for three dimensions)
\begin{equation}
\int r^3 \frac{{\rm d}W(q,h)}{{\rm d} r} = -\frac{3}{4 \pi}.
\end{equation}
Therefore, substituting into equation~\ref{eq:kappa_alpha_full} yields $\kappa_\alpha = 1/10$, and so equation \ref{eq:contraction} becomes
\begin{equation}
\frac{{\rm d}v_x}{{\rm d}t} \bigg |_{\alphaSPH} \approx  \frac{1}{20} \alphaSPH c_{\rm s} h \frac{{\rm d}^2 v_x}{{\rm d}y^2}.
\end{equation}
We note that the same constant is achieved if the integration in the $\phi$-direction is done over all space and simply divided by two.

\subsection{Evaluating the constant of proportionality, $q_{\alpha}$, with the recent form of the SPH artificial viscosity}
\label{sec:new_alpha}

For the more recent form of SPH artificial viscosity (equation \ref{eq:mu_2}), the procedure is identical, but the expansion and, thus, the integral is slightly different.  The required expansion is
\begin{equation}
{\bf v}_{ij} \cdot {\bf \hat{r}}_{ij} =  \frac{1}{r} \left [ \Delta x^p \frac{\partial v_i}{\partial x^p} + \frac{\Delta x^p \Delta x^q}{2} \frac{\partial^2 v_i}{\partial x^p \partial x^q} + \frac{\Delta x^p \Delta x^q \Delta x^a}{6} \frac{\partial^3 v_i}{\partial x^p \partial x^q \partial x^a} \right ] \Delta x^r .
\end{equation}
so that equation \ref{eq:Falpha_simplified} becomes
\begin{equation}
\frac{{\rm d}{v^k_i}}{{\rm d}t} \approx - \frac{\alphaSPH c_{\rm s}}{2}  \frac{\partial^2 v_i^r}{\partial x^p \partial x^q}   \int_{\rm Kernel} \Delta x^p \Delta x^r \Delta x^k \Delta x^q \frac{1}{r^2} \frac{{\rm d}W_{ij}}{{\rm d}r} {\rm d}^3 x.
\label{eq:Falpha_simplified_2}
\end{equation}
Note that $h$ is missing from this equation and the integral differs by a factor of $r$.  Thus, the equivalent of equation~\ref{eq:kappa_alpha_full} that needs to be solved is
\begin{equation}
\kappa_{\alpha} = A \int_{\rm Kernel} r^4 \frac{{\rm d}W_{ij}}{{\rm d}r} ~{\rm d}r
\end{equation}
where $A = (-2 \pi)/15$ (ensuring that we account for the fact that the viscosity is only applied between approaching particles), but this time the integral has a value of $-31 h/(35 \pi)$.  Therefore, we obtain
\begin{equation}
\frac{{\rm d}v_x}{{\rm d}t} \bigg |_{\alphaSPH} \approx \frac{31}{525} \alphaSPH c_{\rm s} h \frac{{\rm d}^2 v_x}{{\rm d}y^2},
\end{equation}
which as noted above is approximately 18\% larger than the value obtained for the original form of the artificial viscosity.

\section{The dissipation associated with $\alphaSPH$ and $\betaSPH$}
\label{appendixD}

\begin{figure}
\centering
\includegraphics[width=0.3\columnwidth]{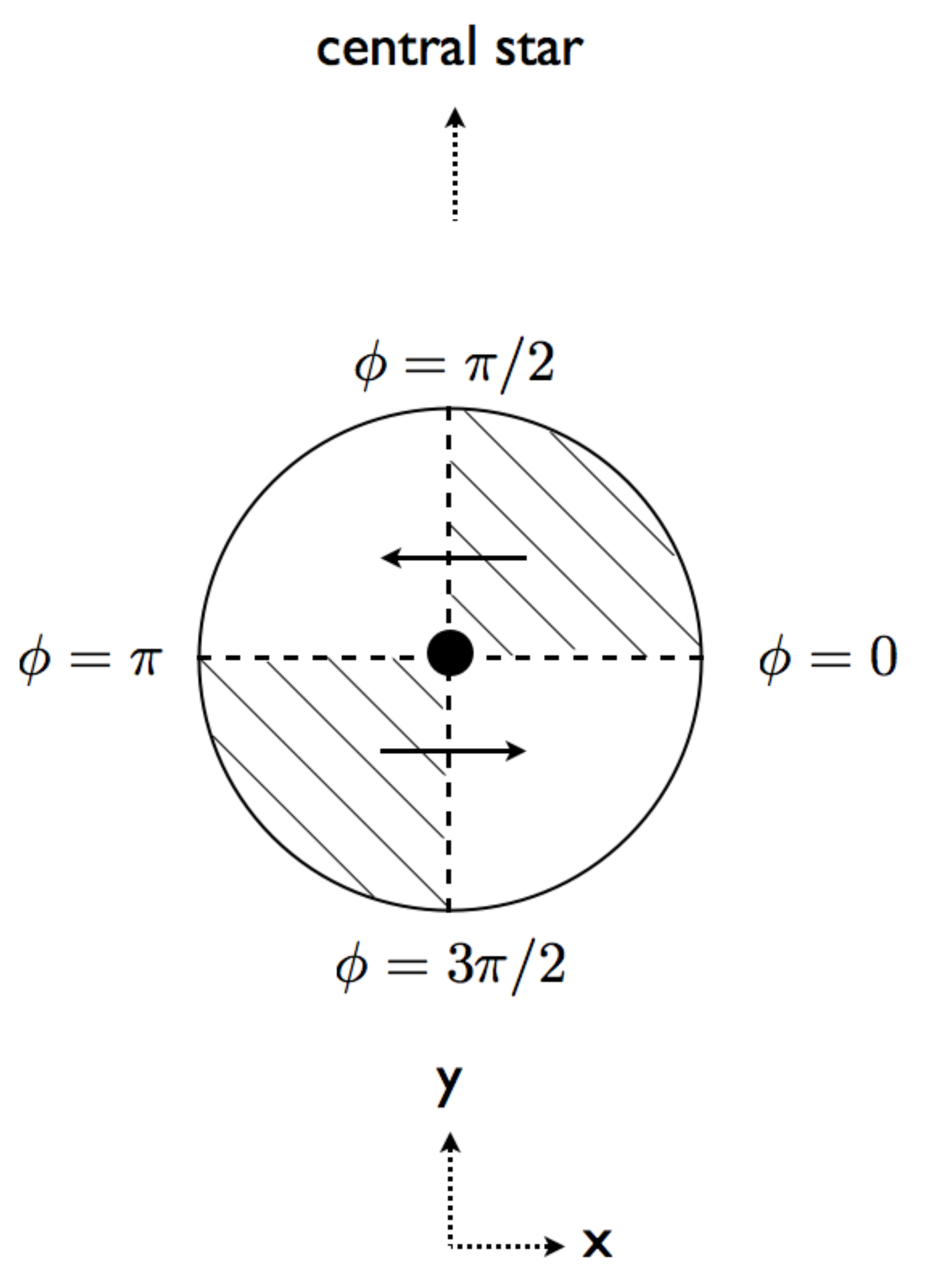}
  \caption{Schematic diagram showing the fluid motion (denoted by solid arrows) surrounding an SPH particle (black dot) in its rest frame, as the gas disc orbits the central star.  In the region of $\phi$-space where $\phi = [0, \pi/2]$ and $\phi = [\pi, (3\pi)/2]$ (shaded region) the fluid is approaching the SPH particle and in this region artificial viscosity is turned on while in all other areas the fluid is receding from the SPH particle and here the artificial viscosity is turned off (see equation~\ref{eq:Pi}).  The origin is the location of the SPH particle.}
\label{fig:schematic}
\end{figure}

For this present paper we are more interested in the magnitude of the thermal dissipation provided by the viscosity than the angular momentum transport as such.  The dissipation rate per unit mass in a viscous accretion disc is given by
\begin{equation}
\frac{{\rm d} u}{{\rm d} t}  = \frac{T_{R\phi}}{\Sigma} \left (R \frac{{\rm d}\Omega}{{\rm d} R} \right ) = \nu R \frac{{\rm d}\Omega}{{\rm d} R} \left ( R \frac{{\rm d}\Omega}{{\rm d} R} \right ) = \frac{9}{4} \nu \Omega^2,
\label{eq:Keplerian_dissipation}
\end{equation}
where $T_{R \phi}$ is the stress tensor and the final equality assumes a Keplerian disc.  In this case the shear rate is
\begin{equation}
\frac{{\rm d}v_x}{{\rm d}y} = R \frac{{\rm d} \Omega}{{\rm d} R} = - \frac{3}{2} \Omega.
\end{equation}

Rather than derive the dissipation rate from the continuum limit of the SPH momentum equation, we can derive the dissipation rate directly from the continuum limit of the SPH energy equation
\begin{equation}
\frac{{\rm d} u_i}{{\rm d} t} = \frac{1}{2} \sum_{j} {m_j \Pi_{ij} {\bf v}_{ij} \cdot {\bf \nabla}_i W_{ij}}.
\end{equation}
Taking the continuum limit of this equation and using the more recent form of the artificial viscosity gives
\begin{equation}
\frac{{\rm d} u}{{\rm d}t} \approx \frac{1}{2} \int_{\rm Kernel} \left [ - \alphaSPH c_{\rm s} {\bf v} \cdot {\bf \hat{r}} + \betaSPH ({\bf v} \cdot {\bf \hat{r}})^2 \right ] {\bf v} \cdot  \hat{\bf r}~ \frac{{\rm d}W}{{\rm d}r}~ {\rm d}^3x.
\label{eq:cont_diss}
\end{equation}

The simplest way to obtain the dissipation rate due to the artificial viscosity in a Keplerian disc is to consider a small patch of the disc around a particle such the the local velocity field is given by
\begin{equation}
v_x \approx - \frac{3 y \Omega_0}{2},
\end{equation}
where $\Omega_0$ is the angular velocity of the fluid at the radius being considered and $y$ is the displacement in the inward radial direction (see Figure~\ref{fig:schematic}).  Inserting this into equation \ref{eq:cont_diss} and taking $x=r\sin\theta\cos\phi$ and $y=r\sin\theta\sin\phi$, we obtain
\begin{equation}
\frac{{\rm d} u}{{\rm d}t} \approx \frac{9}{4} \left [\frac{31}{525} \alpha_{\rm SPH} c_{\rm s} h \Omega_0^2 + \frac{9}{70 \pi} \beta_{\rm SPH} h^2  \Omega_0^3 \right],
\label{eq:orig_dissipation}
\end{equation}
taking care to integrate only over the ranges $\phi=[0,\pi/2]$ and $\phi=[\pi,3\pi/2]$ (i.e. where the flow is approaching and not receding; see Figure~\ref{fig:schematic}).

For the original form of the artificial viscosity, the dissipation is
\begin{equation}
\frac{{\rm d} u}{{\rm d}t} \approx \frac{9}{4} \left [\frac{1}{20} \alpha_{\rm SPH} c_{\rm s} h \Omega_0^2 + \frac{3}{35 \pi} \beta_{\rm SPH} h^2  \Omega_0^3 \right],
\label{eq:new_dissipation}
\end{equation}
again only taking the integral over the regions where the flow is approaching.

Note that the coefficients preceding $\alpha_{\rm SPH}$ in equations \ref{eq:orig_dissipation} and \ref{eq:new_dissipation} are the same as the coefficients appearing in the the kinematic viscosity as given by equations \ref{eq:orig_nu} and \ref{eq:new_nu}, respectively, as is expected from equation \ref{eq:Keplerian_dissipation}.

Using $H = c_s/\Omega$, the ratio of the dissipation rates associated with the linear and quadratic artificial viscosity terms using the recent form of the SPH artificial viscosity is given by
\begin{equation}
\frac{D_{\alpha}}{D_{\beta}} = \frac{62 \pi}{135} \frac{\alphaSPH}{\betaSPH} \frac{H}{h}.
\end{equation}
Typical values of $\alphaSPH$ and $\betaSPH$ are frequently within a factor of 2 of each other (with $\betaSPH = 2 \alphaSPH$).  Therefore, the important variable that determines the relative magnitude of the dissipation associated with the linear and quadratic SPH artificial viscosity terms is the ratio of the smoothing length to disc scaleheight, $h/H$.  In a well resolved disc, $h/H \ll 1$ such that $D_{\alpha} \gg D_{\beta}$.  However, if the disc is poorly resolved and/or $\beta_{\rm SPH} \gg \alpha_{\rm SPH}$ the dissipation associated with the quadratic viscosity may be significant. Finally, note that the above equations assume that the only viscous dissipation comes from the shear in a purely Keplerian disc.

\end{document}